\documentclass[11pt]{article} 
\usepackage[paper=a4paper,margin=1.0in]{geometry}

\usepackage[utf8]{inputenc} 
\usepackage{enumitem}
\usepackage{graphicx} 
\usepackage{mathrsfs}
\usepackage{amsmath,amsthm,amssymb}
\usepackage[italicdiff]{physics}
\usepackage{dsfont}
\usepackage{subcaption}
\usepackage{slashed}
\usepackage{framed}
\usepackage{color}
\usepackage{tcolorbox}
\usepackage{simpler-wick}
\usepackage{etoolbox}
\usepackage{float}
\usepackage{tikz}
\usepackage[compat=1.1.0]{tikz-feynman}
\usetikzlibrary{positioning}
\usepackage{cancel}
\usepackage[normalem]{ulem} 
\usepackage[colorlinks=true,linkcolor=blue,citecolor=red]{hyperref}

\numberwithin{equation}{section}

\newcommand{\fdiag}[1]{%
\mathord{\vcenter{\hbox{\scalebox{0.60}{#1}}}}%
}

\newcommand{\DiagramOneLHS}{%
\begin{tikzpicture}[baseline={(current bounding box.center)}]
	\begin{feynman}
		\vertex (a) ;
		\vertex [below right=2em and 2em of a] (b) ;
		\vertex [above right=2em and 2em of b] (c) ;
            \vertex [below left=0.5em and 0.5em of a] (d) ;
            \vertex [below right=2.1em and 2.1em of d] (e);
            \vertex [below=2em of e] (f);
            \vertex [right=0.8em of f] (g);
            \vertex [above=2em of g] (h);
            \vertex [above right=2em and 2em of h] (i);
            \vertex [above right=0.1em and -0.3em of i] (k) {\(\nu\)};
            \vertex [below right=0.1em and -0.3em of f] (j) {\(\rho\)};
            \vertex [above left=-0.6em and 0.1em of a] (m) {\(\mu\)};
		\diagram{
            {[edges=plain]
        (a) -- (b) -- (c),
        (d) -- (e),
        (h) -- (i),
        (g) -- (h),
        (e) -- (f),
        }
  };
\end{feynman}
\end{tikzpicture}%
}

\newcommand{\DiagramOneRHSa}{%
\begin{tikzpicture}[baseline={(current bounding box.center)}]
	\begin{feynman}
		\vertex (a) {$a_1$};
		\vertex [below right=2em and 2em of a] (b) ;
		\vertex [above right=1.5em and 1.5em of b] (c) {$b_2$};
            \vertex [below left=0.5em and 0.5em of a] (d) {$b_1$};
            \vertex [below right=2.1em and 2.1em of d] (e);
            \vertex [below=2em of e] (f) {$a_3$};
            \vertex [right=0.9em of f] (g) {$b_3$};
            \vertex [above=2.6em of g] (h);
            \vertex [above right=1.3em and 1.3em of h] (i) {$a_2$};
		\diagram{
            {[edges=fermion]
        (a) -- (b) -- (c),
        (e) -- (d),
        (i) -- (h),
        (h) -- (g),
        (f) -- (e),
        }
  };
\end{feynman}
\end{tikzpicture}%
}

\newcommand{\DiagramOneRHSb}{%
\begin{tikzpicture}[baseline={(current bounding box.center)}]
	\begin{feynman}
		\vertex (a) {$b_1$};
		\vertex [below right=2em and 2em of a] (b) ;
		\vertex [above right=1.3em and 1.3em of b] (c) {$a_2$};
            \vertex [below left=0.5em and 0.5em of a] (d) {$a_1$};
            \vertex [below right=2.1em and 2.1em of d] (e);
            \vertex [below=2em of e] (f) {$b_3$};
            \vertex [right=0.9em of f] (g) {$a_3$};
            \vertex [above=2.6em of g] (h);
            \vertex [above right=1.35em and 1.35em of h] (i) {$b_2$};
		\diagram{
            {[edges=fermion]
        (c) -- (b) -- (a),
        (d) -- (e),
        (h) -- (i),
        (g) -- (h),
        (e) -- (f),
        }
  };
\end{feynman}
\end{tikzpicture}
}

\newcommand{\DiagramTwoLHS}{%
\begin{tikzpicture}[baseline={(current bounding box.center)}]
	\begin{feynman}
		\vertex (a) ;
		\vertex [below right=2em and 2em of a] (b) ;
		\vertex [above right=2em and 2em of b] (c) ;
            \vertex [below left=0.5em and 0.5em of a] (d) ;
            \vertex [below right=2.1em and 2.1em of d] (e);
            \vertex [below=2em of e] (f);
            \vertex [right=0.8em of f] (g);
            \vertex [above=2em of g] (h);
            \vertex [above right=2em and 2em of h] (i);
            \vertex [below right=0.1em and -0.3em of f] (j) {\(\mu\)};
		\diagram{
            {[edges=scalar]
        (a) -- (b) -- (c),
        (d) -- (e),
        (h) -- (i),
        (g) -- [plain] (h),
        (e) -- [plain] (f),
        }
  };
\end{feynman}
\end{tikzpicture}%
}

\newcommand{\DiagramTwoRHSa}{%
\begin{tikzpicture}[baseline={(current bounding box.center)}]
	\begin{feynman}
		\vertex (a) {$a_1$};
		\vertex [below right=2em and 2em of a] (b) ;
		\vertex [above right=1.5em and 1.5em of b] (c) {$b_2$};
            \vertex [below left=0.5em and 0.5em of a] (d) {$b_1$};
            \vertex [below right=2.1em and 2.1em of d] (e);
            \vertex [below=2em of e] (f) {$a_3$};
            \vertex [right=0.9em of f] (g) {$b_3$};
            \vertex [above=2.6em of g] (h);
            \vertex [above right=1.3em and 1.3em of h] (i) {$a_2$};
		\diagram{
            {[edges=charged scalar]
        (a) -- (b) -- (c),
        (e) -- (d),
        (i) -- (h),
        (h) -- [fermion] (g),
        (f) -- [fermion] (e),
        }
  };
\end{feynman}
\end{tikzpicture}%
}

\newcommand{\DiagramTwoRHSb}{%
\begin{tikzpicture}[baseline={(current bounding box.center)}]
	\begin{feynman}
		\vertex (a) {$b_1$};
		\vertex [below right=2em and 2em of a] (b) ;
		\vertex [above right=1.3em and 1.3em of b] (c) {$a_2$};
            \vertex [below left=0.5em and 0.5em of a] (d) {$a_1$};
            \vertex [below right=2.1em and 2.1em of d] (e);
            \vertex [below=2em of e] (f) {$b_3$};
            \vertex [right=0.9em of f] (g) {$a_3$};
            \vertex [above=2.6em of g] (h);
            \vertex [above right=1.35em and 1.35em of h] (i) {$b_2$};
		\diagram{
            {[edges=charged scalar]
        (c) -- (b) -- (a),
        (d) -- (e),
        (h) -- (i),
        (g) -- [fermion] (h),
        (e) -- [fermion] (f),
        }
  };
\end{feynman}
\end{tikzpicture}%
}

\newcommand{\DiagramThreeLHS}{%
\begin{tikzpicture}[baseline={(current bounding box.center)}]
	\begin{feynman}
		\vertex (a) ;
		\vertex [below right=2em and 2em of a] (b) ;
		\vertex [below left=2em and 2em of b] (c) ;
            \vertex [below right=0.5em and 0.5em of c] (d) ;
            \vertex [above right=2em and 2em of d] (e);
            \vertex [below right=2em and 2em of e] (f);
            \vertex [above right=0.5em and 0.5em of f] (g);
            \vertex [above left=2em and 2em of g] (h);
            \vertex [above right=2em and 2em of h] (i);
            \vertex [above left=0.5em and 0.5em of i] (j) ;
            \vertex [below left=2em and 2em of j] (k);
            \vertex [above left=2em and 2em of k] (l);
            \vertex [below left=-1.2em and -0.5em of a] (m) {\(\mu\)};
            \vertex [above left=-1.2em and -0.5em of c] (n) {\(\sigma\)};
            \vertex [below right=-0.5em and -0.2em of f] (o) {\(\rho\)};
            \vertex [below right=-0.6em and -0.1em of j] (p) {\(\nu\)};
		\diagram{
            {[edges=plain]
        (a) -- (b) -- (c),
        (d) -- (e) -- (f),
        (g) -- (h) -- (i),
        (j) -- (k) -- (l),
      }
  };
	\end{feynman}
\end{tikzpicture}%
}

\newcommand{\DiagramThreeRHSa}{%
\begin{tikzpicture}[baseline={(current bounding box.center)}]
	\begin{feynman}
		\vertex (a) {$b_1$};
		\vertex [below right=2em and 2em of a] (b) ;
		\vertex [below left=1.5em and 1.5em of b] (c) {$a_4$};
            \vertex [below right=0.65em and 0.65em of c] (d) {$b_4$};
            \vertex [above right=2.2em and 2.2em of d] (e) ;
            \vertex [below right=1.5em and 1.5em of e] (f) {$a_3$};
            \vertex [above right=0.5em and 0.5em of f] (g) {$b_3$};
            \vertex [above left=2.2em and 2.2em of g] (h);
            \vertex [above right=1.5em and 1.5em of h] (i) {$a_2$};
            \vertex [above left=0.65em and 0.65em of i] (j) {$b_2$};
            \vertex [below left=2.2em and 2.2em of j] (k);
            \vertex [above left=1.5em and 1.5em of k] (l) {$a_1$};
		\diagram{
            {[edges=fermion]
        (c) -- (b) -- (a),
        (f) -- (e) -- (d),
        (i) -- (h) -- (g),
        (l) -- (k) -- (j),
      }
  };
	\end{feynman}
\end{tikzpicture}%
}

\newcommand{\DiagramThreeRHSb}{%
\begin{tikzpicture}[baseline={(current bounding box.center)}]
	\begin{feynman}
		\vertex (a) {$a_1$};
		\vertex [below right=2em and 2em of a] (b) ;
		\vertex [below left=1.5em and 1.5em of b] (c) {$b_4$};
            \vertex [below right=0.65em and 0.65em of c] (d) {$a_4$};
            \vertex [above right=2.2em and 2.2em of d] (e) ;
            \vertex [below right=1.5em and 1.5em of e] (f) {$b_3$};
            \vertex [above right=0.5em and 0.5em of f] (g) {$a_3$};
            \vertex [above left=2.2em and 2.2em of g] (h);
            \vertex [above right=1.5em and 1.5em of h] (i) {$b_2$};
            \vertex [above left=0.65em and 0.65em of i] (j) {$a_2$};
            \vertex [below left=2.2em and 2.2em of j] (k);
            \vertex [above left=1.5em and 1.5em of k] (l) {$b_1$};
		\diagram{
            {[edges=fermion]
        (a) -- (b) -- (c),
        (d) -- (e) -- (f),
        (g) -- (h) -- (i),
        (j) -- (k) -- (l),
      }
  };
	\end{feynman}
\end{tikzpicture}%
}

\newcommand{\DiagramFourLHS}{%
\begin{tikzpicture}[baseline={(current bounding box.center)}]
	\begin{feynman}
		\vertex (a) ;
		\vertex [below right=2em and 2em of a] (b) ;
		\vertex [below left=2em and 2em of b] (c) ;
            \vertex [below right=0.5em and 0.5em of c] (d) ;
            \vertex [above right=2em and 2em of d] (e);
            \vertex [below right=2em and 2em of e] (f);
            \vertex [above right=0.5em and 0.5em of f] (g);
            \vertex [above left=2em and 2em of g] (h);
            \vertex [above right=2em and 2em of h] (i);
            \vertex [above left=0.5em and 0.5em of i] (j) ;
            \vertex [below left=2em and 2em of j] (k);
            \vertex [above left=2em and 2em of k] (l);
            \vertex [below left=-1.2em and -0.5em of a] (m) ;
            \vertex [above left=-1.2em and -0.5em of c] (n) {\(\nu\)};
            \vertex [below right=-0.5em and -0.2em of f] (o) ;
            \vertex [below right=-0.6em and -0.1em of j] (p) {$\mu$};
		\diagram{
            {[edges=scalar]
        (a) -- (b) --[plain] (c),
        (d) --[plain] (e) -- (f),
        (g) -- (h) --[plain] (i),
        (j) --[plain] (k) -- (l),
      }
  };
	\end{feynman}
\end{tikzpicture}%
}

\newcommand{\DiagramFourRHSa}{%
\begin{tikzpicture}[baseline={(current bounding box.center)}]
	\begin{feynman}
		\vertex (a) {$b_1$};
		\vertex [below right=2em and 2em of a] (b) ;
		\vertex [below left=1.5em and 1.5em of b] (c) {$a_4$};
            \vertex [below right=0.65em and 0.65em of c] (d) {$b_4$};
            \vertex [above right=2.2em and 2.2em of d] (e) ;
            \vertex [below right=1.5em and 1.5em of e] (f) {$a_3$};
            \vertex [above right=0.5em and 0.5em of f] (g) {$b_3$};
            \vertex [above left=2.2em and 2.2em of g] (h);
            \vertex [above right=1.5em and 1.5em of h] (i) {$a_2$};
            \vertex [above left=0.65em and 0.65em of i] (j) {$b_2$};
            \vertex [below left=2.2em and 2.2em of j] (k);
            \vertex [above left=1.5em and 1.5em of k] (l) {$a_1$};
		\diagram{
            {[edges=charged scalar]
        (c) --[fermion] (b) -- (a),
        (f) -- (e) --[fermion] (d),
        (i) --[fermion] (h) -- (g),
        (l) -- (k) --[fermion] (j),
      }
  };
	\end{feynman}
\end{tikzpicture}%
}

\newcommand{\DiagramFourRHSb}{%
\begin{tikzpicture}[baseline={(current bounding box.center)}]
	\begin{feynman}
		\vertex (a) {$a_1$};
		\vertex [below right=2em and 2em of a] (b) ;
		\vertex [below left=1.5em and 1.5em of b] (c) {$b_4$};
            \vertex [below right=0.65em and 0.65em of c] (d) {$a_4$};
            \vertex [above right=2.2em and 2.2em of d] (e) ;
            \vertex [below right=1.5em and 1.5em of e] (f) {$b_3$};
            \vertex [above right=0.5em and 0.5em of f] (g) {$a_3$};
            \vertex [above left=2.2em and 2.2em of g] (h);
            \vertex [above right=1.5em and 1.5em of h] (i) {$b_2$};
            \vertex [above left=0.65em and 0.65em of i] (j) {$a_2$};
            \vertex [below left=2.2em and 2.2em of j] (k);
            \vertex [above left=1.5em and 1.5em of k] (l) {$b_1$};
		\diagram{
            {[edges=charged scalar]
        (a) -- (b) --[fermion] (c),
        (d) --[fermion] (e) -- (f),
        (g) -- (h) --[fermion] (i),
        (j) --[fermion] (k) -- (l),
      }
  };
	\end{feynman}
\end{tikzpicture}
}

\newcommand{\DiagramFiveLHS}{%
\begin{tikzpicture}[baseline={(current bounding box.center)}]
	\begin{feynman}
		\vertex (a) ;
		\vertex [below right=2em and 2em of a] (b) ;
		\vertex [below left=2em and 2em of b] (c) ;
            \vertex [below right=0.5em and 0.5em of c] (d) ;
            \vertex [above right=2em and 2em of d] (e);
            \vertex [below right=2em and 2em of e] (f);
            \vertex [above right=0.5em and 0.5em of f] (g);
            \vertex [above left=2em and 2em of g] (h);
            \vertex [above right=2em and 2em of h] (i);
            \vertex [above left=0.5em and 0.5em of i] (j) ;
            \vertex [below left=2em and 2em of j] (k);
            \vertex [above left=2em and 2em of k] (l);
            \vertex [below left=-1.2em and -0.5em of a] (m) ;
            \vertex [above left=-1.2em and -0.5em of c] (n) {\(\nu\)};
            \vertex [below right=-0.5em and -0.2em of f] (o) {\(\mu\)};
            \vertex [below right=-0.6em and -0.1em of j] (p) ;
		\diagram{
            {[edges=scalar]
        (a) -- (b) --[plain] (c),
        (d) --[plain] (e) --[plain] (f),
        (g) --[plain] (h) -- (i),
        (j) -- (k) -- (l),
      }
  };
	\end{feynman}
\end{tikzpicture}%
}

\newcommand{\DiagramFiveRHSa}{%
\begin{tikzpicture}[baseline={(current bounding box.center)}]
	\begin{feynman}
		\vertex (a) {$b_1$};
		\vertex [below right=2em and 2em of a] (b) ;
		\vertex [below left=1.5em and 1.5em of b] (c) {$a_4$};
            \vertex [below right=0.65em and 0.65em of c] (d) {$b_4$};
            \vertex [above right=2.2em and 2.2em of d] (e) ;
            \vertex [below right=1.5em and 1.5em of e] (f) {$a_3$};
            \vertex [above right=0.5em and 0.5em of f] (g) {$b_3$};
            \vertex [above left=2.2em and 2.2em of g] (h);
            \vertex [above right=1.5em and 1.5em of h] (i) {$a_2$};
            \vertex [above left=0.65em and 0.65em of i] (j) {$b_2$};
            \vertex [below left=2.2em and 2.2em of j] (k);
            \vertex [above left=1.5em and 1.5em of k] (l) {$a_1$};
		\diagram{
            {[edges=charged scalar]
        (c) --[fermion] (b) -- (a),
        (f) --[fermion] (e) --[fermion] (d),
        (i) -- (h) --[fermion] (g),
        (l) -- (k) -- (j),
      }
  };
	\end{feynman}
\end{tikzpicture}%
}

\newcommand{\DiagramFiveRHSb}{%
\begin{tikzpicture}[baseline={(current bounding box.center)}]
	\begin{feynman}
		\vertex (a) {$a_1$};
		\vertex [below right=2em and 2em of a] (b) ;
		\vertex [below left=1.5em and 1.5em of b] (c) {$b_4$};
            \vertex [below right=0.65em and 0.65em of c] (d) {$a_4$};
            \vertex [above right=2.2em and 2.2em of d] (e) ;
            \vertex [below right=1.5em and 1.5em of e] (f) {$b_3$};
            \vertex [above right=0.5em and 0.5em of f] (g) {$a_3$};
            \vertex [above left=2.2em and 2.2em of g] (h);
            \vertex [above right=1.5em and 1.5em of h] (i) {$b_2$};
            \vertex [above left=0.65em and 0.65em of i] (j) {$a_2$};
            \vertex [below left=2.2em and 2.2em of j] (k);
            \vertex [above left=1.5em and 1.5em of k] (l) {$b_1$};
		\diagram{
            {[edges=charged scalar]
        (a) -- (b) --[fermion] (c),
        (d) --[fermion] (e) --[fermion] (f),
        (g) --[fermion] (h) -- (i),
        (j) -- (k) -- (l),
      }
  };
	\end{feynman}
\end{tikzpicture}%
}
\begin{document}

\title{Nonperturbative Stabilization of D-Instantons \\in the Bosonic IIB Matrix Model}

\author{
    Yu-An Chen$^{a}$\thanks{b09202059@ntu.edu.tw},~
    Hikaru Kawai$^{b,c}$\thanks{kawai.hikaru0@gmail.com},~
    Henry Liao$^{a}$\thanks{henryliao.physics@gmail.com}~
    and Cheng-Tsung Wang$^{d,e}$\thanks{ctwang@post.kek.jp}
    \\[\smallskipamount]
    \small ${}^a$\emph{Department of Physics and Center for Theoretical Physics,}\\
    \small \emph{National Taiwan University, Taipei 106, Taiwan}\\
    \small ${}^b$\emph{Nambu Yoichiro Institute of Theoretical and Experimental Physics (NITEP),}\\
    \small \emph{Osaka Metropolitan University, 3-3-138, Sugimoto, Sumiyoshi-ku, Osaka, 558-8585, Japan}\\
    \small ${}^c$\emph{Department of Physics, National Sun Yat-Sen University, Kaohsiung 80424, Taiwan}\\
    \small ${}^d$\emph{KEK Theory Center, Institute of Particle and Nuclear Studies,}\\
    \small \emph{High Energy Accelerator Research Organization,}\\
    \small \emph{1-1 Oho, Tsukuba, Ibaraki 305-0801, Japan}\\
    \small ${}^e$\emph{Graduate Institute for Advanced Studies, SOKENDAI,}\\
    \small \emph{1-1 Oho, Tsukuba, Ibaraki 305-0801, Japan}
}

\date{}

\maketitle

\begin{abstract}
{%
We study the bosonic type IIB (IKKT) matrix model and the fate of the D-instanton positions $p^{(i)}_\mu$, the diagonal components of the $d$ Hermitian matrices, whose one-loop effective potential infamously drives them to a single point. 
We show that this collapse is an artifact of the leading (one-loop) truncation, while the actual non-collapse of the $p^{(i)}_\mu$ is a nonperturbative effect: it is invisible at one loop but already present in the \emph{exact} (all-loop) two-body interaction. 
Since the two-body sector of the $N\times N$ model factorizes into copies of $N=2$, this interaction is captured exactly by the $\mathrm{U}(2)$ model, and we find that the two D-instantons do not collapse onto each other. 
To set up the computation, we gauge-fix the $\mathrm{U}(N)$ symmetry in a way that keeps the diagonal and off-diagonal sectors distinct, and we handle the residual $\mathrm{U}(1)^N$ symmetry with an auxiliary-ghost BRST construction.
This construction generates a new ghost four-leg vertex; the resulting Faddeev--Popov determinant admits a systematic large-separation expansion that organizes the effective potential into a many-body decomposition, separating the interaction into two-body, three-body, and higher-body contributions.
The exact $N=2$ partition function is finite at finite separation; its naive Lorenz-gauge form develops a negative region at separations of order one, which we trace to a Gribov ambiguity of the Lorenz gauge and resolve with the maximal diagonal gauge---the classical frame containing the perturbative vacuum---where the short-distance force is finite and repulsive, so that the two-body potential develops a stable minimum at finite separation. These results are consistent with a stable, non-collapsed distribution of the $p^{(i)}_\mu$; establishing the detailed distribution and
full $N$-body non-collapse requires the higher-body potentials and is left to future work.%
}
\end{abstract}

\section{Introduction}\label{sec:intro}

A central question in quantum gravity is how the classical metric $g_{\mu\nu}$ describing our universe emerges from a more fundamental theory.
In string theory, for instance, our universe may be realized as a nonperturbative object---a D-brane---whose dynamics calls for a nonperturbative formulation of the theory.
The type IIB or IKKT matrix model \cite{Ishibashi:1996xs} is a promising candidate for such a formulation.

The bosonic part of the IKKT matrix model consists of $d$ Hermitian matrices $A_\mu$, whose diagonal components $p^{(i)}_\mu\equiv(A_\mu)_{ii}$ admit an interpretation as the positions of D-instantons in $d$-dimensional spacetime.
A D-brane is then described by a collection of D-instantons, and its shape---possibly that of our universe---is encoded in the distribution of the $p^{(i)}_\mu$.
This distribution has accordingly been studied over the past decades, both analytically and numerically~\cite{Hotta:1998en,Kim:2011cr,Anagnostopoulos:2020xai,Hatakeyama:2019jlf,Krauth:1999qw}.

A notorious difficulty in this program is the instability of the $p^{(i)}_\mu$ at one loop in the absence of supersymmetry \cite{Aoki:1998vn}.
Concretely, in the bosonic IKKT matrix model\footnote{It is also known as the Hoppe model~\cite{Hoppe1982,deWit:1988ig} or the Yang--Mills matrix integral~\cite{Krauth:1998xh,Krauth:1998yu}, names appearing in different contexts.} the one-loop effective potential for the $p^{(i)}_\mu$ around the classical saddle $(A_\mu)_{ij}=p^{(i)}_\mu\delta_{ij}$ reads
\begin{align}
    V_{ \text{1-loop}}^{\mathrm{B}}=\sum_{ i\neq j}\frac{d-2}{2} \log\left(p^{(i)}_\mu-p^{(j)}_\mu\right)^2.
\end{align}
Minimizing this potential forces $p^{(i)}_\mu=p^{(j)}_\mu$: all D-instantons coincide at a single point and spacetime collapses, which cannot be the final answer.
A well-known resolution is to include the fermionic sector, in which case the one-loop effective potential becomes
\begin{align}
    V_{ \text{1-loop}}^{\mathrm{IKKT}}=\sum_{ i\neq j}\frac{d-2-d_f/2}{2} \log\left(p^{(i)}_\mu-p^{(j)}_\mu\right)^2,
\end{align}
where $d_f$ denotes the number of fermionic degrees of freedom.
The potential vanishes identically for $d=10$ and $d_f=16$ (Majorana--Weyl fermions in ten-dimensional spacetime), the field content of superstring theory.
Moreover, with this matter content there is considerable evidence that the full IKKT matrix model reproduces superstring theory in various limits, so this resolution is commonly adopted.

A second motivation for supersymmetry in the conventional picture is the spontaneous symmetry breaking (SSB) of the global symmetry $\mathrm{SO}(10)$ (or $\mathrm{SO}(1,9)$ in the Lorentzian model).
We regard the reduction $\mathrm{SO}(10)\to \mathrm{SO}(3)$ or $\mathrm{SO}(4)$ as a phenomenological requirement: a realistic vacuum should describe a $(3+1)$-dimensional universe, so an acceptable configuration of the $p^{(i)}_\mu$ must single out four extended directions among the ten.
Since SSB was shown not to occur in the purely bosonic sector~\cite{Hotta:1998en}, this provides another natural reason to introduce the fermionic sector, whose interaction with the bosonic sector makes SSB possible \cite{Nishimura:2000ds,Anagnostopoulos:2020xai}.

Both arguments, however, admit alternative readings.



First, regarding the one-loop instability, Ref.~\cite{Hotta:1998en} pointed out that the one-loop description cannot be extrapolated to the region where the eigenvalues approach one another. In particular, it was shown there that higher-loop contributions become comparable to the one-loop contribution at sufficiently short distances, and on this basis a finite characteristic extent of the eigenvalue distribution, together with its large-$N$ scaling, was predicted. Monte Carlo simulations showed that this perturbative estimate is actually saturated, while the $1/D$ expansion provided independent analytical support for the corresponding large-$N$ behavior. Thus, Ref.~\cite{Hotta:1998en} established that the eigenvalue distribution does not collapse and that higher-loop effects are essential at short distances.
In the present work, we investigate how this stabilization is realized explicitly in the interaction between individual D-instantons and in terms of the diagonal components rather than eigenvalues under careful gauge-fixing.

Second, the original arguments against SSB are formulated at finite $N$, with the large-$N$ theory defined through the limit.
At $N\to\infty$, however, the bosonic IKKT matrix model possesses saddles that cannot be reached by this procedure; in particular, there exists a saddle closely related to the $\mathfrak{so}(1,3)$ algebra~\cite{Liao:2025so13saddle}.
Our universe could presently be described by such a meta-stable saddle, expanding in size while it slowly decays toward the global minimum of the path integral~\cite{Ho:2025htr}.
The bosonic IKKT matrix model therefore remains worth investigating as a physical model of our universe.


In this work we focus on the microscopic structure of this nonperturbative stabilization. In particular, we ask whether it can already be identified at the level of the interaction between individual D-instantons as the diagonal components, which requires proper gauge fixing. We answer this question by computing the full two-body interaction exactly to all orders.
We make this concrete by isolating the two-body sector.
As we show in Sec.~\ref{sec:body_expansion}, the two-body contributions of the $N\times N$ model are precisely multiple copies of the $N=2$ case; that is, the two-body sector factorizes into independent pairs.
The two-body interaction can therefore be computed \emph{exactly} in the $\mathrm{U}(2)$ model of Sec.~\ref{sec:n=2}, and the result shows that two D-instantons do not collapse onto each other.
 The mechanism is the competition between the attractive one-loop tail and a short-distance repulsion generated by the Gribov horizon of the \emph{maximal diagonal gauge}---the classical frame of the theory---which produces a stable minimum of the two-body potential at a finite separation.
We do not claim that this determines the full equilibrium distribution of the $p^{(i)}_\mu$: that distribution generically places some of the $p^{(i)}_\mu$ at separations of order $\mathcal{O}(1)$, where the higher-body potentials become important.
Determining the detailed distribution therefore requires those higher-body potentials and is left to future work.

Methodologically, we proceed as follows.
In Sec.~\ref{sec:gauge_fixing_ikkt} we gauge-fix the $\mathrm{U}(N)$ symmetry of the IKKT matrix model in the Lorenz gauge while keeping the roles of the diagonal and off-diagonal components of $A_\mu$ distinct.
In the process we find a novel contribution that originates from the residual $\mathrm{U}(1)^N$ subgroup of $\mathrm{U}(N)$.
With the gauge-fixed action, we compute the effective potential for the $p^{(i)}_\mu$ by integrating out first the ghosts and then the off-diagonal components of $A_\mu$.
Since the general form of the Faddeev--Popov determinant is rather complicated, we expand it in the large-distance limit, $(p^{(i)}_\mu-p^{(j)}_\mu)^2\gg 1$.
The leading order of this expansion reproduces the one-loop potential above, while the subleading terms, once the off-diagonal components of $A_\mu$ are integrated out, constitute corrections that become important as the $p^{(i)}_\mu$ approach one another.
The structure of these corrections grows richer with the order of the expansion: at the $k$-th order the summations run over $2,3,\dots,(k+1)$ internal indices, and upon integration over the off-diagonal components they yield $2$-, $3$-, $\dots$, $(k+1)$-body potentials for the $p^{(i)}_\mu$.

\paragraph{Summary of results and plan of the paper.}
In Sec.~\ref{sec:gauge_fixing_ikkt} we gauge-fix the $\mathrm{U}(N)$ symmetry in the Lorenz gauge, keeping the diagonal and off-diagonal components distinct, and carry out the BRST quantization, which introduces a novel four-leg ghost vertex absent from the standard treatment.
In Sec.~\ref{sec:body_expansion} we expand the resulting Faddeev--Popov determinant at large distance and organize it as a many-body decomposition, in which the two-body sector of the $N\times N$ model factorizes into independent copies of $N=2$.
In Sec.~\ref{sec:n=2} we evaluate the $N=2$ ($\mathrm{U}(2)$) model exactly and obtain a finite two-body interaction, whose naive Lorenz-gauge partition function develops a negative region at separations of order one. In Sec.~\ref{sec:gribov-ambiguity} we trace this negativity to the Gribov ambiguity of the Lorenz gauge and resolve it with the maximal diagonal gauge---the classical frame of the theory---in which the short-distance interaction is finite and repulsive and the two-body potential develops a stable minimum: the two D-instantons do not collapse onto each other. Section~\ref{sec:general-N} collects the conjectural extrapolations to general $N$ together with the open questions.
Section~\ref{sec:conclusion} contains our conclusions and outlook.


\section{Gauge Fixing and BRST Quantization}\label{sec:gauge_fixing_ikkt}

We start with the bosonic part of the Euclidean IKKT matrix model in general dimension $d$,
\begin{equation}
    S_\text{IKKT}=-\frac{1}{4}\textrm{tr}\left[A_\mu,A_\nu\right]^2,
\end{equation}
where $\mu,\nu=1,\dots ,d$ and the $A_\mu$ are Hermitian $N\times N$ matrices.
The model possesses a $\mathrm{U}(N)$ symmetry: $S$ is invariant under
\begin{align}
    A_\mu \mapsto U A_\mu U^\dagger \,,
\end{align}
for any $U\in \mathrm{U}(N)$.

The equation of motion reads
\begin{align}
    \left[A_\nu,\left[A_\nu,A_\mu\right]\right]=0.
\end{align}
A simple class of solutions consists of mutually commuting matrices, $\left[A_\mu,A_\nu\right]=0$, which can be diagonalized simultaneously; we denote the diagonalized classical saddles by $P_\mu$.

\subsection{Gauge Fixing}

Since the $P_\mu$ can be interpreted as the positions of D-instantons, or spacetime points \cite{Ishibashi:1996xs,Aoki:1998vn}, it is crucial to understand how the model governs them, in particular once quantum effects from the off-diagonal components are taken into account.
This picture motivates the separation of the diagonal and off-diagonal components of $A_\mu$,
\begin{align}
    A_\mu=P_\mu+\tilde{A}_\mu\,,
\end{align}
where $P_\mu$ is the diagonal matrix of the diagonal components of $A_\mu$, and $\tilde{A}_\mu$ collects the off-diagonal components.
Explicitly, in components,
\begin{align}
    (P_\mu)_{ij}=p_\mu^{(i)}\delta_{ij},\,\,\,\,
    (\tilde{A}_\mu)_{ij}=\tilde{a}_{\mu,ij}
\end{align}
with $\tilde{a}_{\mu,ii}=0$.

The diagonal components $p_\mu^{(i)}$, however, are not invariant under a gauge transformation,
\begin{equation}
\begin{split}
    A_\mu &\mapsto A_\mu' = A_\mu+i[\,\xi\,,\,A_\mu\,]\,,\\
    P_\mu &\mapsto P_\mu' = P_\mu +i[\,\xi\,,\,A_\mu\,]_D\,,
\end{split}
\end{equation}
where $\xi\in\mathfrak{u}(N)$ and the subscript $D$ denotes the diagonal part.
To identify them unambiguously as spacetime points, we must fix the $\mathrm{U}(N)$ gauge symmetry.
A natural choice is the Lorenz gauge, which in field theories reads $\partial_\mu W^\mu(x)=0$.
In matrix model language, this translates through the large-$N$ reduction to
\begin{equation}\label{eq:gauge-cond}
    [P_\mu,A_\mu]=0\,,
\end{equation}
where $[P_\mu,\,\cdot\,]$ is reminiscent of the derivative in large-$N$ reduced models\footnote{See the textbook of Yuri Makeenko for a thorough review~\cite{Makeenko:2002uj}.}~\cite{Eguchi:1982nm,Parisi:1982gp,Gross:1982at}.

This condition, however, does not fix the gauge completely, since its diagonal components ($i=j$) are trivial.
More concretely, in components the condition reads
\begin{equation}
    [P_\mu,A_\mu]_{ij}=\left( p_\mu^{(i)}-p_\mu^{(j)} \right)A_{\mu,ij},
\end{equation}
whose right-hand side vanishes identically for $i=j$.\footnote{The same gauge condition and its residual gauge are also discussed for BFSS matrix quantum mechanics in~\cite{Lin:2014wka}.}
A residual gauge transformation $\xi_0$ is one that leaves the gauge condition unchanged, i.e.
\begin{equation}\label{eq:residual_gauge_eq}
    0=[P_\mu,A_\mu]=[P_\mu',A_\mu']\,,
\end{equation}
with
\begin{align}
    A_\mu &\mapsto A_\mu' = A_\mu+i[\,\xi_0\,,\,A_\mu\,]\,,\\
    P_\mu &\mapsto P_\mu' = P_\mu +i[\,\xi_0\,,\,A_\mu\,]_D\,.    
\end{align}
Equivalently, we can rewrite \eqref{eq:residual_gauge_eq} as
\begin{equation}\label{eq:gf_p_xi0}
    \left[\,i[\xi_0\,,\,A_\mu]_{D}\,,A_\mu\,\right]+\left[\,P_\mu\,,i[\xi_0\,,\,A_\mu]\,\right]=0\,.
\end{equation}
Using the Jacobi identity for the second term,
\begin{equation} \label{eq:jacobi}
    \left[\,P_\mu\,,[\xi_0\,,\,A_\mu]\,\right]=-\left[\,\xi_0\,,[A_\mu\,,\,P_\mu]\,\right]-\left[\,A_\mu\,,[P_\mu\,,\,\xi_0]\,\right]\,,
\end{equation}
one finds that, with the gauge-fixing condition \eqref{eq:gauge-cond} imposed, the first term vanishes.
Substituting \eqref{eq:jacobi} back into \eqref{eq:gf_p_xi0}, we obtain
\begin{equation}\label{eq:combined}
    \big[\,A_\mu\,,\,[\,A_\mu\,,\xi_0\,]_D-[\,P_\mu\,,\xi_0\,]\;\big]=0\,.
\end{equation}
For a generic configuration $A_\mu$, this condition requires the inner bracket to vanish,
\begin{equation}
    [\,A_\mu\,,\xi_0\,]_D-[\,P_\mu\,,\xi_0\,]=0.
\end{equation}
Since the first term is diagonal and the second purely off-diagonal---the two live in orthogonal subspaces of $\mathfrak{u}(N)$ with respect to the trace inner product---each must vanish independently.
In components, the off-diagonal condition reads $(p_\mu^{(i)}-p_\mu^{(j)})(\xi_0)_{ij}=0$; for generic, non-degenerate $p_\mu^{(i)}$ it forces $(\xi_0)_{ij}=0$ for $i\neq j$, after which the diagonal condition is satisfied automatically. The zero modes are therefore the diagonal matrices $\xi_0=\mathrm{diag}( \Xi_1,\Xi_2,\cdots,\Xi_N)$ in the Cartan subalgebra. Here we use that the $p^{(i)}_\mu$ are generically non-degenerate; the measure-zero locus of coincident eigenvalues, on which the residual gauge symmetry is enhanced, is excluded.

This gauge choice is further justified by the observation that the zero modes $\xi_0$ leave $P_\mu$ invariant,
\begin{equation}
    P_\mu \mapsto P_\mu' = P_\mu +i[\,\xi_0\,,\,A_\mu\,]_{D}\, = P_\mu \,,
\end{equation}
which endows $P_\mu$ with an unambiguous physical meaning. Since $P_\mu$ is invariant under residual gauge transformations and itself lives in the Cartan subalgebra, this can be stated as
\begin{equation}
    0=\delta_{\xi_0}P_\mu=i[\xi_0,P_\mu]\,,
\end{equation}
which makes manifest the covariance of the gauge condition under residual gauge transformations,
\begin{equation}
\label{zero mode covraince gauge condition}
    \delta_{\xi_0}[P_\mu,A_\mu]=\left[\delta_{\xi_0}P_\mu,A_\mu  \right]+\left[P_\mu,\delta_{\xi_0}A_\mu \right]=i\left[\xi_0,[P_\mu,A_\mu]\right]\,.
\end{equation}
Note that this equation holds even before the gauge-fixing condition $[P_\mu,\tilde{A}_\mu]=0$ is imposed.

On the Lorenz-gauge slice $[P_\mu,\tilde{A}_\mu]=0$, the original gauge group $\mathrm{U}(N)$ is therefore reduced to the $N$ residual $\mathrm{U}(1)$ phase transformations,
\begin{equation}
    \mathrm{U}(N)\rightarrow \mathrm{U}(1)^{N}\,,
\end{equation}
which must be dealt with in the BRST quantization below.

\subsection{BRST Quantization}

When quantizing a system with gauge symmetry, only the configurations satisfying the gauge condition should be integrated over in the partition function. Although this singles out the physical degrees of freedom, the calculation is not thereby simplified: the symmetry is lost, and the Faddeev--Popov determinant enters nonperturbatively, leaving the usual perturbative expansion inaccessible. By introducing ghosts, BRST quantization implements the gauge fixing without giving up perturbation theory. The construction works because the ghosts carry the wrong spin-statistics and count as negative degrees of freedom: once assigned to fluctuations along the gauge orbit, they cancel the unphysical modes and leave a system with essentially only physical modes.

To employ BRST quantization, we construct a graded algebra whose generator $\delta_\text{BRST}$ is nilpotent up to a residual $\mathrm{U}(1)^N$ transformation,
\begin{equation}
    \delta_{\text{BRST}}^2=\delta_{\xi_0}\,.
\end{equation}
To close the algebra, the zero modes must be included.
At the same time, the ghosts should fluctuate only along the gauge orbit, since the fluctuation along the gauge slice is not preferred~\cite{Henneaux:1992ig}. 
One canonical approach introduces a ghost-for-ghost, as in the Batalin--Vilkovisky (BV) formalism~\cite{Batalin:1981jr,Batalin:1983ggl}. We adopt a simpler implementation of the same idea, removing the extra ghosts by means of additional auxiliary fields; our procedure closely follows~\cite{Pestun:2007rz}.

We first introduce the BRST algebra,
\begin{equation}
\begin{split}
    &\delta_{\text{BRST}}\,A_\mu=i\left[c,A_\mu\right]\,,\quad \delta_{\text{BRST}}\,c=a_0+ic^2\,,\quad\delta_{\text{BRST}}\,a_0=0\,,\\
    &\delta_{\text{BRST}}\,b=B\,,\quad\delta_{\text{BRST}}\,B=i[a_0,b]\,,\\
    &\delta_{\text{BRST}}\,\tilde{a}_0=c_0\,,\quad\delta_{\text{BRST}}\,c_0=i[a_0,\tilde{a}_0]\,,\\
    &\delta_{\text{BRST}}\,B_0=b_0\,,\quad\delta_{\text{BRST}}\,b_0=i[a_0,B_0]\,.\\
\end{split}
\end{equation}
Here $A_\mu$ are the original fields, and the BRST transformation acts as a gauge transformation with Grassmann parameter $c$. The ghost fields are
\begin{equation}
    \text{ghost}=\{c,c_0,b,b_0,B,B_0,a_0,\tilde{a}_0\}\,,
\end{equation}
where we introduce auxiliary ghost zero modes living in the Cartan subalgebra,
\begin{equation}
    \text{ghost}_0=\{c_0,b_0,B_0,a_0,\tilde{a}_0\}\,,
\end{equation}
whose role is to remove the extra ghosts in the original BRST multiplets $\{c,b,B\}$. Since $\delta_{\text{BRST}}$ is fermionic, $a_0$ and $\tilde{a}_0$ are bosonic. There are three BRST doublets: the usual $\{b,B\}$, together with $\{\tilde{a}_0,c_0\}$ and $\{B_0,b_0\}$ involving the zero modes. The transformation $\delta_{\text{BRST}}$ is nilpotent up to a residual gauge transformation,
\begin{equation}
    \delta_{\text{BRST}}^2\,\left(\text{fields}\right)=i\left[a_0,\text{fields}\right]=\delta_{a_0}\left(\text{fields}\right)\,.
\end{equation}
As an explicit check, apply $\delta_{\text{BRST}}$ twice to the ghost $c$, using $\delta_{\text{BRST}}c=a_0+ic^2$, $\delta_{\text{BRST}}a_0=0$, and that $\delta_{\text{BRST}}$ acts as a graded (fermionic) derivation, so $\delta_{\text{BRST}}(c^2)=(\delta_{\text{BRST}}c)\,c-c\,(\delta_{\text{BRST}}c)=[\delta_{\text{BRST}}c\,,\,c]$,
\begin{equation}
    \delta_{\text{BRST}}^2\,c=\delta_{\text{BRST}}\left(a_0+ic^2\right)=0+i\,[\,a_0+ic^2\,,\,c\,]=i\,[a_0,c]-[c^2,c]=i\,[a_0,c]\,,
\end{equation}
where $[c^2,c]=0$ since any matrix commutes with its own square. This reproduces $\delta_{a_0}c$, confirming the nilpotency up to the residual gauge transformation.

We next establish the invariance of the partition function under the addition of any BRST-exact potential,
\begin{equation}
    S\rightarrow S+\delta_{\text{BRST}}V\,.
\end{equation}
It suffices to prove the equivalent statement
\begin{equation}
    0=\frac{d}{dt}Z_t=\frac{d}{dt}\int d\Phi \,e^{-S+t\delta_{\text{BRST}}V}\,,
\end{equation}
i.e.\ that for any given $V$ the partition function is independent of $t$, $Z_t=Z_0$, and hence unchanged by any BRST-exact deformation. The proof parallels the standard BRST argument; only the zero modes require extra care. Applying the $t$-derivative gives
\begin{equation}
    \int d\Phi\left(\delta_{\text{BRST}}V\right) \,e^{-S+t\delta_{\text{BRST}}V}\,,
\end{equation}
and the strategy is to move $\delta_{\text{BRST}}$ in front of the whole integrand and use $\delta_{\text{BRST}}\,(\text{const.})=0$. The measure is BRST invariant by the usual reasoning: the BRST (super-)transformation matrix has no diagonal elements, so its supertrace vanishes. To see this, we pass to the color basis by writing
\begin{equation}
    \Phi=\Phi_a\,t^a\,,
\end{equation}
where the $t^a$ form a basis of the Lie algebra $\mathfrak{u}(N)$, satisfying $[t^a,t^b]=if^{abc}t^c$ with real structure constant s $f^{abc}$. 
The potentially dangerous contributions come from $A_\mu$ and $c$; the remaining fields manifestly contribute no diagonal elements. For these two, the BRST transformations can be rewritten as
\begin{equation}
\begin{split}
    &\delta_{\text{BRST}}A_\mu = ic^aA_\mu^b\,[t^a,t^b]=-c^aA_\mu^b\,f^{abc}\,t^c\,,\\
    &\delta_{\text{BRST}}\,c = a_0+ic^ac^b\,t^at^b=a_0+\frac{i}{2}c^ac^b[t^a,t^b]=a_0-\frac{1}{2}c^ac^bf^{abc}t^c\,.
\end{split}
\end{equation}
Projecting both sides onto the basis $t^c$,
\begin{equation}
\begin{split}
    &\delta_{\text{BRST}}A_\mu^c = -c^aA_\mu^b\,f^{abc}\,,\\
    &\delta_{\text{BRST}}\,c^c = 2\text{tr}\left(a_0 t^c \right)-\frac{1}{2}c^ac^bf^{abc}\,,
\end{split}
\end{equation}
we see that $\frac{\partial  \delta_{\text{BRST}}A_\mu^c}{\partial A_\mu^c}=0$ and $\frac{\partial \delta_{\text{BRST}}c^c}{\partial c^c}=0$. Putting the pieces together, the off-diagonal multiplet $\{\tilde{c},\tilde{b},\tilde{B}\}$ transforms with vanishing supertrace---this is exactly the content of the structure-constant computation above.
The integration measure is therefore BRST invariant.

The action satisfies $\delta_{\text{BRST}}\,S=0$, as explained above. The BRST-exact term is not gauge invariant, owing to the gauge condition; it is, however, invariant under the residual gauge transformations, since the gauge condition is zero-mode covariant, Eq.~\eqref{zero mode covraince gauge condition}. We therefore have
\begin{equation}
    \delta_{\text{BRST}}^{\,2}V=\delta_{a_0}V=0\,.
\end{equation}
The invariance of the partition function then follows:
\begin{equation}
    \int d\Phi\left(\delta_{\text{BRST}}V\right) \,e^{-S+t\delta_{\text{BRST}}V}
    =\int d\Phi\left(\delta_{\text{BRST}}\,V e^{-S+t\delta_{\text{BRST}}V}\right)
    =\delta_{\text{BRST}}\int d\Phi\,V e^{-S+t\delta_{\text{BRST}}V}=0\,.
\end{equation}

We now come to the main point: the elimination of the extra ghosts. The trick is to introduce the potential
\begin{equation}
    V=i\text{tr}\left(b[P_\mu,A_\mu]\right)+i\text{tr}\left(bB_0\right)+i\text{tr}\left(c\tilde{a}_0\right)\,,
\end{equation}
which, acted on by $\delta_{\text{BRST}}$, becomes
\begin{equation}
\label{ghost potential}
\begin{split}
        \delta_{\text{BRST}}V
        &=i\text{tr}\left(B[P_\mu,A_\mu]\right)-i\text{tr}\left(b\,\delta_{\text{BRST}}[P_\mu,A_\mu]\right)\\
        &+i\text{tr}\left(BB_0\right)-i\text{tr}\left(bb_0\right)\\
        &+i\text{tr}\left((a_0+ ic^2)\tilde{a}_0\right)-i\text{tr}\left(cc_0\right)\,,
\end{split}
\end{equation}
which contains several bilinears pairing ghosts with their zero modes. We first verify that the components residing in the Cartan subalgebra indeed decouple from the system; the auxiliary ghost zero modes then remove them from the measure. It is convenient to separate the ghost zero modes by writing
\begin{equation}
    c=c_D+\tilde{c}\,,\quad b=b_D+\tilde{b}\,,\quad B=B_D+\tilde{B}\,,
\end{equation}
where the diagonal elements are the zero modes and the off-diagonal elements carry a tilde. We must check that $\{c_D,b_D,B_D\}$ drop out of the first line of (\ref{ghost potential}). The first term of (\ref{ghost potential}) simplifies because the gauge condition has vanishing diagonal elements,
\begin{equation}
    \text{tr}\left(B[P_\mu,A_\mu]\right)=\text{tr}\left(\tilde{B}[P_\mu,A_\mu]\right)\,,
\end{equation}
and $B_D$ therefore decouples. The second term can be expanded as follows,
\begin{equation}
\begin{split}\label{eq:bdiag}
    -i\text{tr}\left(b\,\delta_{\text{BRST}}[P_\mu,A_\mu]\right)=\text{tr}\left(b[\,[c,A_\mu]_D\,,\,A_\mu]\right)+\text{tr}\left(b[P_\mu\,,\,[c,A_\mu]\,]\right)\,.
\end{split}
\end{equation}
Both terms on the right-hand side of Eq.~\eqref{eq:bdiag} take the form $b_{ij}T_{ji}$ where $T_{ji}$ represents the double commutator in each term.
Since $T_{ji}$ has vanishing diagonal, that is,
\begin{equation}
    [\text{diagonal  matrix}\,,\text{general matrix}]_D=0\,,
\end{equation}
the diagonal part of $b$ does not appear in Eq.~\eqref{eq:bdiag}.
Hence, $b_D$ decouples. We now turn to $c$. 
The first term in Eq.~\eqref{eq:bdiag} contains $c$ in the same manner as the case of $b$, after moving the commutators to enclose $b$ instead of $c$.
Hence, $c_D$ does not appear. The second term can be rewritten using the Jacobi identity,
\begin{equation}
    \text{tr}\left(b[P_\mu\,,\,[c,A_\mu]\,]\right)=-\text{tr}\left(b[c\,,\,[A_\mu,P_\mu]\,]\right)-\text{tr}\left(b[A_\mu\,,\,[P_\mu,c]\,]\right)\,.
\end{equation}
The argument that $c_D$ drops out then proceeds in three steps.
\emph{(i)} Integrating out $\tilde{B}$ in the gauge-fixing term $i\text{tr}(\tilde{B}[P_\mu,A_\mu])$ produces a delta function on the off-diagonal components, imposing the strict (Landau) form of the Lorenz gauge condition
\begin{equation}
    \delta^{N^2-N}\left([P_\mu,A_\mu]\right)\,,
\end{equation}
i.e.\ $[P_\mu,A_\mu]=0$, which sets the first term, $-\text{tr}(b[c\,,\,[A_\mu,P_\mu]])=\text{tr}(b[c\,,\,[P_\mu,A_\mu]])$, to zero.
\emph{(ii)} For the remaining second term, the Jacobi identity used above has already isolated the inner commutator $[P_\mu,c]$, leaving $-\text{tr}(b[A_\mu\,,\,[P_\mu,c]\,])$; splitting $c=c_D+\tilde{c}$ isolates the diagonal contribution $[P_\mu,c_D]$.
\emph{(iii)} The diagonal $c_D$ commutes with the diagonal $P_\mu$, so
\begin{equation}
    [P_\mu\,,c_D]=0\,,
\end{equation}
in the same way as we recognize the zero modes above.

Hence $c_D$ decouples from this term as well, and we have finished the proof that $\{c_D,b_D,B_D\}$ decouples from the ghost potential,
\begin{equation}
\begin{split}
    &i\text{tr}\left(B[P_\mu,A_\mu]\right)-i\text{tr}\left(b\,\delta_{\text{BRST}}[P_\mu,A_\mu]\right)\\
    =&i\text{tr}\left(\tilde{B}[P_\mu,A_\mu]\right)+\text{tr}\left(\tilde{b}[\,[\tilde{c},A_\mu]_D\,,\,A_\mu]\right)+\text{tr}\left(\tilde{b}[P_\mu\,,\,[\tilde{c},A_\mu]\,]\right)\,.
\end{split}
\end{equation}
It remains to show how the auxiliary ghosts remove the decoupled zero modes. Integrating out the auxiliary ghosts $\{c_0,b_0,B_0\}$ produces delta functions for the zero modes $\{c_D,b_D,B_D\}$,
\begin{equation}
\begin{split}
    &i\text{tr}\left(BB_0\right)-i\text{tr}\left(bb_0\right)-i\text{tr}\left(cc_0\right)\\
    =&i\text{tr}\left(B_DB_0\right)-i\text{tr}\left(b_Db_0\right)-i\text{tr}\left(c_Dc_0\right)\,,
\end{split}
\end{equation}
and the zero modes are thereby safely eliminated. Concretely, the delta functions $\delta(c_D)\delta(b_D)\delta(B_D)$ set the diagonal ghosts to zero with unit Jacobian, leaving the measure $d\tilde{c}\,d\tilde{b}\,d\tilde{B}$ over the $3N(N-1)$ off-diagonal ghost degrees of freedom. The last piece, $i\text{tr}\left((a_0+ ic^2)\tilde{a}_0\right)$, likewise produces a delta function $\delta(a_0+ ic^2)$ upon integrating out $\tilde{a}_0$, after which $a_0$ is removed by direct integration.

Let us summarize the procedure. To obtain a meaningful effective action for the diagonal components $P_\mu$, the gauge group $\mathrm{U}(N)$ must be fixed. Faddeev--Popov gauge fixing leads to the gauge-fixed partition function
\begin{equation}
\begin{split}
        Z_{\text{IKKT}}
        &=\int dA_\mu\,e^{-S_{\text{IKKT}}(A_\mu)}\\
        &=\text{vol}\left(\frac{\mathrm{U}(N)}{\mathrm{U}(1)^{N}}\right)\int dA_\mu \, \delta'([P_\mu,A_\mu]) \, \text{det}'\left(\frac{\partial \left[P_\mu^\xi,A_\mu^\xi\right]}{\partial \xi}\right)e^{-S_{\text{IKKT}}}\,,
\end{split}
\end{equation}
where $\delta'$ and $\text{det}'$ indicate the care required by the residual gauge. This integral is well-defined but intrinsically nonperturbative, which hinders the loop expansion. We therefore resort to BRST quantization, introducing the ghost multiplet $\{c,b,B\}$. This alone does not treat the zero modes, i.e.\ the residual gauge, properly. One might be tempted to introduce only the off-diagonal ghosts $\{\tilde{c},\tilde{b},\tilde{B}\}$, since the gauge fixing only concerns $\mathrm{U}(N)/\mathrm{U}(1)^{N}$; this fails because $\mathrm{U}(N)/\mathrm{U}(1)^{N}$ is not a group. While more sophisticated schemes such as the BV formalism could handle this issue, we may equally keep $\{c,b,B\}\in \mathfrak{u}(N)$ and introduce auxiliary ghosts $\{c_0,b_0,B_0,a_0,\tilde{a}_0\}$ in the Cartan subalgebra of $\mathfrak{u}(N)$, which cancel the zero modes $\{c_D,b_D,B_D\}$. We thus rewrite the partition function,
\begin{equation}
\begin{split}
    Z_{\text{IKKT}}
        &=\int dA_\mu\,e^{-S_{\text{IKKT}}}\\
        &=\text{vol}\left(\frac{\mathrm{U}(N)}{\mathrm{U}(1)^{N}}\right)\int dA_\mu\,dc\,db\,dB\,dc_0\,db_0\,dB_0\,da_0\,d\tilde{a}_0\,e^{-S_{\text{IKKT}}+\delta_{\text{BRST}}V}\,,
\end{split}
\end{equation}
and plug in the ghost potential (\ref{ghost potential}). After integrating out $\{c_0,b_0,B_0,a_0,\tilde{a}_0\}$, the integral becomes
\begin{equation}
\begin{split}
    &\int dA_\mu\,dc\,db\,dB\,dc_0\,db_0\,dB_0\,da_0\,d\tilde{a}_0\,e^{-S_{\text{IKKT}}+\delta_{\text{BRST}}V}\\
    =&\int dA_\mu\,dc\,db\,dB\,\delta(c_D)\delta(b_D)\delta(B_D)\,e^{-S_{\text{IKKT}}-S_{\text{gauge fixing}}-S_{\text{ghost}}}\\
    =&\int dA_\mu\,d\tilde{c}\,d\tilde{b}\,d\tilde{B}\,e^{-S_{\text{IKKT}}-S_{\text{gauge fixing}}-S_{\text{ghost}}}\,,
\end{split}
\end{equation}
where
\begin{equation}
    S_{\text{gauge fixing}}=-i\,\text{tr}\left(\tilde{B}[P_\mu,A_\mu]\right)\,,
\end{equation}
and
\begin{equation}
\begin{split}
    S_{\text{ghost}}
    &=-\text{tr}\left(\tilde{b}[\,[\tilde{c},A_\mu]_D\,,\,A_\mu]\right)-\text{tr}\left(\tilde{b}[P_\mu\,,\,[\tilde{c},A_\mu]\,]\right)\\
    &=-\text{tr}[P_\mu,\tilde{b}][A_\mu,\tilde{c}]+\text{tr}[\tilde{b},A_\mu]\,[\tilde{c},A_\mu]_D\\
    &=-\text{tr}[P_\mu,\tilde{b}][A_\mu,\tilde{c}]+\text{tr}[\tilde{b},A_\mu]_D[\tilde{c},A_\mu]_D\\
    &=-\text{tr}[P_\mu,\tilde{b}][A_\mu,\tilde{c}]+\text{tr}[\tilde{b},\tilde{A}_\mu]_D[\tilde{c},\tilde{A}_\mu]_D\,.
\end{split}
\end{equation}
Let us spell out how the second, four-leg term arises, as it is the methodological novelty of this construction. It originates entirely from the first piece of \eqref{eq:bdiag}, $\text{tr}\left(\tilde{b}[\,[\tilde{c},A_\mu]_D\,,\,A_\mu]\right)$, into which we substitute $A_\mu=P_\mu+\tilde{A}_\mu$. The inner diagonal projection simplifies because $[\tilde{c},P_\mu]$ is the commutator of an off-diagonal matrix with a diagonal one and is therefore purely off-diagonal, so its diagonal projection vanishes,
\begin{equation}
\label{eq:inner_proj_collapse}
    [\tilde{c},A_\mu]_D=[\tilde{c},P_\mu]_D+[\tilde{c},\tilde{A}_\mu]_D=[\tilde{c},\tilde{A}_\mu]_D\,,
\end{equation}
which is already diagonal. The outer commutator then splits as
\begin{equation}
\label{eq:outer_split}
    \text{tr}\left(\tilde{b}[\,[\tilde{c},\tilde{A}_\mu]_D\,,\,A_\mu]\right)
    =\text{tr}\left(\tilde{b}[\,[\tilde{c},\tilde{A}_\mu]_D\,,\,P_\mu]\right)
    +\text{tr}\left(\tilde{b}[\,[\tilde{c},\tilde{A}_\mu]_D\,,\,\tilde{A}_\mu]\right)\,,
\end{equation}
and the first term vanishes since $[\tilde{c},\tilde{A}_\mu]_D$ is diagonal and $[\text{diagonal},P_\mu]=0$ (both factors are diagonal and commute). Only the second term survives. Using cyclicity of the trace and the fact that $[\tilde{c},\tilde{A}_\mu]_D$ is diagonal---so that contracting it against $[\tilde{b},\tilde{A}_\mu]$ projects out only the diagonal part of the latter---we obtain
\begin{equation}
\label{eq:four_leg_collapse}
    \text{tr}\left(\tilde{b}[\,[\tilde{c},\tilde{A}_\mu]_D\,,\,\tilde{A}_\mu]\right)
    =-\,\text{tr}\left([\tilde{b},\tilde{A}_\mu]_D\,[\tilde{c},\tilde{A}_\mu]_D\right)\,.
\end{equation}
With the leading minus sign carried from \eqref{eq:bdiag} into $S_{\text{ghost}}$, this is precisely the four-leg term $+\text{tr}[\tilde{b},\tilde{A}_\mu]_D[\tilde{c},\tilde{A}_\mu]_D$ above. It is manifestly $\mathcal{O}(\tilde{A}^2)$ in the off-diagonal gauge field and is generated purely by the consistent gauge fixing with the residual $\mathrm{U}(1)^{N}$; it is absent in the field theory, where the ghost vertex is only the three-leg term $-\text{tr}[P_\mu,\tilde{b}][A_\mu,\tilde{c}]$.

Physically, this term couples the residual diagonal sector to the off-diagonal fluctuations and, unlike the standard three-leg ghost vertex $-\text{tr}[P_\mu,\tilde{b}][A_\mu,\tilde{c}]$, acts as a four-leg vertex; as shown in Sec.~\ref{sec:body_expansion}, it is precisely what truncates the large-distance ghost-determinant expansion at $\mathcal{O}(\tilde{A}^2)$. The explicit four-leg Feynman vertex factor, with its spacetime/color indices and diagram, is given in Appendix~\ref{sec:feyn_rules_det}.

Despite the many intermediate steps, the final form is remarkably simple. The partition function now reads
\begin{equation}
\begin{split}\label{eq:z_ikkt_gf}
        Z_{\text{IKKT}}
        &=\int dA_\mu\,e^{-S_{\text{IKKT}}}\\
        &=\text{vol}\left(\frac{\mathrm{U}(N)}{\mathrm{U}(1)^{N}}\right)\int dA_\mu\,d\tilde{c}\,d\tilde{b}\,d\tilde{B}\,e^{-S_{\text{IKKT}}-S_{\text{gauge fixing}}-S_{\text{ghost}}}\\
        &=\text{vol}\left(\frac{\mathrm{U}(N)}{\mathrm{U}(1)^{N}}\right)\int dA_\mu\,d\tilde{c}\,d\tilde{b}\;\delta\!\left([P_\mu,A_\mu]\right)\,e^{-S_{\text{IKKT}}-S_{\text{ghost}}}\,,
\end{split}
\end{equation}
where $A_\mu\in\mathfrak{u}(N)$ and $\{\tilde{c},\tilde{b},\tilde{B}\}$ have only off-diagonal components; the delta function in the last line represents the Lorenz gauge in its strict (Landau) form, which has support only on the $N^2-N$ off-diagonal directions.\footnote{
We here comment on the $R_\xi$ (Feynman-type) gauges, obtained by imposing
\begin{equation}
    [P_\mu,A_\mu]=h
\end{equation}
and integrating over $h$ with a Gaussian weight $\exp({-\alpha\,\text{tr}\,h^2})$, or equivalently by adding $\alpha\,\text{tr}(bB)$ to $V$. These gauges do not work here, and the reason is most transparent in the Faddeev--Popov language: $[P_\mu,A_\mu]=h$ with $h\neq0$ cannot serve as a gauge-fixing condition because it does not slice through every gauge orbit. The diagonal configurations $A_\mu=P_\mu$, for instance, can never satisfy it. The underlying reason is simple: the configuration space possesses a center under the $\mathrm{U}(N)$ action---in our case the diagonal matrices---and the center must satisfy any admissible gauge condition, since otherwise physical configurations would be excluded.} This ghost action and the Faddeev--Popov determinant will play a central role in the following sections.


\section{Ghost Determinant and Large Distance Expansion}\label{sec:body_expansion}

Sec.~\ref{sec:gauge_fixing_ikkt} furnishes the gauge-fixed action of the IKKT matrix model, from which we now derive the distribution of the diagonal components, i.e.\ the spacetime points.
The main idea is as follows.

In the path-integral picture, \eqref{eq:z_ikkt_gf} integrates over all matrix degrees of freedom, $A_\mu, \tilde{c},\tilde{b}$ and $\tilde{B}$.
Since our main interest is the distribution of the diagonal components of $A_\mu$, namely $P_\mu$, we integrate out $\tilde{A}_\mu, \tilde{c}, \tilde{b}$ and $\tilde{B}$ to obtain an effective potential for the diagonal components.
Concretely, dropping henceforth the constant volume factor $\text{vol}\big(\mathrm{U}(N)/\mathrm{U}(1)^{N}\big)$,
\begin{align}
    Z_\text{IKKT}
    &=\int dP_\mu d\tilde{A}_\mu\,d\tilde{c}\,d\tilde{b}\,d\tilde{B}\,e^{-S_{\text{IKKT}}-S_{\text{gauge fixing}}-S_{\text{ghost}}}\\
    &=\int dP_\mu e^{-V(P_\mu)},
\end{align}
where 
\begin{align}
\label{eq:effective action}
    e^{-V(P_\mu)}\equiv\int d\tilde{A}_\mu\,d\tilde{c}\,d\tilde{b}\,d\tilde{B}\,e^{-S_{\text{IKKT}}-S_{\text{gauge fixing}}-S_{\text{ghost}}}.
\end{align}
The most probable distribution of the spacetime points $p_\mu^{(i)}$ is then obtained by minimizing $V$, or equivalently maximizing $e^{-V}$, with respect to the $p_\mu^{(i)}$.

Owing to the shift symmetry of the theory, $P_\mu\mapsto P_\mu+c_\mu\mathds{1}$ for constants $c_\mu$, $V$ depends only on the differences of the $p_{\mu}^{(i)}$.
Its general form is thus
\begin{align}
    V(P_\mu)
    &=\sum_{ij:\text{ distinct}}f_2\left(\left(p^{(i)}_\mu-p^{(j)}_\mu\right)^2\right)\\
    &+\sum_{ijk:\text{ distinct}}f_3\left( \left(p^{(i)}_\mu-p^{(j)}_\mu\right),\left(p^{(i)}_\mu-p^{(k)}_\mu\right)\right)\\
    &+\sum_{ijkl:\text{ distinct}}f_4\left(\left(p^{(i)}_\mu-p^{(j)}_\mu\right),\left(p^{(i)}_\mu-p^{(k)}_\mu\right),\left(p^{(i)}_\mu-p^{(l)}_\mu\right)\right)+ \dots,
\end{align}
where each $f_a$ involves $a$ of the $p_\mu^{(i)}$ and represents the $a$-body interaction among them.

In general, $V$ contains interactions among the $p_\mu^{(i)}$ ranging from two- to $N$-body.
While the classification by $a$-body interactions offers a clear physical picture, generic multi-body problems are intractable.
We therefore concentrate on the exact, nonperturbative $N=2$ computation and resolve its technical subtleties to extract the physical picture.
This is done in two steps.
In this section we extract the two-body sector from the general large-distance expansion; the detailed analysis of the $N=2$ case, the basic building block of the two-body sector, is carried out in Sec.~\ref{sec:n=2}.

\subsection{Large Distance Expansion for Ghost Determinant}
\label{sec:large_dist_exp_for_ghost_det}

To investigate the large-distance regime ($\Delta_{ij}^2\gg1$) at general $N$, where $\Delta_{ij}$ denotes the separation $d$-vector $\Delta_{\mu,ij}\equiv p_\mu^{(i)}-p_\mu^{(j)}$ and $\Delta_{ij}^2=\Delta_{ij}\cdot\Delta_{ij}$, we expand the ghost determinant perturbatively at large $\Delta_{ij}^2$.
Recall that the Faddeev--Popov ghost determinant takes the form
\begin{align}
\label{eq:FP_1}
    \Delta_{\text{FP}}=\int d\tilde{b}d\tilde{c}~\exp\left\{-S_{\text{ghost}}[\tilde{b},\tilde{c},A]\right\}=\int d\tilde{b}d\tilde{c} ~ \exp\left\{\tr([P_\mu,\tilde{b}][A_\mu,\tilde{c}])-\tr([\tilde{b},\tilde{A}_\mu]_D[\tilde{c},\tilde{A}_\mu]_D)\right\},
\end{align}
which can be written as
\begin{align}
    \Delta_{\text{FP}}\equiv\int d\tilde{b}d\tilde{c} ~ \exp\left\{-\sum_{\substack{i,j,k,\ell=1\\i\neq j, k\neq \ell}}^NM_{ijk\ell}\tilde{b}_{ij}\tilde{c}_{k\ell}\right\}=\det(M),
\end{align}
where the determinant is taken over the $N(N-1)$-dimensional space labeled by the bi-index $[ij]$ of $\tilde{b}$ (or $[k\ell]$ of $\tilde{c}$).
We then expand this $N(N-1)\times N(N-1)$ matrix perturbatively at large $\Delta_{ij}^2$.
Direct computation of \eqref{eq:FP_1} gives
\begin{align}
    M_{ijk\ell} = & ~ 
    \Delta_{ij}^2\delta_{i\ell}\delta_{jk}+\Delta_{ij}\cdot\left(\tilde{a}_{\ell i}\delta_{jk}-\tilde{a}_{jk}\delta_{i\ell}\right)+\left(\tilde{a}_{ji}\cdot \tilde{a}_{\ell k}\right)\left(\delta_{ik}+\delta_{j\ell}-\delta_{i\ell}-\delta_{jk}\right)
\end{align}
where the dot denotes the spacetime inner product, i.e., $\tilde{a}_{ji}\cdot\tilde{a}_{\ell k}=\sum_{\mu=1}^d\tilde{a}^\mu_{ji}\tilde{a}^\mu_{\ell k}$.
For later convenience, we define
\begin{align}
    (M_0)_{ijk\ell} &\equiv \Delta_{ij}^2\delta_{i\ell}\delta_{jk},\\
    (M_1)_{ijk\ell} &\equiv \Delta_{ij}\cdot\left(\tilde{a}_{\ell i}\delta_{jk}-\tilde{a}_{jk}\delta_{i\ell}\right),\\
    (M_2)_{ijk\ell} &\equiv \left(\tilde{a}_{ji}\cdot \tilde{a}_{\ell k}\right)\left(\delta_{ik}+\delta_{j\ell}-\delta_{i\ell}-\delta_{jk}\right).
\end{align}
Since $M_0\sim\mathcal{O}(\Delta^2)$, we expand $\Delta_{\text{FP}}$ as follows.
\begin{align}
    \Delta_{\text{FP}}
    =&\ \det(M)
    =\det(M_0)\cdot\det\!\left[\mathds{1}
    +(M_0^{-1}M_1)
    +(M_0^{-1}M_2)\right]\notag\\
    =&\ \left[\prod_{\substack{i,j=1\\i\neq j}}^N\Delta_{ij}^2\right]
    \cdot\Bigg\{
    1+\tr\left(M_0^{-1}M_2\right)
    -\frac{1}{2}\tr\left[\left(M_0^{-1}M_1\right)^2\right]
    -\tr\left(M_0^{-1}M_1M_0^{-1}M_2\right)
    \nonumber\\
    &\quad
    +\frac{1}{3}\tr\left[\left(M_0^{-1}M_1\right)^3\right]
    +\frac{1}{2}\left[\tr\left(M_0^{-1}M_2\right)\right]^2
    -\frac{1}{2}\tr\left[\left(M_0^{-1}M_2\right)^2\right]
    \nonumber\\
    &\quad
    -\frac{1}{2}
    \tr\left[\left(M_0^{-1}M_1\right)^2\right]
    \tr\left(M_0^{-1}M_2\right)
    +\tr\left[\left(M_0^{-1}M_1\right)^2M_0^{-1}M_2\right]
    \nonumber\\
    &\quad
    +\frac{1}{8}
    \left[\tr\left(M_0^{-1}M_1\right)^2\right]^2
    -\frac{1}{4}
    \tr\left[\left(M_0^{-1}M_1\right)^4\right]
    +\mathcal{O}\!\left(\tilde{A}^5\right)
    \Bigg\},\label{eq:det_expansion}
\end{align}
where $\tr(\cdot)$ sums over the $N(N-1)$-dimensional bi-index space, and we have used 
\begin{align}
    \tr(M_0^{-1}M_1)=\sum_{\substack{ijk\ell\\i\neq j\,,\,k\neq \ell}}^N\left(M_0^{-1}\right)_{ijk\ell}\left(M_1\right)_{k\ell ij}=\sum_{\substack{ijk\ell\\i\neq j\,,\,k\neq \ell}}^N\Delta_{ij}^{-2}\delta_{i\ell}\delta_{kj}\left[\Delta_{k\ell}\cdot\left(\tilde{a}_{jk}\delta_{\ell i}-\tilde{a}_{\ell i}\delta_{kj}\right)\right]=0.
\end{align}
For the derivation of this determinant expansion formula, see Appendix~\ref{sec:det_exp_deriv}.


\begin{figure}[!htb]
    \centering
    \small
    \begin{subequations}
    \renewcommand{\theequation}{0.\arabic{equation}}

    \begin{minipage}{0.48\textwidth}
    \begin{equation*}
        \begin{aligned}
            \fdiag{\DiagramOneLHS} & =
            -\delta_{\mu\rho}\left[\Delta^{a_1a_3}_{\nu}\fdiag{\DiagramOneRHSa}-\Delta^{a_1a_2}_{\nu}\fdiag{\DiagramOneRHSb}\right],
        \end{aligned}
    \end{equation*}
    \end{minipage}
    \hfill
    \begin{minipage}{0.48\textwidth}
    \begin{equation*}
        \begin{aligned}
                \;\;\fdiag{\DiagramTwoLHS} & =
            -\left[\Delta^{a_1a_3}_{\mu}\fdiag{\DiagramTwoRHSa}-\Delta^{a_1a_2}_{\mu}\fdiag{\DiagramTwoRHSb}\right],
        \end{aligned}
    \end{equation*}
    \end{minipage}

    \vspace{1em}

    \begin{minipage}{0.48\textwidth}
    \begin{equation*}
        \begin{aligned}
            \fdiag{\DiagramThreeLHS} & =
            -\frac12\left[\delta_{\mu\nu}\delta_{\rho\sigma}\fdiag{\DiagramThreeRHSa}-\delta_{\mu\rho}\delta_{\nu\sigma}\fdiag{\DiagramThreeRHSb}\right],
        \end{aligned}
    \end{equation*}
    \end{minipage}
    \hfill
    \begin{minipage}{0.48\textwidth}
    \begin{equation*}
        \begin{aligned}
            \;\;\fdiag{\DiagramFourLHS} & =
            \delta_{\mu\nu}\left[\delta_{a_2a_4}\fdiag{\DiagramFourRHSa}+\delta_{a_1a_3}\fdiag{\DiagramFourRHSb}\right],
        \end{aligned}
    \end{equation*}
    \end{minipage}

    \vspace{1em}

    \begin{minipage}{0.98\textwidth}
    \begin{equation*}
        \begin{aligned}
            \fdiag{\DiagramFiveLHS} & =
            -\delta_{\mu\nu}\left[\delta_{a_1a_3}\fdiag{\DiagramFiveRHSa}+\delta_{a_2a_4}\fdiag{\DiagramFiveRHSb}\right].
        \end{aligned}
    \end{equation*}
    \end{minipage}

    \end{subequations}
    \caption{Feynman rules for the gauge-fixed model collected in one place: the ghost propagator $\langle\tilde{b}\tilde{c}\rangle$ ($M_0^{-1}$) and the gauge-field propagator $\langle\tilde{A}\tilde{A}\rangle$, the three-leg vertices (the ghost--gauge vertex from $M_1$, with one $\Delta\cdot\tilde{A}$ gauge leg attached to a ghost line, and the gauge cubic vertex $S_3$ from $S_\text{IKKT}$), and the \emph{new} four-leg vertex from $M_2$, i.e.\ the four-leg ghost term $\tr[\tilde{b},\tilde{A}_\mu]_D[\tilde{c},\tilde{A}^\mu]_D$ (together with the gauge quartic $S_4$). The explicit vertex factors are derived in Appendix~\ref{sec:feyn_rules_det}, Eqs.~\eqref{eq:gauge_cubic}--\eqref{eq:ghost_M2_vertex}.}
    \label{fig:Feynman-rules}
\end{figure}


For brevity, we summarize the expression for $\Delta_{\text{FP}}$ derived in Appendix~\ref{sec:det_exp_deriv} as
\begin{align}\label{eq:ts}
    \Delta_{\text{FP}}=\left[\prod_{\substack{i,j=1\\i\neq j}}^N\Delta_{ij}^2\right]\cdot\left\{1+\mathcal{T}_{2}+\mathcal{T}_{2\times2}+\mathcal{T}_{2\times3}+\mathcal{T}_{3}+\mathcal{T}_{3\times3}+\mathcal{T}_{4}+\mathcal{T}_5+\mathcal{O}(\tilde{A}^5)\right\}.
\end{align}
The subscripts of $\mathcal{T}_{x\times y\times \dots}$ label the index structure: the numbers $x,y,\dots$ count the distinct indices in each summation, and $\times$ indicates a product of such sums.
For example, $\mathcal{T}_{2\times 3}$ carries the summations $\sum_{i,j,\,i\neq j}\sum_{k,\ell,m\,\text{distinct}}$ and thus contains terms involving three, four or five distinct indices, depending on whether $i,j$ coincide with $k,\ell,m$.

Each term in this expansion can be understood as a Feynman diagram: $M^{-1}_0$ provides the ghost propagator, while $M_1$ and $M_2$ provide three- and four-leg vertices, respectively.
For instance, the vanishing of $\tr(M_0^{-1}M_1)$ reflects the impossibility of forming a closed diagram from a single ghost propagator and a single three-leg vertex.
In diagrammatic language, treating the residual gauge thus introduces a novel four-leg vertex into the Feynman rules.
The rules are collected in Fig.~\ref{fig:Feynman-rules}; concrete examples and details are given in Appendix~\ref{sec:explicit_ghost_det}.

We now extract the two-body sector by selecting from \eqref{eq:ts} the terms whose subscripts contain only $2$'s, namely $1$, $\mathcal{T}_2$ and $\mathcal{T}_{2\times 2}$.
One might worry that terms such as $\mathcal{T}_{2\times 2\times 2}$, $\mathcal{T}_{2\times 2\times 2\times 2}$, $\dots$ lurk at higher orders.
Such terms do appear at higher orders, but only as products over \emph{distinct} pairs: a repeated factor from the same pair would require multiple powers of the same $bc$ ghost component, e.g.\ $\tilde{b}_{12}^{\,2}$, which vanish identically in a zero-dimensional theory. In fact, since every entry of $M_1$ and $M_2$ connecting two different pairs requires a shared index, the two-body sector of $\det(M)$ is exactly the \emph{product} over pairs of the corresponding $2\times2$ block determinants---each equal to the $N=2$ expression below. Its expansion reproduces $1+\mathcal{T}_2+\mathcal{T}_{2\times2}$ at $\mathcal{O}(\tilde{A}^4)$ and generates the distinct-pair products $\mathcal{T}_{2\times2\times2},\dots$ at higher orders; this product structure is the precise statement that the two-body sector factorizes into independent pairs.
The two-body sector is therefore, through $\mathcal{O}(\tilde{A}^4)$,
\begin{align}
    \Delta^{\text{2-body}}_{\text{FP}}=\left[\prod_{\substack{i,j=1\\i\neq j}}^N\Delta_{ij}^2\right]\cdot\left\{1+\mathcal{T}_{2}+\mathcal{T}_{2\times2}\right\}.
\end{align}

This two-body sector consists of multiple copies of the $N=2$ case, in which only the terms $\mathcal{T}_2$ and $\mathcal{T}_{2\times2}$ survive.
Explicitly, for $N=2$ the determinant reads
\begin{align}
    \Delta_{\text{FP}}^{(N=2)} 
    &=\left[\prod_{\substack{i,j=1\\i\neq j}}^2\Delta_{ij}^2\right]\cdot\left\{1+\mathcal{T}_{2}+\mathcal{T}_{2\times2}\right\}\\
    &= \left[\prod_{\substack{i,j=1\\i\neq j}}^2\Delta_{ij}^2\right]\cdot\left\{1-2\sum_{\substack{i,j=1\\i\neq j}}^2\Delta_{ij}^{-2}\left(\tilde{a}_{ij}\cdot \tilde{a}_{ji}\right)-2\sum_{\substack{i,j=1\\i\neq j}}^2\Delta_{ij}^{-4}\left[\tilde{a}_{ij}^2\tilde{a}_{ji}^2+\left(\tilde{a}_{ij}\cdot \tilde{a}_{ji}\right)^2\right]\right.\notag\\
    & ~ 
    \left.+2\sum_{\substack{i,j=1\\i\neq j}}^2\sum_{\substack{k,\ell=1\\k\neq \ell}}^2\Delta_{ij}^{-2}\Delta_{k\ell}^{-2}\left(\tilde{a}_{ij}\cdot \tilde{a}_{ji}\right)\left(\tilde{a}_{k\ell}\cdot \tilde{a}_{\ell k}\right)\right\}\notag\\
    &=
    \Delta_{12}^4-4\Delta_{12}^2\left(\tilde{a}_{12}\cdot \tilde{a}_{21}\right)-4\left[\tilde{a}_{12}^2\tilde{a}_{21}^2-\left(\tilde{a}_{12}\cdot \tilde{a}_{21}\right)^2\right].
\end{align}
Since $\tilde{A}_\mu$ is Hermitian, we can decompose its off-diagonal entry as
\begin{align}
    \tilde{a}_{\mu,12}=\tilde{a}^R_\mu+i\,\tilde{a}^I_\mu~,~\tilde{a}_{\mu,21}=\tilde{a}^R_\mu-i\,\tilde{a}^I_\mu,
\end{align}
where $\tilde{a}^R,\tilde{a}^I\in\mathds{R}^d$.
We thus obtain the explicit form to be used in Sec.~\ref{sec:n=2},
\begin{align}\label{eq:delta_fp_n2}
    \Delta_{\text{FP}}^{(N=2)}=\Delta_{12}^4-4\Delta_{12}^2\left(\tilde{a}^R\cdot \tilde{a}^R+\tilde{a}^I\cdot \tilde{a}^I\right)+16\left(\tilde{a}^R\right)^2\left(\tilde{a}^I\right)^2-16\left(\tilde{a}^R\cdot \tilde{a}^I\right)^2.
\end{align}

 Let us make precise the sense in which the $N=2$ computation captures the two-body sector of the $N\times N$ model. Define $f_2$ constructively as the part of $V$ that survives when all other D-instantons are removed to infinity: the couplings of the pair $\{i,j\}$ to any spectator $k$ enter only through the off-diagonal blocks $\tilde{a}_{ik}$, $\tilde{a}_{jk}$, whose propagators fall off as $1/\Delta_{ik}^2$, so all $f_{a\geq3}$ vanish in this limit and $f_2(\Delta_{ij})$ is well defined. Every term of $S_\text{IKKT}+S_\text{gauge fixing}+S_\text{ghost}$ that involves only the pair indices $\{i,j\}$ is exactly the $\mathrm{U}(2)$ action of the pair's $2\times2$ block (the cubic vertex requires three distinct indices and drops out); the Lorenz-gauge delta function and the measure factorize per block, and the two-body sector of the Faddeev--Popov determinant is the product of pair-block determinants exhibited above. Hence the connected contributions built solely from the pair's block fields resum, to all orders, into the exact $\mathrm{U}(2)$ partition function: $f_2$ is computed exactly by the $N=2$ model of Sec.~\ref{sec:n=2}.

For the higher-body contributions the number of terms grows rapidly with the number of bodies, and their analysis is beyond the scope of this paper.
They are nevertheless essential for the detailed distribution of the $p^{(i)}_\mu$, and extracting a clear physical picture from them remains a worthwhile problem  (see Sec.~\ref{sec:general-N} for their status at general $N$).

\section{\texorpdfstring{$N=2$}{N=2} Exact Result}\label{sec:n=2}

We now present an explicit all-loop computation of the two-body interaction at $N=2$, recovering in particular \eqref{eq:delta_fp_n2}.
After obtaining the exact partition function $Z(p)$, where $p\equiv\sqrt{\Delta_{12}^2}$ is the separation of the two D-instantons, and analyzing its IR and UV behavior, we find that the naive Lorenz-gauge result develops a negative region at $p\sim\mathcal{O}(1)$.
We then show that this is a gauge artifact---the Gribov ambiguity of the Lorenz gauge---and that the maximal diagonal gauge resolves it, yielding a finite partition function and a repulsive short-distance potential.

\subsection{Off-Diagonal Integration}
To compute the effective action, Eq.~\eqref{eq:effective action}, we first assemble the action in component form explicitly. The IKKT part of the action in the diagonal--off-diagonal decomposition reads
\begin{equation}
    S_\text{IKKT} = -\frac{1}{4}\tr[A_\mu,A_\nu]^2 = -\frac{1}{2}\tr[P_\mu,\tilde A_\nu]^2+\frac{1}{2}\tr[P_\mu,\tilde A_\mu]^2-\tr[P_\mu,\tilde A_\nu]\left[\tilde A_\mu,\tilde A_\nu\right]-\frac{1}{4}\tr[\tilde A_\mu,\tilde A_\nu]^2\,,
\end{equation}
where one sees a familiar form of the Yang--Mills action by recognizing $[P_\mu,\,\cdot\,]\sim\partial_\mu$. They are, in order, the kinetic, gauge, three-leg and four-leg terms. One can find the details in App.~\ref{sec:feyn_rules_det}. For our case of interest here, $N=2$, this action simplifies because its three-leg interaction vanishes,
\begin{equation}
    S^{\mathrm{U}(2)}_\text{IKKT} = -\frac{1}{2}\tr[P_\mu,\tilde A_\nu]^2+\frac{1}{2}\tr[P_\mu,\tilde A_\mu]^2-\frac{1}{4}\tr[\tilde A_\mu,\tilde A_\nu]^2\,.
\end{equation}
This can be easily checked by plugging in the component below.\footnote{Since the degrees of freedom are only off-diagonal, the two ends of the propagator must carry different indices, cf.\ Eq.~\eqref{eq.propagator}, e.g.\ indices $i$ and $j$ with $i\neq j$. This makes the Feynman diagram a graph coloring problem as the edge (propagator) can only have two different colors on two sides. In the case of $N=2$, it is a two -color problem, which cannot support a three-leg vertex where three edges merge.}

We expand the matrices in the Pauli basis, $\{\mathbb{I}_2\,,\sigma_1\,,\sigma_2\,,\sigma_3\}$. Since $\tilde b\,,\tilde c$ are off-diagonal, they have support only on $\sigma_1$ and $\sigma_2$. For $A_\mu=P_\mu+\tilde A_\mu$, it has support on all four basis elements. However, $A_\mu$ enters the action only through commutators; the identity part decouples. We can simplify the component form by involving only the following
\begin{equation}
\begin{split}
A_\mu&=
\begin{pmatrix}
        \frac{1}{2}\left(p^{(1)}_\mu-p^{(2)}_\mu\right) & \tilde{a}^R_{\mu}+i\,\tilde{a}^I_{\mu}\\
        \tilde{a}^R_\mu-i\,\tilde{a}^I_\mu & -\frac{1}{2}\left(p^{(1)}_\mu-p^{(2)}_\mu\right)
\end{pmatrix}
=\tilde{a}^R_\mu\,\sigma_1-\tilde{a}^I_\mu\,\sigma_2+\frac{1}{2}\Delta_{\mu,12}\,\sigma_3,
\\
\quad \tilde b&=
\begin{pmatrix}
        0 & b^R+i\,b^I\\
        -b^R+i\,b^I & 0
\end{pmatrix}
= ib^I\,\sigma_1 + ib^R\,\sigma_2,\\
\quad \tilde c&=
\begin{pmatrix}
        0 & c^R+i\,c^I\\
        c^R-i\,c^I & 0
\end{pmatrix}
=c^R\,\sigma_1-c^I\,\sigma_2,
\end{split}
\end{equation}
where the off-diagonal gauge vector components and the $bc$ ghosts have been split into real and imaginary parts, labeled $R$ and $I$, respectively. $\Delta_{\mu,12}=p^{(1)}_\mu-p^{(2)}_\mu$ denotes the separation of the two eigenvalues. 

The Lorenz-gauge delta function in Eq.~\eqref{eq:z_ikkt_gf} eliminates the gauge term in the action and yields
\begin{align}
S^{\mathrm{U}(2)}_{\text{gauge}}
=&-\frac{1}{2}\tr\left[P_\mu,\tilde{A}_\nu\right]^2
-\frac{1}{4}\tr\left[\tilde{A}_\mu,\tilde{A}_\nu\right]^2 \notag\\
=&\,\Delta_{12}^2\left(
\tilde{a}^R \cdot \tilde{a}^R
+
\tilde{a}^I \cdot \tilde{a}^I
\right)+4\left[
\left(\tilde{a}^R\cdot \tilde{a}^R\right)\left(\tilde{a}^I\cdot \tilde{a}^I\right)
-\left(\tilde{a}^R\cdot \tilde{a}^I\right)^2
\right]\;.
\end{align}
The Lorenz-gauge delta function in component form reads
\begin{equation}
    \delta([P_\mu,A_\mu]) = \delta\left(\Delta_{12}\cdot \tilde{a}^R\right)\,\delta\left(\Delta_{12}\cdot \tilde{a}^I\right)\,,
\end{equation}
which eliminates one vector component each. To make it explicit, we can use $\mathrm{SO}(d)$ to rotate the diagonal vector to $\Delta_{\mu,12}=|\Delta_{12}|\delta_{1\mu}$. The delta function now becomes
\begin{equation}
    \delta\left(\Delta_{12}\cdot \tilde{a}^R\right)\,\delta\left(\Delta_{12}\cdot \tilde{a}^I\right) = \frac{1}{\Delta_{12}^2}\,\delta\left(\tilde{a}^R_1\right)\delta\left( \tilde{a}^I_1\right)\,,
\end{equation}
and the integration of $\tilde{a}^R_1$ and $ \tilde{a}^I_1$ is then trivially performed. These two gauge-boson vectors $\tilde{\mathbf{a}}^R$ and $\tilde{\mathbf{a}}^I$ now live in $\mathbb{R}^{d-1}$.

For the ghost term, we have
\begin{equation}
\begin{split}
    S^{\mathrm{U}(2)}_{\text{gh}}
    &=-\text{tr}\left([P_\mu,\tilde{b}][P^\mu,\tilde{c}]\right)+\text{tr}\left([\tilde{A}_\mu,\tilde{b}]_{D}[\tilde{A}^\mu,\tilde{c}]_{D}\right)\\
    &=2i
\begin{pmatrix}
    b^R&b^I
\end{pmatrix}
\begin{pmatrix}
    -4\,\tilde{\mathbf{a}}^R\cdot\tilde{\mathbf{a}}^I & -\Delta_{12}^2+4\,\tilde{\mathbf{a}}^R\cdot\tilde{\mathbf{a}}^R\\
    \Delta_{12}^2-4\,\tilde{\mathbf{a}}^I\cdot\tilde{\mathbf{a}}^I & 4\,\tilde{\mathbf{a}}^R\cdot\tilde{\mathbf{a}}^I
\end{pmatrix}
\begin{pmatrix}
    c^R\\
    c^I
\end{pmatrix}\;.
\end{split}
\end{equation}
Integrating over the $bc$ ghosts yields the Faddeev--Popov determinant,
\begin{equation}
\label{eq:FP-det}
    \begin{split}
        \Delta_{\text{FP}}^{(N=2)}&=\text{det}
    \begin{pmatrix}
        -4\,\tilde{\mathbf{a}}^R\cdot\tilde{\mathbf{a}}^I & -\Delta_{12}^2+4\,\tilde{\mathbf{a}}^R\cdot\tilde{\mathbf{a}}^R\\
        \Delta_{12}^2-4\,\tilde{\mathbf{a}}^I\cdot\tilde{\mathbf{a}}^I & 4\,\tilde{\mathbf{a}}^R\cdot\tilde{\mathbf{a}}^I
    \end{pmatrix}\\
    &=\Delta_{12}^4-4\Delta_{12}^2\,\left(\tilde{\mathbf{a}}^R\cdot\tilde{\mathbf{a}}^R+\tilde{\mathbf{a}}^I\cdot\tilde{\mathbf{a}}^I\right)+16\left(\tilde{\mathbf{a}}^R\cdot\tilde{\mathbf{a}}^R\right)\left(\tilde{\mathbf{a}}^I\cdot\tilde{\mathbf{a}}^I\right)-16\left(\tilde{\mathbf{a}}^R\cdot\tilde{\mathbf{a}}^I\right)^2\;,
\end{split}
\end{equation}
up to an overall constant, recovering the form \eqref{eq:delta_fp_n2}.

Collecting all the terms and ignoring the overall numerical constant, the partition function reads
\begin{equation}
\label{eq:Z-Delta12}
    Z(\Delta_{12})=\int d\tilde{\mathbf{a}}^R\,d\tilde{\mathbf{a}}^I \, \frac{1}{\Delta_{12}^2}\left[ \Delta_{12}^4-4\Delta_{12}^2\left(\left(\tilde{\mathbf{a}}^R\right)^2+\left(\tilde{\mathbf{a}}^I\right)^2\right)+16\left(\tilde{\mathbf{a}}^R\right)^2\left(\tilde{\mathbf{a}}^I\right)^2-16\left(\tilde{\mathbf{a}}^R\cdot\tilde{\mathbf{a}}^I\right)^2 \right]\,e^{-S^{\mathrm{U}(2)}_{\text{gauge}}}.
\end{equation}
Note that $\tilde{\mathbf{a}}^R$ and $\tilde{\mathbf{a}}^I$ each have $(d-1)$ components, so the integral runs over two $(d-1)$-dimensional vectors. It is convenient to rescale
$$\mathbf{X}=\sqrt{\Delta_{12}^2}\,\tilde{\mathbf{a}}^R\equiv p\,\tilde{\mathbf{a}}^R \;\;,\;\;\mathbf{Y}=\sqrt{\Delta_{12}^2}\,\tilde{\mathbf{a}}^I\equiv p\,\tilde{\mathbf{a}}^I,$$
so as to normalize the Gaussian factor in $S^{\mathrm{U}(2)}_{\text{gauge}}$. The partition function then becomes
\begin{equation}
\begin{split}
    Z(p)=\left(\frac{1}{p}\right)^{2(d-1)+2}&\int d\mathbf{X} d\mathbf{Y}\,\left( p^4-4(\mathbf{X}\cdot\mathbf{X}+\mathbf{Y}\cdot\mathbf{Y})+\frac{16}{p^4}\left[(\mathbf{X}\cdot\mathbf{X})(\mathbf{Y}\cdot\mathbf{Y})-(\mathbf{X}\cdot\mathbf{Y})^2\right]\right)\\
    &\times \text{exp}\left\{-\left( \mathbf{X}\cdot \mathbf{X} + \mathbf{Y}\cdot \mathbf{Y} +\frac{4}{p^4}\left[(\mathbf{X}\cdot\mathbf{X})(\mathbf{Y}\cdot\mathbf{Y})-(\mathbf{X}\cdot\mathbf{Y})^2\right]\right) \right\}\;.
\end{split}\label{eq:zpp}
\end{equation}
Apart from the quartic interactions, this is a standard two-vector Gaussian integral. By the $\mathrm{SO}(d-1)$ invariance we may choose $\hat{\mathbf{X}}$ as the polar axis of spherical coordinates, so that $\mathbf{X}\cdot\mathbf{Y}=xy\cos\theta$:
\begin{equation}
\label{eq:Z[p]-1}
\begin{split}
    Z(p)
    =\left(\frac{1}{p}\right)^{2(d-1)+2}&\int_0^\infty\int_0^\infty dx dy\,x^{d-2}y^{d-2}\int_0^\pi d\theta\,\sin^{d-3}\theta\left( p^4 - 4\left(x^2+y^2\right) + \frac{16}{p^4}x^2 y^2 \sin^2\theta \right)\\
    &\times \text{exp}\left\{-\left(x^2+y^2+\frac{4}{p^4}x^2y^2 \sin^2\theta\right)\right\}\;,
\end{split}
\end{equation}
where the Jacobian $\sin^{d-3}\theta$ is the volume of the $(d-3)$-sphere of radius $\sin\theta$. Performing the $\theta$ integration first renders the $x,y$ integrals tractable in terms of special functions; the derivation is given in Appendix~\ref{app:int-XY}. The result is
\begin{align}
\label{eq:Z[p]}
Z(p) &= \frac{\pi\, 2^{1-2d} \Gamma(d-2)}{p^4} \Bigg[
p^4 \left(p^4 - 4d + 4\right)
U\!\left(\frac{d}{2}, \frac{3}{2}, \frac{p^4}{4}\right) \nonumber \\
&\qquad\qquad\qquad\qquad
+\, 2(d-1)\left(p^4 + 2d - 8\right)
U\!\left(\frac{d}{2}, \frac{1}{2}, \frac{p^4}{4}\right)
\Bigg] \, .\footnotemark
\end{align}
\footnotetext{The closed form has been cross-checked against a direct numerical evaluation of the triple integral~\eqref{eq:Z[p]-1}.}
The $U(a,b,c)$ denotes the Kummer confluent hypergeometric function and since the Gamma function $\Gamma(d-2)$ requires $d>2$, the partition function is well defined only above two dimensions. We now analyze its behavior as follows.

\subsection{IR and UV Behavior}
In the infrared (IR) regime the diagonal components are well-separated, $p=\sqrt{\Delta_{12}^2}\gg1$, and the partition function is approximately
\begin{align}
\label{eq:two-body-large-distance}
    Z(p \gg 1) \simeq p^{-2d}\,2^{1-d}\,\pi\left[p^4\,\Gamma\left(d-2\right)-\left(d-1\right)\left(d+2\right)\,\Gamma(d-2)+\frac{d^2+9d+16}{2p^4}\Gamma(d)+\mathcal{O}\left(\frac{1}{p^8}\right)\right]\;.
\end{align}
The leading order, $p^{-2d+4}$, reproduces the one-loop behavior; the $p^{-2d}$ term corresponds to the two-loop amplitude, and so forth.
Note that the expansion parameter here is $p^{-4}$, as can be seen from the interaction term in Eq.~\eqref{eq:zpp}.
Since the Gamma functions in \eqref{eq:two-body-large-distance} are finite for $d>2$, the partition function vanishes as $p\rightarrow\infty$, a prerequisite for the finiteness of the total partition function $Z_{\mathrm{U}(2)}=\int d^dp\,Z(p)$.
In this regime we expect the physics to be perturbative, i.e.\ amenable to a Feynman-diagram interpretation.

In the ultraviolet (UV) regime the diagonal components approach each other, $p=\sqrt{\Delta_{12}^2}\ll1$, and the partition function is approximately
\begin{align}
    Z(p \ll 1) \simeq & ~
    \frac{2^{1-d}\,\pi\,(d-4)\,\Gamma\!\left(\frac{d}{2}-1\right)}{p^4}
    -\frac{2^{2-d}\,\pi\,\Gamma\!\left(\frac{d+1}{2}\right)}{p^2}\notag\\
    &~
    +\frac{\pi\,\Gamma\!\left(\frac{d}{2}-1\right)\left[(d^2-4d+2)\,\Gamma\!\left(\frac{d-1}{2}\right)+8\,\Gamma\!\left(\frac{d+1}{2}\right)\right]}{2^{d+1}\,\Gamma\!\left(\frac{d-1}{2}\right)}
    +\mathcal{O}\left(p^2\right).
\end{align}
The leading term, of order $p^{-4}$, diverges as $p\rightarrow0^+$. The physics of this regime is expected to be nonperturbative and is not yet clear; we shall probe it through the UV limit of the gauge-fixed integrand in the maximal diagonal gauge of Sec.~\ref{sec:max-diag-gauge}.
Note that the leading UV coefficient carries an overall factor $(d-4)$, so the $p^{-4}$ singularity disappears precisely at $d=4$. This is tied to the convergence threshold of the gauge-fixed integral: near the origin $Z(p)\sim p^{-4}$, so the radial integrand $p^{d-1}Z(p)\sim p^{d-5}$ and convergence at small $p$ requires $d>4$~\cite{Krauth:1998xh}.

 A quick consistency check is provided by the convergence of $Z_{\mathrm{SU}(2)}$, known to hold only for $d\geq5$~\cite{Krauth:1998xh,Krauth:1998yu}. 
 Since $Z(p)$ is regular away from the origin and infinity, only the two asymptotic limits need to be examined:
 \begin{equation}
 \begin{split}
     Z_{\mathrm{U}(2)}&=\int\,d^dp\,Z(p)\\
     &=\text{vol}(S_{d-1})\int_0^\infty\,dp\,p^{d-1}\,Z(p)
 \end{split}
 \end{equation}
 As $p\rightarrow0$ the integrand is $\mathcal{O}(p^{d-5})$, requiring $d>4$ for convergence (at $d=4$ the small-$p$ singularity is in fact absent, owing to the $(d-4)$ factor noted above). 
 As $p\rightarrow\infty$ it is $\mathcal{O}(p^{-d+3})$, again requiring $d>4$, with $d=4$ marginally (logarithmically) divergent. 
 Furthermore, the decomposition of $\mathrm{U}(2)$ contains the center-of-mass modes from $[x_\mu\mathbb{I}_2,A_\nu]=0$ and the $\mathrm{SU}(2)$ part, indicating that the $p$-behaviors for $\mathrm{U}(2)$ and $\mathrm{SU}(2)$ are the same.
 The asymptotics are thus consistent with the known result.

It is worth clarifying the two distinct $d$-thresholds that appear here. The bound $d>4$ above is the convergence condition of the \emph{gauge-fixed} two-body integrand: once the Lorenz gauge has been imposed and the radial measure $p^{d-1}dp$ included, both endpoints of the $p$-integral converge for $d>4$. This is not in conflict with the statement that the \emph{full} $\mathrm{SU}(2)$ matrix integral $Z_{\mathrm{SU}(2)}$ converges only for $d\geq5$~\mbox{\cite{Krauth:1998xh,Krauth:1998yu}}: the latter is the convergence of the complete (unreduced) bosonic integral, which is more restrictive because it retains the flat directions and the angular/measure factors that our gauge-fixed, off-diagonal-integrated reduction has already stripped away. In other words, $d>4$ is the threshold for our reduced integrand, while $d\geq5$ is the threshold for the underlying $\mathrm{SU}(2)$ partition function; the gap between them reflects the gauge-fixing and the accompanying measure factors.

\subsection{Negative Region of \texorpdfstring{$Z(p)$}{Z(p)}}\label{sec:negative-region}

\begin{figure*}[!htb]
    \centering
    \begin{subfigure}{0.48\textwidth}
        \centering
        \includegraphics[width=\linewidth]{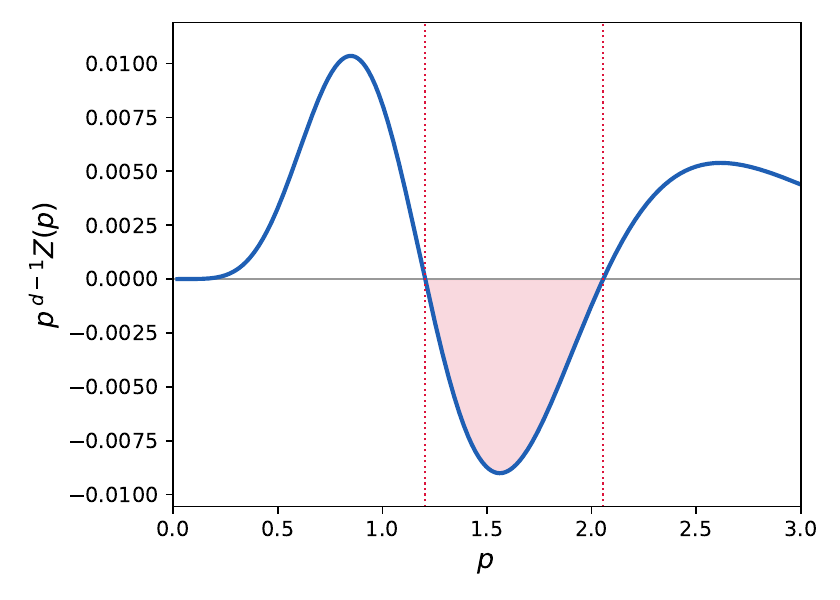}
        \caption{The integrand $p^{d-1}Z(p)$ for $d=10$ over $p\in(0,3)$, computed from the closed form~\eqref{eq:Z[p]}. The curve dips below zero in the window $1.204<p<2.055$ (dotted lines), signaling the Gribov ambiguity diagnosed in Sec.~\ref{sec:gribov-ambiguity}.}
        \label{fig:N=2_Z_1}
    \end{subfigure}
    \hfill
    \begin{subfigure}{0.48\textwidth}
        \centering
        \includegraphics[width=\linewidth]{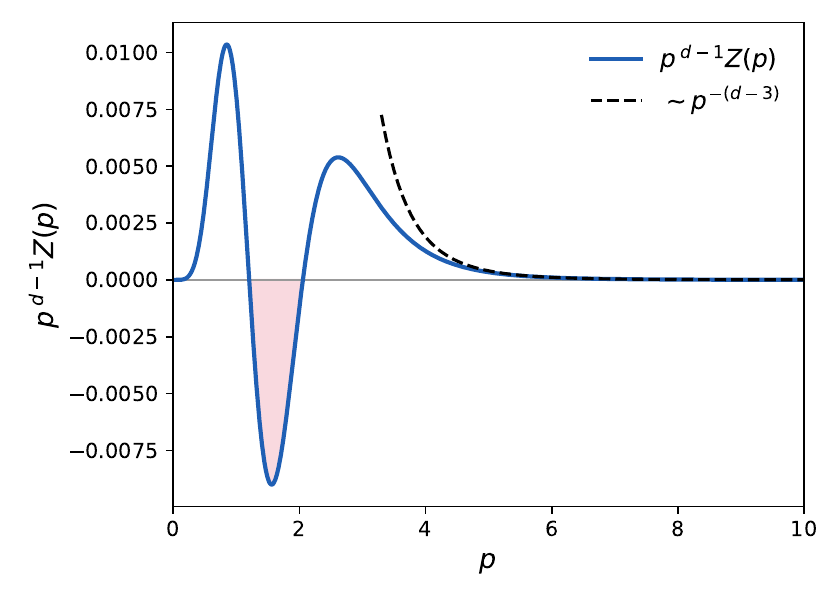}
        \caption{The same integrand $p^{d-1}Z(p)$ for $d=10$ over the wider range $p\in(0,10)$, showing the approach to the perturbative fall-off $p^{d-1}Z(p)\sim p^{-d+3}$ at large $p$.}
        \label{fig:N=2_Z_10}
    \end{subfigure}

    \caption{The integrand $p^{d-1}Z(p)$ for $d=10$ computed from the closed form~\eqref{eq:Z[p]}. Left: zoomed view highlighting the negative region associated with the Gribov ambiguity. Right: wider momentum range showing the asymptotic perturbative behavior.
    \label{fig:N=2_Z_combined}}
\end{figure*}

There is a region around $p\sim\mathcal{O}(1)$ where $Z(p)$ becomes negative; for example, at $d=10$ it is clearly shown in Figs.~\ref{fig:N=2_Z_1}--\ref{fig:N=2_Z_10}.
Although the integrated partition function $Z_{\mathrm{U}(2)}$ remains finite, a negative $Z(p)$ cannot be written as $e^{-V(p)}$ for any real potential $V$, so the interpretation of $Z(p)$ as an effective potential between the two D-instantons fails in this region.
The origin of this non-positivity can be read off directly from the exact Faddeev--Popov determinant~\eqref{eq:FP-det}. While the leading term $\Delta_{12}^4=p^4$ is positive definite, the determinant also contains the indefinite pieces
\begin{equation*}
    -4\Delta_{12}^2\left(\tilde{\mathbf{a}}^R\!\cdot\!\tilde{\mathbf{a}}^R+\tilde{\mathbf{a}}^I\!\cdot\!\tilde{\mathbf{a}}^I\right)
    +16\left[\left(\tilde{\mathbf{a}}^R\!\cdot\!\tilde{\mathbf{a}}^R\right)\left(\tilde{\mathbf{a}}^I\!\cdot\!\tilde{\mathbf{a}}^I\right)-\left(\tilde{\mathbf{a}}^R\!\cdot\!\tilde{\mathbf{a}}^I\right)^2\right]\,,
\end{equation*}
the first of which is manifestly negative. 
In the regime of large diagonal separation, $\Delta_{12}=p\gg1$, the off-diagonal fluctuations are suppressed, scaling as $\tilde{\mathbf{a}}^{R,I}\sim 1/p$.
Consequently, the positive $p^4$ contribution dominates, ensuring that $\langle\Delta_{\text{FP}}^{(N=2)}\rangle > 0$.
Once $p\sim\mathcal{O}(1)$, however, the off-diagonal fluctuations become comparable to $p$, the term $-4\Delta_{12}^2(\tilde{\mathbf{a}}^R\,\cdot\!\tilde{\mathbf{a}}^R+\tilde{\mathbf{a}}^I\!\cdot\!\tilde{\mathbf{a}}^I)$ (together with the positive semi-definite quartic term) is no longer subdominant, and the Gaussian average $\langle\Delta_{\text{FP}}^{(N=2)}\rangle$ can turn negative. This is precisely the negative region visible in Figs.~\ref{fig:N=2_Z_1}--\ref{fig:N=2_Z_10}. As we show in Sec.~\ref{sec:gribov-ambiguity}, this sign-indefinite contribution can be traced to a Gribov copy of the Lorenz gauge with negative Faddeev--Popov determinant.

We emphasize a structural point that motivates the rest of this section. The negativity is present already in the \emph{exact}, all-loop two-body result~\eqref{eq:Z[p]}: it is not an artifact of truncating the large-distance expansion~\eqref{eq:two-body-large-distance}, since the closed-form $Z(p)$ was obtained without any perturbative approximation. Being exact, the result leaves only one possible origin for the negativity: the gauge slice itself. This forces us to scrutinize the Lorenz gauge condition, and, as we show next, the negativity is traced directly to the Gribov ambiguity of that gauge  and resolved by passing to the maximal diagonal gauge.

\section{Gribov Ambiguity \texorpdfstring{and the Classical Frame}{and the Classical Frame}}\label{sec:gribov-ambiguity}

\subsection{Gauge Slice as the Critical-Point Equation}

To diagnose the negativity, we re-examine the Lorenz gauge condition \eqref{eq:gauge-cond}; the negativity enters through the ghost determinant, i.e.\ the second term in Eq.~\eqref{eq:FP-det}. The basic requirement on any gauge condition is the existence of a gauge transformation $U\in \mathrm{U}(N)$ for every matrix configuration $A_\mu$ such that the condition can be satisfied,
\begin{equation}
    \left[P_\mu^U,A_\mu^U\right]=0\,,
\end{equation}
where $A_\mu^U=UA_\mu U^\dagger$ and $P_\mu^U=\left(UA_\mu U^\dagger\right)_D$. This was taken for granted in the previous sections. Here we carry out the proof carefully and show that the negativity originates directly from the Gribov ambiguity of this gauge condition.

The goal is to prove that such $U$ exists for every matrix configuration $A_\mu$. This is a non-linear equation for $U$, so constructing solutions explicitly is impractical. An existence proof can instead be given by recognizing the Lorenz gauge condition \eqref{eq:gauge-cond} as the critical-point equation of the following height function on the group manifold,
\begin{equation}
\label{eq:fA(U)}
    f_A(U) = \tr P_\mu^UP_\mu^U\,,
\end{equation}
where $P_\mu^U$ is the diagonal part of $A_\mu^U$. This scalar function is simply the sum of the squared norms of the diagonal components and is therefore non-negative. It is also bounded by the total inertia, or matrix norm,
\begin{equation}
\label{eq:fA-bounds}
    0\leq f_A(U) \leq \tr A_\mu A_\mu\,.
\end{equation}
Bounded both from below and above, it serves as a smooth height function on $\mathrm{U}(N)$.

We can now make the existence statement rigorous. The map $U\mapsto f_A(U)$ is a continuous (indeed real-analytic) function of $U$, and the gauge group $\mathrm{U}(N)$ is a \emph{compact} manifold. A continuous function on a compact set attains its supremum and infimum (the extreme-value theorem), and by~\eqref{eq:fA-bounds} both are finite. Hence $f_A$ possesses at least one global maximum and one global minimum on $\mathrm{U}(N)$, giving \emph{at least two} extrema for every configuration $A_\mu$. As shown below, an extremum of $f_A$ is exactly a critical point at which the Lorenz gauge condition $[P_\mu^U,A_\mu^U]=0$ holds, so the existence of a gauge representative is guaranteed for every $A_\mu$ with no further assumption.

The first variation of $f_A$ under an infinitesimal gauge transformation reads
\begin{equation}
    \delta_\xi f_A(U) = 2 \tr P_\mu^U \delta_\xi  P_\mu^U = 2 i\tr P_\mu^U  \left[\xi,A_\mu^U\right]_D = 2 i\tr P_\mu^U  \left[\xi,A_\mu^U\right] = -2i\tr\xi\left[P_\mu^U,A_\mu^U\right]\,.
\end{equation}
Since $\delta_\xi f_A(U)$ vanishes for all variations $\xi$ precisely when $[P_\mu^U,A_\mu^U]=0$, the critical points of $f_A$ on the group manifold are exactly the solutions of the Lorenz gauge along the orbit of $A_\mu$. The maximum and minimum guaranteed above are therefore genuine gauge representatives, which completes the existence proof. For generic $A_\mu$ these critical points are isolated modulo the residual gauge: the directions $\xi\in\text{Cartan}$ leave $f_A(U)$ invariant and appear as the flat $\mathrm{U}(1)^N$ directions, while along the remaining $N(N-1)$ directions transverse to the residual gauge the Hessian of $f_A$ is non-degenerate. The configurations for which the transverse Hessian develops a zero mode form a measure-zero set and do not affect the gauge-fixed integral.\footnote{``Isolated'' is therefore always understood modulo the flat $\mathrm{U}(1)^N$ directions.}

It now becomes clear that this gauge fixing encounters the Gribov ambiguity~\cite{Gribov:1977wm,Vandersickel:2012tz,Singer:1978dk}. A proper gauge fixing should select a single $U$ (up to residual gauge transformations) for each $A_\mu$, such that $A_\mu^U$ uniquely satisfies the gauge condition. However, the existence argument above already produces at least a global maximum and a global minimum, and generically $f_A(U)$ possesses additional saddle critical points (for $N=2$ we exhibit exactly one in Sec.~\ref{sec:three-extrema}), so the function carries multiple critical points with positive-, negative- and mixed-signature Hessians. 
The Lorenz-gauge slice therefore intersects each gauge orbit multiple times, where the number of intersections equals that of critical points.\footnote{Similar discussion for the background gauge, $\left[A_\mu^\text{bg},A_\mu\right]=0$, can be found in~\cite{Steinacker:2024unq}, where $A^\text{bg}$ is a constant matrix and a similar scalar function can also be constructed.}

We now make the sign structure of the Faddeev--Popov determinant precise, since it is the source of the negativity diagnosed in Sec.~\ref{sec:negative-region}. The Faddeev--Popov operator of the Lorenz gauge is the Jacobian of the gauge-fixing function $[P_\mu^U,A_\mu^U]$ with respect to the gauge parameter. By the relation $\delta_\xi f_A(U)=-2i\tr\xi[P_\mu^U,A_\mu^U]$, this gauge-fixing function is itself the gradient of $f_A$, so the Faddeev--Popov operator is precisely the \emph{Hessian} of $f_A$ at the critical point, and the Faddeev--Popov determinant is the determinant of that Hessian~\cite{Gribov:1977wm,Vandersickel:2012tz}. When the configuration space is sliced by the bare Lorenz condition $\delta[P_\mu,A_\mu]$, every Gribov copy (every critical point of $f_A$ on the orbit) is retained, and the standard Faddeev--Popov insertion $\sum_{\text{copies}} \det(\text{Hessian})$ replaces the single-copy factor; the gauge-fixed integrand is thus a \emph{sum} of Hessian determinants over the critical points rather than a single positive factor~\cite{Neuberger:1986vv,Neuberger:1986xz}.

The Hessian acts on the $N(N-1)$-dimensional tangent space transverse to the residual $\mathrm{U}(1)^N$ gauge, which is of \emph{even} dimension. Consequently a maximum or a minimum, where the Hessian is sign-definite (all eigenvalues of one sign), has a manifestly \emph{positive} determinant, whereas a saddle with $n_-$ negative Hessian eigenvalues has a determinant of sign $(-1)^{n_-}$ and can therefore be \emph{negative} for odd $n_-$. Since the negativity of $Z(p)$ is exact, hence intrinsic to the gauge slice rather than to any perturbative truncation, it is resolved here as the negative Faddeev--Popov determinant of a saddle Gribov copy.

\subsection{Three Extrema in \texorpdfstring{$N=2$}{N=2}}\label{sec:three-extrema}

Here we use the previous calculation to exhibit the Gribov ambiguity in the $N=2$ matrix model. Recall that, in component form, the Lorenz gauge condition \eqref{eq:gauge-cond} consists of two orthogonality conditions,
\begin{equation}
    0=\Delta_{12}\cdot \tilde{a}^R=\Delta_{12}\cdot \tilde{a}^I\,.
\end{equation}
In the color basis defined by $A_\mu=A^a_\mu\,\sigma_a/2$---so that, in the parametrization above, $A^3_\mu=\Delta_{\mu,12}$, $A^1_\mu=2\tilde{a}^R_\mu$ and $A^2_\mu=-2\tilde{a}^I_\mu$---this requires
\begin{equation}
    0=A_\mu^3A_\mu^1 = A_\mu^3A_\mu^2\,.
\end{equation}
If this were a unique gauge-fixing condition, a unique $A_\mu^3$ (or $\Delta_\mu$) would be obtained. To see that this is not the case, we consider the following $3\times3$ real symmetric matrix,
\begin{equation}
    M_{ab} = A_{a\mu}A_{b\mu} = \mathbf{A}_a\cdot \mathbf{A}_b\,,
\end{equation}
where $M_{33}$ evaluates to $\Delta^2$. The two orthogonality conditions are exactly the statement that the off-diagonal entries in the ``$3$''-row of $M$ vanish, $M_{13}=M_{23}=0$, while the diagonal entry $M_{33}=\mathbf{A}_3\cdot\mathbf{A}_3=\Delta^2$ sitting in the ``$3$'' slot is proportional to the height function of Eq.~\eqref{eq:fA(U)}, $f_A(U)=\tr P_\mu P_\mu=M_{33}/2$. Our goal is to find a $U\in \mathrm{SU}(2)$ that rotates $M$ so that these conditions are satisfied.

Since $\mathrm{SU}(2)\simeq\mathrm{SO}(3)$, the gauge rotation $U$ acts on the color indices of $A_\mu^a$, and hence on $M_{ab}=\mathbf{A}_a\cdot\mathbf{A}_b$ as a real $\mathrm{SO}(3)$ similarity rotation $M\mapsto O\,M\,O^T$. Because $M$ is a real symmetric $3\times3$ matrix, it is diagonalized by such a rotation, $O\,M\,O^T=\text{diag}(\lambda_1,\lambda_2,\lambda_3)$ with real eigenvalues $\lambda_a$, and in the diagonal frame \emph{all} off-diagonal entries---in particular $M_{13}$ and $M_{23}$---vanish identically. Diagonalizing $M$ therefore automatically solves the Lorenz conditions, with $M_{33}$ holding one of the three eigenvalues. The ordering of the eigenvalues, however, can be chosen freely: which eigenvalue is rotated into the ``$3$'' slot, i.e.\ which one is identified with $\Delta^2=M_{33}$, is not fixed by $M_{13}=M_{23}=0$ alone. The three assignments $\Delta^2 = \lambda_1\,,\lambda_2\,,\lambda_3$ thus give three distinct $\mathrm{SO}(3)$ rotations, all satisfying the Lorenz gauge. As a consequence, for $N=2$ the Lorenz-gauge slice intersects the gauge orbit three times---these are the three Gribov copies~\cite{Gribov:1977wm,Vandersickel:2012tz}.

These three copies are precisely the critical points of $f_A(U)$. Since $f_A(U)=M_{33}/2$, with $M_{33}$ the eigenvalue placed in the ``$3$'' slot, ordering $\lambda_1>\lambda_2>\lambda_3$, the assignment $\Delta^2=\lambda_1$ (largest) maximizes $f_A$ along the orbit and is the \emph{maximum}; $\Delta^2=\lambda_3$ (smallest) is the \emph{minimum}; and $\Delta^2=\lambda_2$ (middle) is the intermediate critical point, the \emph{saddle}. By the Hessian-sign analysis of Sec.~\ref{sec:gribov-ambiguity}, the tangent space transverse to the residual gauge has the even dimension $N(N-1)=2$, so the maximum and the minimum, where the Hessian is sign-definite, carry a positive Faddeev--Popov determinant, whereas the saddle, having a single negative Hessian direction, carries a \emph{negative} Faddeev--Popov determinant. The middle eigenvalue $\lambda_2$ is thus the source of the negative contribution.

The previous $N=2$ Lorenz-gauge partition function obtained above therefore actually composes three Gribov copies,
\begin{equation}
    Z_\text{Lorenz gauge}^{N=2}(p) = \sum_{i\in\text{Gribov copies}}(-1)^{n_-^i} \, I_i(p) = I_\text{max}(p)-I_\text{saddle}(p)+I_\text{min}(p)\,,
\end{equation}
where $I_\text{max}$, $I_\text{saddle}$ and $I_\text{min}$ are the contributions of the maximum ($n_-^\text{max}=2$), the saddle ($n_-^\text{saddle}=1$) and the minimum ($n_-^\text{min}=0$), respectively; note that $n_-$ counts the negative eigenvalues of the Hessian of $f_A$, which is negative definite at the maximum. In other words, each $I_k$ is the partition function gauge-fixed by the condition that $\tr P_\mu P_\mu$ be the maximum, the saddle or the minimum of $f_A$ along the orbit, and each such condition selects exactly one configuration from each gauge orbit.

\begin{figure}[t]
  \centering
  \includegraphics[width=\textwidth]{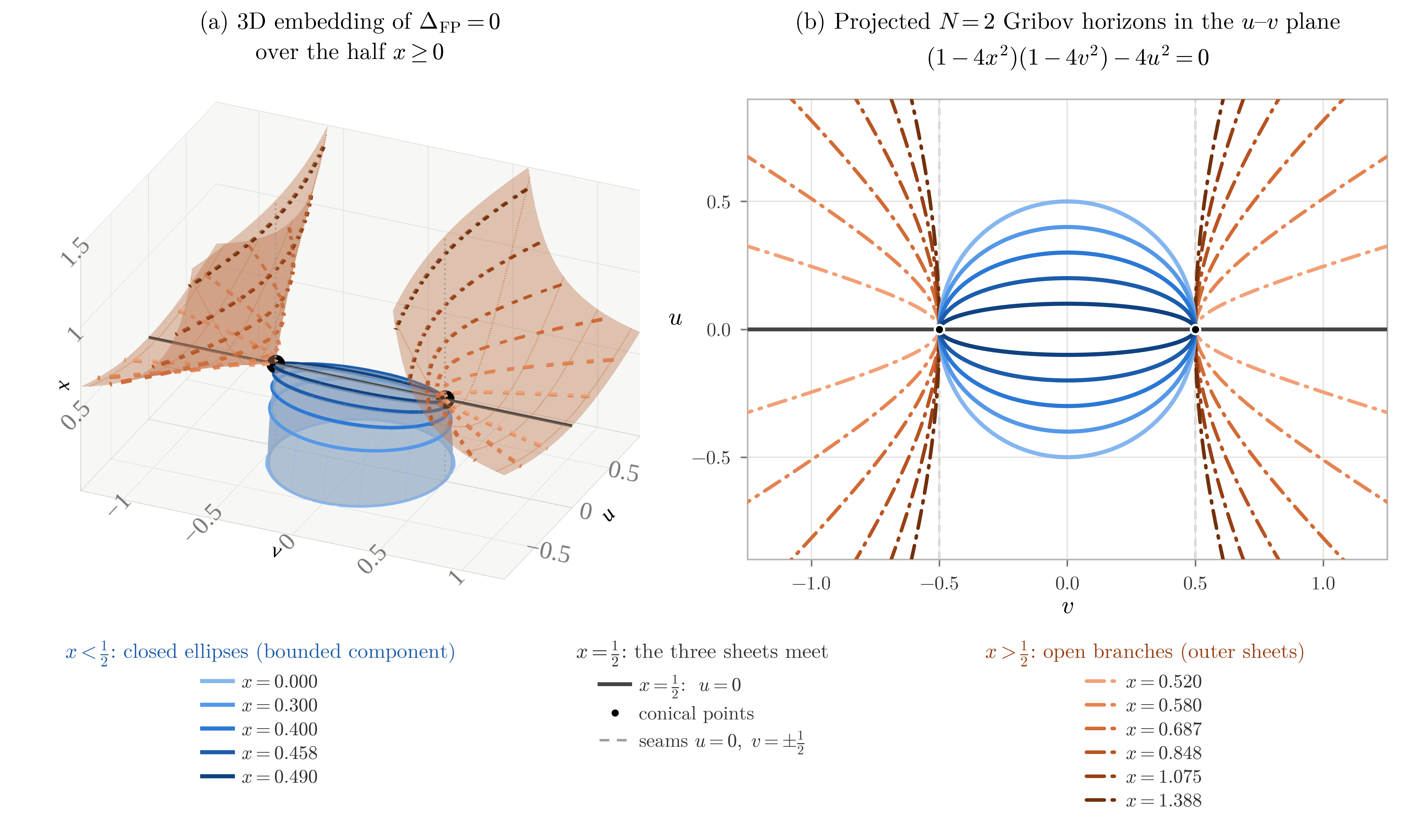}
  \caption{%
    The Lorenz-gauge Gribov horizon for $N=2$, shown as a two-dimensional
    surface in $(u,v,x)$ together with various fixed-$x$ sections in the $(u,v)$
    plane. The horizon satisfies $(1-4x^{2})(1-4v^{2})-4u^{2}=0$. In (a) the
    half $x\ge 0$ is drawn with $x$ vertical, so that the sections are
    horizontal cuts, each in the colour it carries in (b); the half $x<0$ is the
    mirror image. For $x<1/2$ the horizon is the bounded component (blue), whose
    sections are closed ellipses shrinking from the circle $4u^{2}+4v^{2}=1$ at
    $x=0$ to the segment $\{u=0,\,|v|\le 1/2\}$ at $x=1/2$. For $x>1/2$ it
    becomes two open sheets (orange), with hyperbolic sections growing out of
    the rays $\{u=0,\,|v|\ge 1/2\}$. At $x=1/2$ the horizon degenerates to the
    full line $u=0$, along which the three sheets meet; the two conical points
    $(u,v)=(0,\pm 1/2)$, through which every section passes, are marked. The
    bounded component separates the maximal-diagonal region from the saddle
    region, while the outer branches separate the saddle region from the
    minimal-diagonal regions.%
  }
  \label{fig:gribov-horizon}
\end{figure}

However, it is not easy to write these gauge-fixing conditions explicitly in terms of Lorenz-gauge delta functions: on top of the Lorenz condition $\delta[P_\mu,A_\mu]$, one would need to restrict the path-integral domain to configurations with a fixed signature of the Hessian, i.e.\ to a single Gribov region, bounded by the Gribov horizon on which the Faddeev--Popov determinant, Eq.~\eqref{eq:FP-det}, vanishes. See Fig.~\ref{fig:gribov-horizon} for a simple illustration. The functional form of the horizon is difficult to obtain and the integration is hard to carry out. We will adopt a treatment specialized to the $N=2$ matrix model to circumvent this difficulty and obtain the explicit form of $I_\text{max},\,I_\text{saddle}$ and $I_\text{min}$.


\subsection{Singular Value Decomposition}\label{sec:svd}

For $N=2$, owing to the isomorphism $\mathrm{SU}(2)\simeq \mathrm{SO}(3)$ acting together with the spacetime rotations $\mathrm{SO}(d)$, we can perform a singular value decomposition of $A_\mu^a$,
\begin{equation}
    A_\mu^a =
\begin{pmatrix}
    A_1^1 & A_2^1 & A_3^1 & \cdots & A_d^1\\
    A_1^2 & A_2^2 & A_3^2 & \cdots & A_d^2\\
    A_1^3 & A_2^3 & A_3^3 & \cdots & A_d^3\\
\end{pmatrix}
\to U_{ab}\,A_\nu^b\, R^T_{\,\nu\mu} = 
\begin{pmatrix}
    a & 0 & 0 & \cdots & 0\\
    0 & b & 0 & \cdots & 0\\
    0 & 0 & c & \cdots & 0\\
\end{pmatrix}\,,
\end{equation}
where $a>b>c>0$. From here on the letters $a,b,c$ denote these ordered singular values (the ghosts having already been integrated out); the eigenvalues of the matrix $M_{ab}$ of Sec.~\ref{sec:three-extrema} are their squares, $\lambda_1=a^2$, $\lambda_2=b^2$, $\lambda_3=c^2$, so the three Gribov copies correspond to identifying the diagonal separation with one of $a$, $b$, $c$. With this change of variables, the path integral reduces to a three-variable integral,
\begin{equation}
    Z=\int dA \,e^{-S} = \int_0^\infty da\int_0^a db \int_0^b dc \,J(a,b,c) \,e^{-S}\,,
\end{equation}
with the Jacobian~\cite{Smilga:1984jg}
\begin{equation}
    J(a,b,c) = (abc)^{d-3}\left(a^2-b^2\right)\left(a^2-c^2\right)\left(b^2-c^2\right)\,,
\end{equation}
and action
\begin{equation}
    S(a,b,c)=\frac{1}{4}\left(a^2b^2+b^2c^2+c^2a^2\right)\,.
\end{equation}
The maximal, saddle and minimal diagonal gauge fixings then correspond to
\begin{align}
    &I_\text{max}(a)=\frac{1}{a^{d-1}}\int_0^a db\int_0^b dc \,J\,e^{-S}\,,\\
    &I_\text{saddle}(b)=\frac{1}{b^{d-1}}\int_b^\infty da\int_0^b dc \,J\,e^{-S}\,,\\
    &I_\text{min}(c)=\frac{1}{c^{d-1}}\int_c^\infty da\int_c^a db \,J\,e^{-S}\,,
\end{align}
by treating one singular value as the diagonal component and the other two as off-diagonal components. The non-trivial integration limits are exactly the Gribov horizons for the three gauge choices. Concretely, for a fixed diagonal separation the three gauge choices correspond to identifying it with the largest, middle or smallest singular value: the largest gives $I_\text{max}$ ($a$ diagonal; region $0<c<b<a$), the middle gives $I_\text{saddle}$ ($b$ diagonal; region $0<c<b<a$, with $a$ unbounded above), and the smallest gives $I_\text{min}$ ($c$ diagonal; region $c<b<a$, with $a$ unbounded above). The prefactor $1/(\text{diagonal value})^{d-1}$ in each $I_k$ is the residual-gauge volume divided out, and the remaining Jacobian factors $J/(\text{diagonal value})^{d-3}$ leave the two integration variables raised to the power $d-3$. In each gauge the singular value identified with the diagonal block equals the gauge-fixed separation, $a$ (resp.\ $b$, $c$) $=\sqrt{\Delta_{12}^2}\equiv p$, so below we write all three as functions of the common argument $p$.

\begin{figure}[t]
\centering
\includegraphics[width=\linewidth]{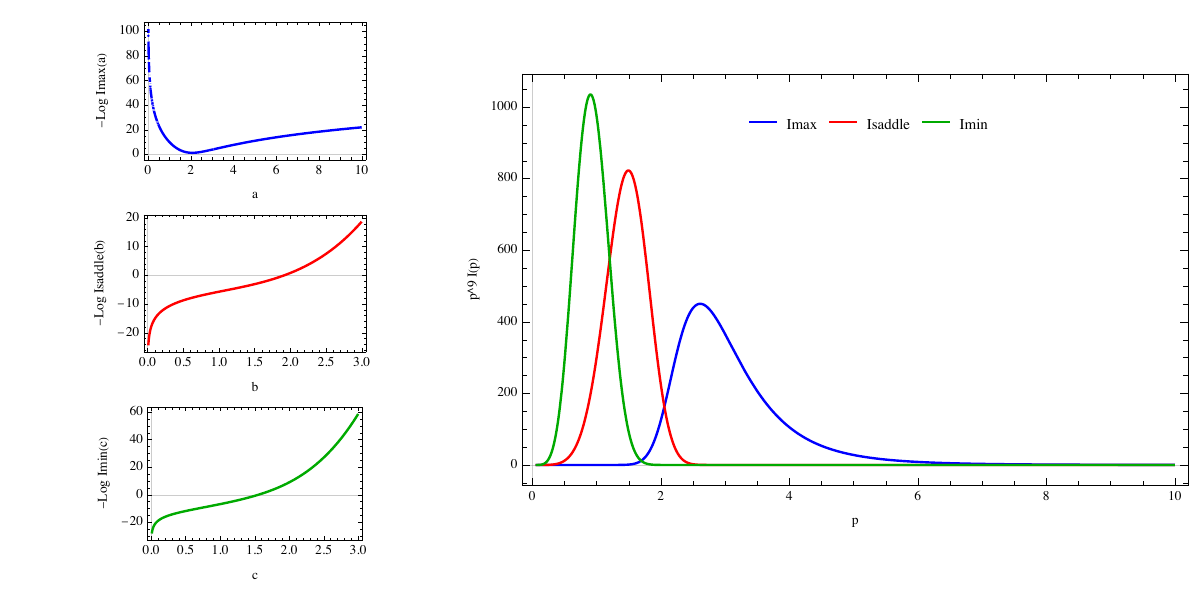}
\caption{The exact $N=2$ partition function in different gauges (left) and the corresponding effective potential (right) as functions of the D-instanton separation  (the $d=10$ closed forms of Appendix~\ref{app:I-integrals}). Only the maximal copy is finite at short distance while carrying the one-loop power law at large distance; the resulting potential rises at both ends and develops the stable minimum discussed in the text.}
\label{fig:partition-veff}
\end{figure}

For general $d$, a single closed form has not been obtained; the $d=10$ closed forms are collected in Appendix~\ref{app:I-integrals}. A sketch of the results is in Fig.~\ref{fig:partition-veff}. The obstruction is structural: performing the innermost Gaussian integral over the $(abc)^{d-3}$ measure produces incomplete-Gamma functions $\gamma\!\big(\tfrac{d-2}{2}+n,\lambda\,(\cdot)^2\big)$ whose order depends on $d$, and these reduce to elementary functions (polynomial$\,\times\,e^{-\kappa p^4}$ for integer order, plus $\operatorname{erf}/\operatorname{erfc}$ for half-integer order) \emph{only} when $\tfrac{d-2}{2}$ is an integer or half-integer; for generic $d$ they do not terminate. The dimension $d=10$ is special precisely because $\tfrac{d-2}{2}=4$ is an integer, so every order encountered after the polynomial weighting is integer or half-integer and the series terminates, yielding the $\operatorname{erf}/\operatorname{erfc}$/exponential/polynomial closed forms of Appendix~\ref{app:I-integrals}. With those $d=10$ closed forms one verifies the equivalence
\begin{equation}\label{eq:I123-equals-Z}
    I_\text{max}(p) -I_\text{saddle}(p)+I_\text{min}(p) = \frac{2^{18}}{\pi}\,Z_\text{Lorenz gauge}(p)\,,
\end{equation}
where the constant $2^{18}/\pi$ is $p$-independent (with $Z(p)$ normalized as in the closed form Eq.~\eqref{eq:Z[p]}). The relative minus sign on the saddle term $I_\text{saddle}$ is the sign assignment of Sec.~\ref{sec:three-extrema}, and it is the unique source of the negative region of $Z(p)$. See Appendix~\ref{app:I-integrals} for the explicit $d=10$ forms and the UV/IR asymptotics and Fig.~\ref{fig:partition-veff} for a sketch.

\subsection{\texorpdfstring{ Classical Frame and }{Classical Frame and }UV Finiteness in the Maximal Diagonal Gauge}\label{sec:max-diag-gauge}

Here we argue that among the three copies at $N=2$ (or the many copies at general $N$), the Gribov copy corresponding to the maximum of the height function, Eq.~\eqref{eq:fA(U)}, is the appropriate choice for the D-instanton interpretation. This choice is not arbitrary; the maximal copy is singled out as the \emph{classical frame} of the theory by the following two properties.

 \emph{(i) The maximal copy is the frame of the classical configuration.} Adopting the standard identification of the diagonal components as the D-instanton positions and the off-diagonal components as their interactions, we see that maximizing $f_A=\tr P_\mu P_\mu$ by definition selects the rotation that places the largest possible share of the conserved matrix ``inertia'' $\tr A_\mu A_\mu$ (the orbit invariant in Eq.~\eqref{eq:fA-bounds}) into the diagonal block $P_\mu$, leaving the smallest possible share in the off-diagonal (``interaction'') block --- precisely the frame in which the diagonal entries are most cleanly read off as D-instanton positions. This \emph{maximal diagonal gauge}---the matrix-model counterpart of the maximal Abelian gauge of lattice gauge theory~\cite{Kronfeld:1987ri}---is the same frame employed in the numerical analysis of~\cite{Hotta:1998en}.\footnote{In Ref.~\cite{Hotta:1998en} this gauge choice is used to define the ``quantumness'' of the emergent spacetime from the matrix model, with numerical simulations performed up to matrix size $N=256$. Although the Gribov horizon hinders perturbative analytic calculations in the UV regime, imposing the maximal diagonal gauge in numerical simulations is a well-established Lie-group optimization problem~\cite{AbsilMahonySepulchre2008}.} Crucially, the purely diagonal configuration $\tilde{A}_\mu=0$ is itself the global maximum of $f_A$, saturating the bound~\eqref{eq:fA-bounds}: the maximal copy is the Gribov region \emph{containing the classical configuration} about which perturbation theory is organized---the matrix-model analogue of the Gribov--Zwanziger restriction to the region of the perturbative vacuum~\cite{Gribov:1977wm,Vandersickel:2012tz}.

 \emph{(ii) In this frame the off-diagonal fluctuation is bounded by the separation.} Since the maximal copy assigns the largest singular value to the diagonal block, the off-diagonal components can never exceed the diagonal separation. For $N=2$ this is explicit in the singular-value domain of Sec.~\ref{sec:svd}, $0<c<b<a=p$: the off-diagonal scale is bounded by $p$ and vanishes as $p\to0$, so the fluctuations stay controlled as the separation shrinks. The saddle and minimal copies, by contrast, integrate the largest singular value over the unbounded range $a\in[b,\infty)$, and it is this that produces their $p^{-4}$ short-distance divergence. This bound is the origin of the UV finiteness derived below.

 A consequence of (ii) is that the maximal copy is also the frame in which the many-body decomposition of Sec.~\ref{sec:body_expansion} is best ordered: with the off-diagonal fluctuations bounded by the separations, the exact $N=2$ two-body interaction is expected to be the leading, representative contribution, and the higher-body pieces enter as corrections. The extent to which they remain subdominant at general $N$ is discussed in Sec.~\ref{sec:general-N}.

This gauge choice yields an interesting scaling law in the UV limit, $p\ll \mathcal{O}(1)$. For $N=2$ the three gauge choices have sharply different IR and UV asymptotics (Appendix~\ref{app:I-integrals}, Fig.~\ref{fig:partition-veff}); their assembly into a single picture is given in Sec.~\ref{sec:large-distance-agreement}, and here we derive the UV scaling of the maximal copy.

The finite UV behavior can be understood through the change of variables $b=ax$, $c=ay$, which gives
\begin{equation}\label{eq:I1-rescaled}
    I^{N=2}_\text{max}(p)=p^{2d}\int_0^1 dx \int_0^x dy \,x^{d-3}y^{d-3}\,(1-x^2)(1-y^2)(x^2-y^2)\,\exp\left(-\frac{p^4}{4}\left(x^2+y^2+x^2y^2\right)\right)\,.
\end{equation}
The scale dependence now appears only as an overall effective coupling in the action, so as $p\to0$ the weight $e^{-S}$ approaches unity and
\begin{equation}
    I^{N=2}_\text{max}(p\to 0) \sim \mathcal{O}(p^{2d} )\,.
\end{equation}

A finite UV result in the vanishing-$p$ limit may seem counterintuitive: the homogeneity of the action naively renders the fluctuations unbounded; for instance, the propagator $\langle \tilde{a}\tilde{a}\rangle\sim1/\Delta^2$ grows without bound. Once the $\Delta$ dependence is scaled out into an overall factor of the action, the system resembles a strong-coupling regime in which every configuration contributes almost equally. 

The width of the fluctuations can be estimated as follows. After scaling out the $\Delta$ dependence, the path-integral weight becomes
\begin{equation}
    \exp\left\{-\mathcal{O}\left(\Delta^4\right) \left[ S_2(\tilde{a}) +S_3(\tilde{a})+S_4(\tilde{a}) \right] \right\}\,,
\end{equation}
where $S_2\,,S_3\,,S_4$ correspond to the quadratic, cubic and quartic terms in the action. The characteristic length is then determined by $S_4\sim\mathcal{O}(\tilde{a}^4)$\footnote{In the fluctuation direction, where the configuration is nearly commuting, the fluctuation scale is determined from $S_2$ to be $\mathcal{O}(1/p^2)$. The UV divergence of $I_\text{saddle}$ and $I_\text{min}$ in $N=2$ can be seen from the contribution of the commuting direction. This large anisotropy does not affect the Gribov-horizon argument for UV finiteness.}, with
\begin{equation}\label{eq:a-action-width}
    ||\tilde{a}||_\text{action}\sim \mathcal{O}(1/p)\,.
\end{equation}

In the maximal diagonal gauge, however, the fluctuation is bounded by the Gribov horizon, located where the Faddeev--Popov determinant first vanishes and changes sign. After the rescaling, every component of the Faddeev--Popov matrix is of $\mathcal{O}(1)$, so we expect the horizon to be located at\footnote{The Gribov horizon may have a non-trivial dependence on $N$. Here we ignore this issue and focus only on the $p$ scaling. The large-$N$ behavior is left for future work.}
\begin{equation}\label{eq:a-gribov-width}
    ||\tilde{a}||_\text{Gribov}\sim \mathcal{O}(1)\,.
\end{equation}

The two scales~\eqref{eq:a-action-width} and~\eqref{eq:a-gribov-width} compete, and this competition is the heart of the non-collapse mechanism. This scale competition is the physical reading of the compact-domain derivation~\eqref{eq:I1-rescaled}---the horizon $x=1$ makes the domain compact and $p$-independent while $e^{-S}\to1$ uniformly on it---rather than an independent estimate. As $p\to0$ the action-set width $\|\tilde{a}\|_\text{action}\sim1/p$ \emph{diverges}: the quartic term becomes a vanishingly weak coupling and the off-diagonal fluctuations are naively unbounded---the strong-coupling regime that produces the spurious $\mathcal{O}(p^{-4})$ divergence in the saddle and minimal copies $I_\text{saddle},I_\text{min}$. The maximal diagonal gauge, however, restricts the off-diagonal fluctuation to the interior of the Gribov horizon, $\|\tilde{a}\|_\text{Gribov}\sim\mathcal{O}(1)$, which stays finite as $p\to0$ (in the rescaled units; in physical units this is precisely the bound of property (ii), off-diagonal $\lesssim p$). In this limit the cutoff $\|\tilde{a}\|_\text{Gribov}$ is much smaller than the would-be action scale $\|\tilde{a}\|_\text{action}\sim1/p$, so the Boltzmann weight $e^{-S}\to1$ is essentially constant over the entire admissible region and the distribution of $\tilde{a}$ inside the horizon is nearly flat. The integration over $\tilde{a}$ then simply measures the finite volume enclosed by the Gribov horizon, which scales with the number of off-diagonal degrees of freedom rather than diverging; at short distance this corresponds to a repulsive, logarithmic potential. Picking the maximal copy thus trades the unbounded fluctuations of the other copies for a finite horizon-enclosed volume, yielding the finite short-distance behavior $Z\sim\mathcal{O}(p^{2d})$ of the maximal copy found above. The extrapolation of this scaling to general $N$, together with the obstructions to establishing it, is collected in Sec.~\ref{sec:general-N}.

\subsection{Large-Distance Agreement and Non-Collapse \texorpdfstring{ in the Maximal Diagonal Gauge}{in the Maximal Diagonal Gauge}}\label{sec:large-distance-agreement}

We can now assemble the two regimes into a single picture, which is the central payoff of this analysis. The decomposition $Z_\text{Lorenz gauge}(p)=I_\text{max}(p)-I_\text{saddle}(p)+I_\text{min}(p)$ of Sec.~\ref{sec:three-extrema} sums the maximal, saddle and minimal Gribov copies, and the verified $d=10$ asymptotics of Appendix~\ref{app:I-integrals} show that these three copies behave very differently in the two limits.

At \emph{large separation}, $\Delta_{ij}^2\gg1$ (the $p\to\infty$ regime of Appendix~\ref{app:I-integrals}), only the maximal copy carries the one-loop power law, $I_\text{max}\simeq 2580480\,p^{-16}+\mathcal{O}(p^{-20})$, i.e.\ $\mathcal{O}(p^{-2d+4})$ for $d=10$, while the saddle and minimal copies are exponentially suppressed, $I_\text{saddle}\sim e^{-p^4/4}$ and $I_\text{min}\sim e^{-3p^4/4}$. Hence in this regime
\begin{equation}\label{eq:Z-large-distance-I1}
    Z(p)=I_\text{max}(p)-I_\text{saddle}(p)+I_\text{min}(p)\;\simeq\; I_\text{max}(p)\,,\qquad p^2\gg1\,,
\end{equation}
up to corrections that are exponentially small in $\Delta^4$. In other words, the naive Lorenz-gauge many-body decomposition of Sec.~\ref{sec:body_expansion} \emph{already} coincides with the maximal-diagonal result at large distance: the exponentially small saddle and minimum contributions cannot disturb the power-law two-body force, and no Gribov care is needed there. 

This match explains why the usual perturbation calculation can safely ignore the Gribov ambiguity. If the fluctuation scale is much smaller than the Gribov horizon, the multiple copies problem can be bypassed as the other copies are not seen in the perturbative regime. Since the fluctuation center here is $\tilde{A}_\mu=0$, which is the global maximum of $f_A$ (property (i) of Sec.~\ref{sec:max-diag-gauge}), it is enclosed by the maximal copy, and the loop expansion never encounters the saddle or minimal copies. This is manifested in the large-distance agreement  above, at the level of the two-body force.

At \emph{short distance}, $\Delta_{ij}^2\ll1$ (the $p\to0$ regime), the hierarchy is inverted: the saddle and minimal copies diverge, $I_\text{saddle},I_\text{min}\sim \mathcal{O}(p^{-4})$, while the maximal copy vanishes, $I_\text{max}\sim\mathcal{O}(p^{2d})$. The negative saddle term $-I_\text{saddle}$ then drives the spurious negativity of $Z(p)$ diagnosed in Sec.~\ref{sec:gribov-ambiguity}, which is an artifact of summing over Gribov copies rather than a genuine feature of the two-instanton force. Here the maximal diagonal gauge is the appropriate choice: retaining the single maximal copy discards the spurious $I_\text{saddle},I_\text{min}$ and leaves the finite, repulsive short-distance behavior $Z\sim\mathcal{O}(p^{2d})$ derived above~\eqref{eq:I1-rescaled}. This short-distance repulsion---a logarithmic potential generated by the finite Gribov-horizon volume rather than by any divergence---is what underlies the non-collapse of the two D-instantons: as their separation shrinks, the partition function vanishes rather than blowing up, so the configurations are pushed apart.

 Combining the two regimes into an effective two-body potential $V(\Delta)=-\log Z(\Delta)$, with $Z=I_\text{max}>0$ in the maximal diagonal gauge, makes the conclusion sharp: the large-separation behavior $Z\sim\Delta^{-2d+4}$ reproduces the attractive one-loop logarithm of Sec.~\ref{sec:intro}, $V\simeq(d-2)\log\Delta^2$, while the short-distance repulsion just described corresponds to $V\simeq-2d\log\Delta\to+\infty$. Rising at both ends, $V(\Delta)$ possesses a minimum at a finite separation $\Delta_\ast\sim\mathcal{O}(1)$: the two D-instantons neither collapse onto one another nor fly apart, but settle at a preferred separation. This stable two-body minimum is the precise statement of the non-collapse established here; the full $N$-body distribution requires the higher-body potentials (Sec.~\ref{sec:general-N}).


\section{General \texorpdfstring{$N$}{N}: Conjectures and Open Questions}\label{sec:general-N}

 The results established so far concern the two-body sector: the exact $N=2$ ($\mathrm{U}(2)$) computation of Secs.~\ref{sec:n=2}--\ref{sec:gribov-ambiguity} is complete and free of approximation. This section discusses the extension to the full $N$-body problem. The guiding picture is the following: in the maximal diagonal gauge the off-diagonal fluctuation of every pair is bounded by that pair's separation, so that at short distance each pair behaves as in the $N=2$ model, and the partition function is expected to reduce to the corresponding product of pair contributions. Part of this picture can be derived---the pairwise bound below holds exactly for any $N$---while the remaining ingredients, namely the exact Gribov horizon, the copy structure and the measure factors at general $N$, are not under analytic control. We therefore first state the extrapolated scalings, clearly labeled as conjectures, then derive the pairwise bound that supports them, and close with the questions left open.

 The maximal-diagonal-gauge partition function vanishes as $Z\sim\mathcal{O}(p^{2d})$ at short distance for $N=2$. Since the maximal diagonal gauge treats all off-diagonal pairs on the same footing, the natural extrapolation to general $N$, with all $p_\mu^{(i)}$ of a common order $\mathcal{O}(\bar{p})$, is
\begin{equation}\label{eq:Z-UV-scaling}
    Z_\text{max. diag. gauge}\left(p_\mu^{(i)}\right)\sim\mathcal{O}\left( \bar{p}^{\,d(N^2-N)}\right)\,,
\end{equation}
 and, resolving the dependence on the individual separations,
\begin{equation}\label{eq:Z-UV-scaling-fine}
    Z_\text{max. diag. gauge}\left(p_\mu^{(i)}\right)\sim\mathcal{O}\left(\prod_{i<j}\left(p^{(i)}-p^{(j)}\right)^{2d}\right)\,.
\end{equation}
 We stress that \eqref{eq:Z-UV-scaling}--\eqref{eq:Z-UV-scaling-fine} are \emph{conjectures} for $N\ge3$, not derived results. The explicit three-extrema and singular-value analysis underlying the $N=2$ horizon-volume counting relied crucially on the accident $\mathrm{SU}(2)\simeq\mathrm{SO}(3)$, which let the gauge rotation act as a real $\mathrm{SO}(3)$ diagonalization of the $3\times3$ matrix $M_{ab}=\mathbf{A}_a\cdot\mathbf{A}_b$. This isomorphism does \emph{not} hold for $N\ge3$, so neither the number and structure of the Gribov copies nor the general-$N$ horizon volume is established; \eqref{eq:Z-UV-scaling}--\eqref{eq:Z-UV-scaling-fine} are written down under the assumption that the maximal diagonal gauge treats all off-diagonal pairs identically.

 The finer form \eqref{eq:Z-UV-scaling-fine} presupposes that the only scale governing each off-diagonal block $\tilde{a}_{ij}$ is its own separation $\Delta_{ij}$. This presupposition can in fact be derived: the $N=2$ bound of Sec.~\ref{sec:max-diag-gauge}---the off-diagonal block never exceeds the separation---extends to \emph{every pair} at general $N$, as follows. Consider the $\mathrm{SU}(2)$ subgroup of $\mathrm{U}(N)$ that rotates only the $(i,j)$ plane. The height function $f_A$ of Eq.~\eqref{eq:fA(U)} is built solely from the diagonal entries; under such a rotation the diagonal entries with $k\neq i,j$ are untouched, while the $i$-th and $j$-th entries transform only through the pair's $2\times2$ sub-block, formed by $p^{(i)}_\mu$, $p^{(j)}_\mu$ and $\tilde{a}_{\mu,ij}$. The restriction of $f_A$ to this sub-orbit is therefore the $N=2$ problem of Sec.~\ref{sec:three-extrema} applied to the sub-block: decomposing $\tilde{a}_{\mu,ij}=\tilde{a}^R_{\mu,ij}+i\,\tilde{a}^I_{\mu,ij}$ and forming the $3\times3$ Gram matrix $M^{(ij)}$ of the three $d$-vectors $\big(2\tilde{a}^R_{ij},\,-2\tilde{a}^I_{ij},\,\Delta_{ij}\big)$, in exact analogy with the matrix $M_{ab}$ there, one finds $f_A=\mathrm{const}+M^{(ij)}_{33}/2$ along the sub-orbit, with $M^{(ij)}_{33}=\Delta_{ij}^2$ and the constant collecting the invariant entries. A configuration in the maximal diagonal gauge maximizes $f_A$ along the full orbit, hence along every such sub-orbit. Under the sub-orbit rotation, $M^{(ij)}_{33}=v^{T}M^{(ij)}v$ with the unit vector $v$ (the third row of the $\mathrm{SO}(3)$ rotation, as in Sec.~\ref{sec:three-extrema}) ranging over the whole sphere, so maximality fixes $\Delta_{ij}^2=M^{(ij)}_{33}=\lambda_{\max}\big(M^{(ij)}\big)$ by the Rayleigh principle---for every pair. Since $M^{(ij)}$ is positive semi-definite with $\tr M^{(ij)}=\Delta_{ij}^2+4\sum_\mu|\tilde{a}_{\mu,ij}|^2$, the two remaining eigenvalues sum to at most $2\Delta_{ij}^2$, giving
\begin{equation}\label{eq:pairwise-bound}
    \sum_\mu\big|\tilde{a}_{\mu,ij}\big|^2\;\le\;\frac{1}{2}\,\Delta_{ij}^2
    \qquad\text{for every pair }\{ij\}\,.
\end{equation}
 In the maximal diagonal gauge no subset of D-instantons can therefore retain off-diagonal fluctuations larger than its internal separations---the $N=2$ mechanism extends pairwise. We stress the division of labor: the bound \eqref{eq:pairwise-bound} is exact and fixes the linear size of each block's domain to $\Delta_{ij}$; the passage to the scaling \eqref{eq:Z-UV-scaling-fine}---filling that domain with the flat short-distance weight and the per-pair measure factors of the $N=2$ computation---is the estimate, and it is there that the conjecture resides.

 The pairwise conditions above are, however, only \emph{necessary} conditions for the maximum. The actual integration region of the maximal copy is bounded by the exact Gribov horizon---the first zero of the full Faddeev--Popov determinant---which involves the complete $N(N-1)$-dimensional Hessian of $f_A$, including directions that mix different pairs, and whose form for $N\ge3$ is unknown. The true region may thus be strictly smaller than the pairwise-bounded one, and hitting the horizon before the pairwise bounds saturate cannot be excluded. Consequently neither the horizon volume---and with it the precise scalings \eqref{eq:Z-UV-scaling}--\eqref{eq:Z-UV-scaling-fine}---nor the suppression of the higher-body potentials is derived here; establishing the latter, and hence the full $N$-body non-collapse, requires the explicit evaluation of the higher-body terms.

 Several questions therefore remain open: (i) the non-collapse of the full $N$-body system without supersymmetry, which needs the higher-body potentials; (ii) the detailed equilibrium distribution of the $p^{(i)}_\mu$, which those potentials fix; (iii) the general-$N$ Gribov-copy structure and horizon volume, absent the $\mathrm{SU}(2)\simeq\mathrm{SO}(3)$ shortcut; and (iv) a numerical study performed directly in the maximal diagonal gauge, and a reformulation of the many-body decomposition in that gauge from the outset rather than in the Lorenz gauge.


\section{Conclusion}\label{sec:conclusion}

We have showed explicitly that the notorious one-loop collapse of the D-instanton positions
$p^{(i)}_\mu$ in the bosonic IKKT matrix model is an artifact of the leading perturbative
truncation, and that the non-collapse of the $p^{(i)}_\mu$ even without supersymmetry is instead a nonperturbative
effect already encoded in the exact two-body interaction.

Two statements are established. First, the one-loop potential is recovered as the leading
large-distance term of the gauge-fixed expansion of the Faddeev--Popov determinant, so the
collapse that it predicts is a statement about that leading truncation only; whether it survives
is decided by the exact two-body result. Second, because the two-body sector of the $N\times N$ model
factorizes into copies of $N=2$, the exact $\mathrm{U}(2)$ computation captures it: the resulting
partition function $Z(p)$ is finite at finite separation, and in the maximal diagonal gauge
the short-distance two-body interaction is \emph{repulsive}, and the two-body potential develops a stable minimum at a finite separation $\Delta_\ast$, so the two D-instantons do
not collapse onto each other. The remaining claims---the non-collapse of the full $N$-body system, the detailed equilibrium distribution of the $p^{(i)}_\mu$, and the general-$N$ UV scaling---are conjectural; they, and the obstructions to establishing them, are collected in Sec.~\ref{sec:general-N}.

Along the way we developed several methodological tools. We gauge-fixed $\mathrm{U}(N)$ while keeping
the diagonal and off-diagonal sectors distinct and handled the residual $\mathrm{U}(1)^N$ with an
auxiliary-ghost BRST construction, producing a new four-leg ghost vertex and a Faddeev--Popov
determinant whose ghost bilinear terminates exactly at $\mathcal{O}(\tilde{A}^2)$ and whose large-distance expansion organizes into a many-body decomposition. We further identified the short-distance negativity of the
naive Lorenz-gauge $N=2$ result as a Gribov ambiguity of multiple intersections between gauge slice and orbit, and showed that
the maximal diagonal gauge---the classical frame containing the perturbative vacuum---selects a single Gribov copy whose short-distance behavior is
finite and repulsive while agreeing with the naive many-body decomposition at large distance
(verified in closed form at $d=10$, with the scaling argument holding for general $d$).

Several concrete directions remain, as collected in Sec.~\ref{sec:general-N}. One should compute the higher-body potentials and from them the detailed distribution of the $p^{(i)}_\mu$; extend the exact analysis beyond $N=2$, where the $\mathrm{SU}(2)\simeq \mathrm{SO}(3)$ shortcut is unavailable---although the pairwise bound of Eq.~\eqref{eq:pairwise-bound} already extends the $N=2$ mechanism to every pair; perform a numerical study directly in
the maximal diagonal gauge; and recast the many-body decomposition itself in the maximal diagonal
gauge from the outset, rather than in the Lorenz gauge. Taken together, our results are
consistent with the conjecture that the off-diagonal fluctuations stabilize a non-collapsed
distribution of D-instantons. Establishing such stabilization analytically, matching the existing numerical study~\cite{Hotta:1998en}, and recognizing such emergent geometry through the covariant derivative interpretation~\cite{Hanada:2005vr,Hattori:2024btt,Hattori:2026hrh} lie beyond
what we have proven here.

It will also be interesting to extend the discussion to the supersymmetric theory, where the D-instantons instead live in superspace~\cite{Green:1996xf,Green:1997tv}. In the supersymmetric theory, the large-distance behavior is no longer dominated by the Vandermonde factor but by the fermionic zero modes~\cite{Aoki:1998vn,Itoyama:2026eqh}, which yields an attractive force. Another parallel line of recent development is the mass -deformed polarized IKKT  model~\cite{Bonelli:2002mb,Hartnoll:2024csr,Komatsu:2024bop}, which turns out to produce a large emergent spacetime~\cite{Komatsu:2024ydh,Chou:2025rwy}. The method of many-body decomposition should apply well in those cases, where the D-instantons tend to sit far from each other.

Another important direction we have not covered is the Lorentzian signature IKKT, in which the emergent spacetime is presumably much more meaningful~\cite{Kim:2011cr,Hirasawa:2024dht,Asano:2026mal,Asano:2024edo}. The methodologies developed in the paper formally work within $e^{iS_\text{IKKT}}$, although the sign problem kicks in and the convergence of the integral still lacks a rigorous mathematical proof. Nonetheless, many pioneering works have been performed and yielded insightful results (see~\cite{Asano:2024def,Chou:2025moy,Anagnostopoulos:2026qvz,Anagnostopoulos:2026utg} and the references therein). In particular, the numerical methods recently proposed for large-$N$ reduced models---the regularized master-field approximation~\cite{Maeta:2026miu} and a matrix bootstrap that dispenses with the positivity constraint~\cite{Maeta:2026oku}---provide a direct numerical route to the eigenvalue distribution, and hence to the distribution of the $p^{(i)}_\mu$. In contrast to the positivity-based matrix bootstrap~\cite{Anderson:2016rcw,Lin:2020mme,Kazakov:2021lel}, which constrains observables through semidefinite moment inequalities, these methods solve the loop equations for a self-consistent distribution and are free of the sign problem, so that they apply directly to the Lorentzian model in which our construction formally operates. It will therefore be interesting to realize the D-instanton picture in the Lorentzian framework, where the background time $A_0$ is an unambiguous setup.

\section*{Acknowledgments}
We would like to thank Chien-Yu Chou, Pei-Ming Ho, Reishi Maeta, Jun Nishimura, Worapat Pensiuk, Tadakatsu Sakai, Wei-Hsiang Shao, and Asato Tsuchiya for enlightening discussions.
H.L. is supported in part by the Ministry of Science and Technology grants 112-2112-M-002-024-MY3 and 112-2628-M-002-003-MY3.
C.T.W. is supported in part by the Ministry of Science and Technology, R.O.C. (NSTC 112-2112-M-002-024-MY3).

\appendix

\section{Feynman Rules and Determinant Expansion}
\label{sec:feyn_rules_det}

\subsection{Feynman Rules}
To derive the Feynman rules of the gauge-fixed model, we first recall the total action,
\begin{align}
\label{eq:S-tot}
    S_{\text{tot}} = & ~
    S_{\text{IKKT}}+S_{\text{gauge fixing}}+S_{\text{ghost}}\notag\\
    = & ~ 
    -\frac{1}{4}\tr\left[A_\mu,A_\nu\right]^2-\frac{1}{2\alpha} \tr\left[P_\mu,\tilde{A}^\mu\right]^2-\tr\left[P_\mu,\tilde{b}\right]\left[A^\mu,\tilde{c}\right]+\tr\left[\tilde{A}_\mu,\tilde{b}\right]_{D}\left[\tilde{A}^\mu,\tilde{c}\right]_{D}\,.
\end{align}
The quadratic part is
\begin{align}
    S_{2,\text{tot}} = & ~ \frac{1}{2}\sum_{i\neq j}\left[\left(p^{(i)}-p^{(j)}\right)^2\left|\tilde{a}^\mu_{ij}\right|^2-\left(1-\frac{1}{\alpha}\right)\left(p_\mu^{(i)}p_\nu^{(i)}+p_\mu^{(j)}p_\nu^{(j)}-p_\mu^{(i)}p_\nu^{(j)}-p_\mu^{(j)}p_\nu^{(i)}\right)\tilde{a}^\mu_{ij}\tilde{a}^\nu_{ji}\right]\notag\\
    &~
    +\sum_{i\neq j}\left(p^{(i)}-p^{(j)}\right)^2\tilde{b}_{ij}\tilde{c}_{ji}.
\end{align}
In the Landau gauge ($\alpha\to0$), this yields the gauge-field and ghost propagators
\begin{align}
\label{eq.propagator}
    \langle\tilde{a}^\mu_{ij}\tilde{a}^\nu_{k\ell}\rangle=2\left(\Delta^{-2}_{ij}\delta^{\mu\nu}-\Delta_{ij}^{-4}\Delta_{ij}^\mu\Delta_{ij}^\nu\right)\begin{tikzpicture}[baseline={(current bounding box.center)}]
	\begin{feynman}
		\vertex (a) {\(i\)};
		\vertex [below=1em of a] (b) {\(j\)};
		\vertex [right=1.5cm of a] (c) {\(l\)};
            \vertex [right=1.5cm of b] (d) {\(k\)};
		\diagram{
            {[edges=fermion]
        (a) -- (c),
        (d) -- (b),
        }
    };
	\end{feynman}
    \end{tikzpicture}~~,~~\langle\tilde{b}_{ij}\tilde{c}_{k\ell}\rangle=\Delta_{ij}^{-2} \begin{tikzpicture}[baseline={(current bounding box.center)}]
	\begin{feynman}
		\vertex (a) {\(i\)};
		\vertex [below=1em of a] (b) {\(j\)};
		\vertex [right=1.5cm of a] (c) {\(l\)};
            \vertex [right=1.5cm of b] (d) {\(k\)};
		\diagram{
            {[edges=charged scalar]
        (a) -- (c),
        (d) -- (b),
        }
    };
	\end{feynman}
    \end{tikzpicture}~.
\end{align}
This is a standard computation in the Landau gauge.

\paragraph{Gauge-field vertices.}
The gauge-field vertices arise from expanding the IKKT action $S_{\text{IKKT}}=-\tfrac14\tr[A_\mu,A_\nu]^2$ around a commuting saddle, $A_\mu=P_\mu+\tilde{A}_\mu$ with $[P_\mu,P_\nu]=0$.
Since $[A_\mu,A_\nu]=[P_\mu,\tilde{A}_\nu]-[P_\nu,\tilde{A}_\mu]+[\tilde{A}_\mu,\tilde{A}_\nu]$, the quartic action organizes by powers of $\tilde{A}$ into a quadratic piece (the propagator above), a cubic three-leg vertex, and a quartic four-leg vertex:
\begin{align}
\label{eq:gauge_cubic}
    S_3 = & ~ -\tr\!\left(\left[P_\mu,\tilde{A}_\nu\right]\left[\tilde{A}^\mu,\tilde{A}^\nu\right]\right),\\
\label{eq:gauge_quartic}
    S_4 = & ~ -\frac{1}{4}\tr\!\left[\tilde{A}_\mu,\tilde{A}_\nu\right]^2,
\end{align}
where in $S_3$ we have used $[P_\mu,P_\nu]=0$ and cyclicity to collect the two equal cross terms.
In the color basis, with $P_\mu$ diagonal one has $\left[P_\mu,\tilde{A}_\nu\right]_{ij}=\left(p_\mu^{(i)}-p_\mu^{(j)}\right)\tilde{a}_{\nu,ij}=\Delta_{\mu,ij}\,\tilde{a}_{\nu,ij}$, where $\Delta_{\mu,ij}=p_\mu^{(i)}-p_\mu^{(j)}$ plays the role of the matrix-model structure constant. Writing out the trace, the cubic vertex therefore carries exactly one factor of the separation $\Delta_{\mu,ij}$ dressing three off-diagonal gauge legs,
\begin{align}
\label{eq:gauge_cubic_comp}
    S_3 = -\sum_{i,j,k=1}^N\Delta_{\mu,ij}\,
    \tilde{a}^\nu_{ij}\left(\tilde{a}^\mu_{jk}\,\tilde{a}^\nu_{ki}-\tilde{a}^\nu_{jk}\,\tilde{a}^\mu_{ki}\right),
\end{align}
where a nonvanishing closed contraction requires the three color indices $\{i,j,k\}$ to be mutually distinct, so the cubic vertex is absent for $N=2$;
\begin{center}
\begin{tikzpicture}[baseline={(current bounding box.center)}]
    \begin{feynman}
        \vertex (v) at (0,0);
        \vertex [above left=1.2cm of v] (a) {\(\tilde{a}_{ij}\)};
        \vertex [below left=1.2cm of v] (b) {\(\tilde{a}_{jk}\)};
        \vertex [right=1.5cm of v] (c) {\(\tilde{a}_{ki}\)};
        \diagram*{
            (a) -- [boson] (v) -- [boson] (b),
            (v) -- [boson] (c),
        };
    \end{feynman}
\end{tikzpicture}
\qquad
\begin{tikzpicture}[baseline={(current bounding box.center)}]
    \begin{feynman}
        \vertex (v) at (0,0);
        \vertex [above left=1.0cm of v] (a) {\(\tilde{a}_{ij}\)};
        \vertex [below left=1.0cm of v] (b) {\(\tilde{a}_{jk}\)};
        \vertex [above right=1.0cm of v] (c) {\(\tilde{a}_{k\ell}\)};
        \vertex [below right=1.0cm of v] (d) {\(\tilde{a}_{\ell i}\)};
        \diagram*{
            (a) -- [boson] (v) -- [boson] (b),
            (c) -- [boson] (v) -- [boson] (d),
        };
    \end{feynman}
\end{tikzpicture}
\end{center}
the left diagram being the cubic vertex \eqref{eq:gauge_cubic} ($\propto\Delta\cdot$\,structure constant, three gauge legs) and the right one the quartic vertex \eqref{eq:gauge_quartic} (four gauge legs, the pure $\tr[\tilde{A}_\mu,\tilde{A}_\nu]^2$ piece). The compact algebraic content is simply \eqref{eq:gauge_cubic} and \eqref{eq:gauge_quartic}; \eqref{eq:gauge_cubic_comp} merely makes explicit that a single power of $\Delta_{\mu,ij}$ is attached to one of the three legs.

\paragraph{Ghost vertices.}
The ghost vertices are read off from the bilinear form $M=M_0+M_1+M_2$ of Sec.~\ref{sec:body_expansion}, whose components were given in the $M_{ijk\ell}$ formula there. The leading piece $M_0$ supplies the ghost propagator $M_0^{-1}$ used above. The linear-in-$\tilde{A}$ piece $M_1$ is the standard three-leg ghost--gauge vertex coming from $-\text{tr} [P_\mu,\tilde{b}][\tilde{A}^\mu,\tilde{c}]$,
\begin{align}
\label{eq:ghost_M1_vertex}
    \left(M_1\right)_{ijk\ell}=\Delta_{ij}\cdot\left(\tilde{a}_{\ell i}\,\delta_{jk}-\tilde{a}_{jk}\,\delta_{i\ell}\right),
\end{align}
which attaches one gauge leg ($\sim\Delta\cdot\tilde{A}$) to a $\tilde{b}\tilde{c}$ ghost line. The genuinely new object is the four-leg ghost vertex carried by the quadratic-in-$\tilde{A}$ piece $M_2$, which originates from the residual-$\mathrm{U}(1)^N$ term $\text{tr}[\tilde{b},\tilde{A}_\mu]_D[\tilde{c},\tilde{A}^\mu]_D$,
\begin{align}
\label{eq:ghost_M2_vertex}
    \left(M_2\right)_{ijk\ell}=\left(\tilde{a}_{ji}\cdot\tilde{a}_{\ell k}\right)\left(\delta_{ik}+\delta_{j\ell}-\delta_{i\ell}-\delta_{jk}\right),
\end{align}
attaching \emph{two} gauge legs to a single $\tilde{b}\tilde{c}$ vertex. This four-leg ghost vertex, absent from the standard treatment, is the novel feature introduced by fixing the redundant residual gauge (see also Fig.~\ref{fig:Feynman-rules}); diagrammatically,
\begin{center}
\begin{tikzpicture}[baseline={(current bounding box.center)}]
    \begin{feynman}
        \vertex (v) at (0,0);
        \vertex [above left=1.2cm of v] (b) {\(\tilde{b}\)};
        \vertex [below left=1.2cm of v] (c) {\(\tilde{c}\)};
        \vertex [right=1.5cm of v] (a) {\(\tilde{A}^\mu\)};
        \diagram*{
            (b) -- [charged scalar] (v) -- [charged scalar] (c),
            (v) -- [boson] (a),
        };
    \end{feynman}
\end{tikzpicture}
\qquad
\begin{tikzpicture}[baseline={(current bounding box.center)}]
    \begin{feynman}
        \vertex (v) at (0,0);
        \vertex [above left=1.0cm of v] (b) {\(\tilde{b}\)};
        \vertex [below left=1.0cm of v] (c) {\(\tilde{c}\)};
        \vertex [above right=1.0cm of v] (a1) {\(\tilde{A}^\mu\)};
        \vertex [below right=1.0cm of v] (a2) {\(\tilde{A}_\mu\)};
        \diagram*{
            (b) -- [charged scalar] (v) -- [charged scalar] (c),
            (v) -- [boson] (a1),
            (v) -- [boson] (a2),
        };
    \end{feynman}
\end{tikzpicture}
\end{center}
the left vertex being the three-leg ghost vertex \eqref{eq:ghost_M1_vertex} ($M_1$, one gauge leg) and the right the new four-leg ghost vertex \eqref{eq:ghost_M2_vertex} ($M_2$, two gauge legs).

\subsection{Derivation of Determinant Expansion}
\label{sec:det_exp_deriv}

Here we derive the determinant expansion \eqref{eq:det_expansion} from the component form of $M=M_0+M_1+M_2$ given in the main text.
Since $M_0\sim\mathcal{O}(\Delta^2)$ is invertible in the large-distance regime $\Delta_{ij}^2\gg1$, we factor out the leading piece,
\begin{align}
\label{eq:det_factor}
    \det(M) = \det(M_0)\cdot\det\left[\mathds{1}+M_0^{-1}M_1+M_0^{-1}M_2\right],
\end{align}
where the overall factor is read off directly from the diagonal (in pair space) form $(M_0)_{ijk\ell}=\Delta_{ij}^2\delta_{i\ell}\delta_{jk}$, giving
\begin{align}
\label{eq:detM0}
    \det(M_0)=\prod_{\substack{i,j=1\\i\neq j}}^N\Delta_{ij}^2,
    \qquad
    \left(M_0^{-1}\right)_{ijk\ell}=\Delta_{ij}^{-2}\,\delta_{i\ell}\delta_{jk}.
\end{align}
To expand the remaining determinant we use the identity $\det(\mathds{1}+X)=\exp\tr\log(\mathds{1}+X)$ together with the series
\begin{align}
\label{eq:logseries}
    \log(\mathds{1}+X)=X-\frac{1}{2}X^2+\frac{1}{3}X^3-\frac{1}{4}X^4+\mathcal{O}(X^5),
    \qquad
    X\equiv \mathcal{A}+\mathcal{B},
\end{align}
where we have introduced the shorthand
\begin{align}
    \mathcal{A}\equiv M_0^{-1}M_1=\mathcal{O}(\tilde{A})~,\qquad
    \mathcal{B}\equiv M_0^{-1}M_2=\mathcal{O}(\tilde{A}^2).
\end{align}
We organize the expansion by the total power of $\tilde{A}$, assigning weight $1$ to $\mathcal{A}$ and weight $2$ to $\mathcal{B}$, and retain all terms through $\mathcal{O}(\tilde{A}^4)$.
Substituting $X=\mathcal{A}+\mathcal{B}$ into \eqref{eq:logseries} and keeping only words of total weight $\leq4$,
\begin{align}
\label{eq:trlog}
    \tr\log(\mathds{1}+X) = & ~ \tr(\mathcal{A})+\tr(\mathcal{B})-\frac{1}{2}\tr(\mathcal{A}^2)
    -\tr(\mathcal{A}\mathcal{B})+\frac{1}{3}\tr(\mathcal{A}^3)\notag\\
    & ~ -\frac{1}{2}\tr(\mathcal{B}^2)+\tr(\mathcal{A}^2\mathcal{B})-\frac{1}{4}\tr(\mathcal{A}^4)
    +\mathcal{O}(\tilde{A}^5),
\end{align}
where we have used cyclicity of the trace to combine terms such as $\tr(\mathcal{A}\mathcal{B})=\tr(\mathcal{B}\mathcal{A})$.
Exponentiating and collecting again by total weight in $\tilde{A}$ up to fourth order,
\begin{align}
\label{eq:exp_pre}
    \det(\mathds{1}+X)
    = & ~ 1+\tr(\mathcal{A})+\Big[\tr(\mathcal{B})-\tfrac{1}{2}\tr(\mathcal{A}^2)+\tfrac{1}{2}\tr(\mathcal{A})^2\Big]\notag\\
    & ~ +\Big[-\tr(\mathcal{A}\mathcal{B})+\tfrac{1}{3}\tr(\mathcal{A}^3)+\tr(\mathcal{A})\tr(\mathcal{B})-\tfrac{1}{2}\tr(\mathcal{A})\tr(\mathcal{A}^2)+\tfrac{1}{6}\tr(\mathcal{A})^3\Big]\notag\\
    & ~ +\Big[\tfrac{1}{2}\tr(\mathcal{B})^2-\tfrac{1}{2}\tr(\mathcal{B}^2)-\tfrac{1}{2}\tr(\mathcal{A}^2)\tr(\mathcal{B})+\tr(\mathcal{A}^2\mathcal{B})+\tfrac{1}{8}\tr(\mathcal{A}^2)^2-\tfrac{1}{4}\tr(\mathcal{A}^4)\notag\\
    & ~ \quad+\big(\text{terms containing }\tr(\mathcal{A})\big)\Big]+\mathcal{O}(\tilde{A}^5),
\end{align}
where the brackets group the contributions at orders $\tilde{A}^1,\tilde{A}^2,\tilde{A}^3,\tilde{A}^4$, respectively.
This is the cumulant expansion before any simplification.

The expansion simplifies dramatically because of the identity
\begin{align}
\label{eq:tra_zero}
    \tr(\mathcal{A})=\tr\!\left(M_0^{-1}M_1\right)=0.
\end{align}
To see this, use \eqref{eq:detM0} and the component form of $M_1$ from the main text,
\begin{align}
    \tr\!\left(M_0^{-1}M_1\right)
    =\sum_{\substack{i,j,k,\ell\\i\neq j,\,k\neq\ell}}^N\left(M_0^{-1}\right)_{ijk\ell}\left(M_1\right)_{k\ell ij}
    =\sum_{\substack{i,j,k,\ell\\i\neq j,\,k\neq\ell}}^N\Delta_{ij}^{-2}\,\delta_{i\ell}\delta_{jk}\,\Delta_{k\ell}\cdot\left(\tilde{a}_{jk}\,\delta_{\ell i}-\tilde{a}_{\ell i}\,\delta_{kj}\right).
\end{align}
The two Kronecker deltas $\delta_{i\ell}\delta_{jk}$ from $M_0^{-1}$ are diagonal in the pair space, so they force $\ell=i$ and $k=j$; the surviving structure then requires the additional delta $\delta_{\ell i}$ or $\delta_{kj}$ to project onto $\tilde{a}_{ii}$, i.e.\ a \emph{diagonal} entry of $\tilde{A}$.
Since $\tilde{A}$ is purely off-diagonal, $\tilde{a}_{ii}=0$, and every term in the sum vanishes.
Geometrically, $M_0^{-1}$ is diagonal in pair space while $M_1$ carries a single power of the off-diagonal $\tilde{A}$, so the trace is forced onto the (vanishing) diagonal of $\tilde{A}$; equivalently, no closed diagram can be built from a single ghost propagator and a single three-leg vertex.

Imposing \eqref{eq:tra_zero} eliminates every term containing a factor of $\tr(\mathcal{A})$ in \eqref{eq:exp_pre}, and the surviving terms are exactly
\begin{align}
    \det(\mathds{1}+X)
    = & ~ 1
    +\tr(\mathcal{B})-\frac{1}{2}\tr(\mathcal{A}^2)
    -\tr(\mathcal{A}\mathcal{B})+\frac{1}{3}\tr(\mathcal{A}^3)\notag\\
    & ~ +\frac{1}{2}\tr(\mathcal{B})^2-\frac{1}{2}\tr(\mathcal{B}^2)-\frac{1}{2}\tr(\mathcal{A}^2)\tr(\mathcal{B})+\tr(\mathcal{A}^2\mathcal{B})+\frac{1}{8}\tr(\mathcal{A}^2)^2-\frac{1}{4}\tr(\mathcal{A}^4)+\mathcal{O}(\tilde{A}^5).
\end{align}
Restoring $\mathcal{A}=M_0^{-1}M_1$ and $\mathcal{B}=M_0^{-1}M_2$ and multiplying by the prefactor \eqref{eq:detM0} reproduces precisely the determinant expansion \eqref{eq:det_expansion} quoted in the main text.
Finally, since the matrix $M$ truncates at $\mathcal{O}(\tilde{A}^2)$, the lowest-weight words dropped from $\tr\log(\mathds{1}+X)$ are the weight-$5$ terms $\tr(\mathcal{A}\mathcal{B}^2)$ from $\tfrac{1}{3}\tr(X^3)$, $\tr(\mathcal{A}^3\mathcal{B})$ from $-\tfrac{1}{4}\tr(X^4)$, and $\tr(\mathcal{A}^5)$ from $\tfrac{1}{5}\tr(X^5)$; all neglected contributions are therefore of order $\mathcal{O}(\tilde{A}^5)$, and the expansion above is exact through $\mathcal{O}(\tilde{A}^4)$.

\subsection{Explicit Form of Ghost Determinant}
\label{sec:explicit_ghost_det}

Evaluating each trace in the determinant expansion \eqref{eq:det_expansion} in terms of $\Delta_{ij}$ and $\tilde{A}_\mu$ produces the terms $\mathcal{T}_2,\dots,\mathcal{T}_5$ of the many-body decomposition, collected below.
Before listing them, we fix two pieces of notation used throughout these expressions.
\begin{itemize}
    \item \textbf{``Diff.''} Under a multi-index sum, the label $\{i,j,k,\dots\}\,\text{Diff.}$ means that all the indices appearing inside the braces are mutually \emph{distinct}, i.e.\ the sum runs only over configurations in which no two of those indices coincide. This isolates the genuine $n$-body contribution carried by $n$ distinct spacetime points.
    \item \textbf{$\left(\tilde{a}_{(\leftrightarrow)}\right)$.} This symbol denotes the additional term obtained from the one written explicitly by swapping the order/labels of the off-diagonal fields $\tilde{A}$ (equivalently, exchanging the two ghost legs attached to a vertex), keeping the $\Delta$ factors fixed. It is shorthand for ``plus the index-swapped partner'' and arises because each three-leg vertex can attach the two ghosts in either order.
\end{itemize}

As a worked example, we derive the two-body term $\mathcal{T}_2$, i.e.\ the part of the expansion that involves only two distinct site indices.
It is convenient to write $\mathcal{A}\equiv M_0^{-1}M_1$ and $\mathcal{B}\equiv M_0^{-1}M_2$ as in Appendix~\ref{sec:det_exp_deriv}.
The leading ($\Delta^{-2}$) piece of $\mathcal{T}_2$ is the genuinely $\mathcal{O}(\tilde{A}^2)$ contribution, which comes from the order-$\tilde{A}^2$ cumulant of \eqref{eq:det_expansion},
\begin{align}
\label{eq:T2_origin}
    \mathcal{T}_2^{(\Delta^{-2})}=\Big[\tr(\mathcal{B})-\tfrac{1}{2}\tr(\mathcal{A}^2)\Big]_{\text{2-body}}.
\end{align}
Here $\tr(\mathcal{A}^2)=\tr[(M_0^{-1}M_1)^2]$ carries a single power of $\Delta$ at each of its two $M_1$ insertions; with $M_0^{-1}$ diagonal in pair space, a nonzero closed contraction requires the two three-leg vertices to share three \emph{distinct} sites, so $\tr(\mathcal{A}^2)$ has \emph{no} two-body part and contributes only to $\mathcal{T}_3$.
Hence the two-body $\mathcal{O}(\tilde{A}^2)$ term is carried entirely by $\tr(\mathcal{B})=\tr(M_0^{-1}M_2)$.
Using $\left(M_0^{-1}\right)_{ijk\ell}=\Delta_{ij}^{-2}\delta_{i\ell}\delta_{jk}$ and the component form of $M_2$ from the main text,
\begin{align}
    \tr\!\left(M_0^{-1}M_2\right)
    = & ~ \sum_{\substack{i,j,k,\ell\\i\neq j,\,k\neq\ell}}^N\Delta_{ij}^{-2}\,\delta_{i\ell}\delta_{jk}\,\left(\tilde{a}_{\ell k}\cdot\tilde{a}_{ji}\right)\left(\delta_{ki}+\delta_{\ell j}-\delta_{kj}-\delta_{\ell i}\right)\notag\\
    = & ~ \sum_{\substack{i,j\\i\neq j}}^N\Delta_{ij}^{-2}\left(\tilde{a}_{ij}\cdot\tilde{a}_{ji}\right)\left(\delta_{ji}+\delta_{ij}-\delta_{jj}-\delta_{ii}\right)
    =-2\sum_{\substack{i,j\\i\neq j}}^N\Delta_{ij}^{-2}\left(\tilde{a}_{ij}\cdot\tilde{a}_{ji}\right),
\end{align}
where setting $\ell=i,\,k=j$ (from $\delta_{i\ell}\delta_{jk}$) leaves the bracket $\delta_{ji}+\delta_{ij}-\delta_{jj}-\delta_{ii}=-2$ on the off-diagonal $i\neq j$.
This reproduces the first term of $\mathcal{T}_2$.

The subleading ($\Delta^{-4}$) piece of $\mathcal{T}_2$ is $\mathcal{O}(\tilde{A}^4)$: it is the two-body part of the connected order-$\tilde{A}^4$ cumulants of \eqref{eq:det_expansion},
\begin{align}
\label{eq:T2_origin4}
    \mathcal{T}_2^{(\Delta^{-4})}=\Big[-\tfrac{1}{2}\tr(\mathcal{B}^2)+\tr(\mathcal{A}^2\mathcal{B})-\tfrac{1}{4}\tr(\mathcal{A}^4)\Big]_{\text{2-body}}.
\end{align}
(The disconnected products $\tfrac{1}{2}\tr(\mathcal{B})^2$, $\tfrac{1}{8}\tr(\mathcal{A}^2)^2$ and $-\tfrac{1}{2}\tr(\mathcal{A}^2)\tr(\mathcal{B})$ factorize into two separate single-loop traces and are collected instead in the $\mathcal{T}_{i\times j}$ terms.)
Restricting each trace to closed contractions on a single pair of sites $\{i,j\}$, every $\Delta$ becomes $\Delta_{ij}$ and the four $\tilde{A}$ factors contract in the two inequivalent ways, giving
\begin{align}
    \mathcal{T}_2^{(\Delta^{-4})}=-2\sum_{\substack{i,j\\i\neq j}}^N\Delta_{ij}^{-4}\left[\tilde{a}_{ij}^2\,\tilde{a}_{ji}^2+\left(\tilde{a}_{ij}\cdot\tilde{a}_{ji}\right)^2\right],
\end{align}
where $\tilde{a}_{ij}^2\equiv\tilde{a}_{ij}\cdot\tilde{a}_{ij}$ and the two scalar structures correspond to the two contraction patterns of the four spacetime indices.
Adding $\mathcal{T}_2^{(\Delta^{-2})}+\mathcal{T}_2^{(\Delta^{-4})}$ reproduces exactly the expression for $\mathcal{T}_2$ recorded below.
The higher-body terms $\mathcal{T}_3,\dots,\mathcal{T}_5$ follow from the same procedure applied to the remaining cumulants of \eqref{eq:det_expansion}, now keeping configurations with three, four, and five distinct indices.

\begin{align}
\label{eq:T2}
    \mathcal{T}_2=-2\sum_{\substack{i,j=1\\i\neq j}}^N\Delta_{ij}^{-2}\left(\tilde{a}_{ij}\cdot\tilde{a}_{ji}\right)-2\sum_{\substack{i,j=1\\i\neq j}}^N\Delta_{ij}^{-4}\left[\tilde{a}_{ij}^2\tilde{a}_{ji}^2+\left(\tilde{a}_{ij}\cdot\tilde{a}_{ji}\right)^2\right].
\end{align}
\begin{align}
\label{eq:T2_2}
    \mathcal{T}_{2\times2}=2\sum_{\substack{i,j=1\\i\neq j}}^N\sum_{\substack{k,\ell=1\\k\neq \ell}}^N\Delta_{ij}^{-2}\Delta_{k\ell}^{-2}\left(\tilde{a}_{ij}\cdot\tilde{a}_{ji}\right)\left(\tilde{a}_{k\ell}\cdot\tilde{a}_{\ell k}\right).
\end{align}
\begin{align}
\label{eq:T2_3}
    \mathcal{T}_{2\times3} = & ~ 
    \sum_{\substack{i,j=1\\i\neq j}}^N\sum_{\substack{k,\ell,m=1\\\{k,\ell,m\}\text{ Diff.}}}^N\Delta_{ij}^{-2}\Delta_{k\ell}^{-2}\Delta_{km}^{-2}\left(\Delta_{k\ell}\cdot\tilde{a}_{\ell m}\right)\left(\Delta_{km}\cdot\tilde{a}_{m\ell}\right)\left(\tilde{a}_{ij}\cdot\tilde{a}_{ji}\right)\notag\\
    & ~ +\sum_{\substack{i,j=1\\i\neq j}}^N\sum_{\substack{k,\ell,m=1\\\{k,\ell,m\}\text{ Diff.}}}^N\Delta_{ij}^{-2}\Delta_{k\ell}^{-2}\Delta_{km}^{-2}\left(\Delta_{k\ell}\cdot\tilde{a}_{m\ell}\right)\left(\Delta_{km}\cdot\tilde{a}_{\ell m}\right)\left(\tilde{a}_{ij}\cdot\tilde{a}_{ji}\right)\notag\\
    \equiv & ~ \sum_{\substack{i,j=1\\i\neq j}}^N\sum_{\substack{k,\ell,m=1\\\{k,\ell,m\}\text{ Diff.}}}^N\Delta_{ij}^{-2}\Delta_{k\ell}^{-2}\Delta_{km}^{-2}\left[\left(\Delta_{k\ell}\cdot\tilde{a}_{\ell m}\right)\left(\Delta_{km}\cdot\tilde{a}_{m\ell}\right)+\left(\tilde{a}_{(\leftrightarrow)}\right)\right]\left(\tilde{a}_{ij}\cdot\tilde{a}_{ji}\right).
\end{align}
\begin{align}
\label{eq:T3}
    \mathcal{T}_3 = & ~
    -\frac{1}{2}\sum_{\substack{i,j,k=1\\\{i,j,k\}\text{ Diff.}}}^N\Delta_{ij}^{-2}\Delta_{ik}^{-2}\left[\left(\Delta_{ij}\cdot \tilde{a}_{jk}\right)\left(\Delta_{ik}\cdot \tilde{a}_{kj}\right)+\left(\tilde{a}_{(\leftrightarrow)}\right)\right]\notag\\
    & ~ 
    -\sum_{\substack{i,j,k=1\\\{i,j,k\}\text{ Diff.}}}^N\Delta_{ij}^{-2}\Delta_{ik}^{-2}\left[\left(\Delta_{ij}\cdot \tilde{a}_{jk}\right)\left(\tilde{a}_{ki}\cdot\tilde{a}_{ij}\right)+\left(\tilde{a}_{(\leftrightarrow)}\right)\right]\notag\\
    & ~
    -\frac{1}{2}\sum_{\substack{i,j,k=1\\\{i,j,k\}\text{ Diff.}}}^N\Delta_{ij}^{-2}\Delta_{ik}^{-2}\left[\left(\tilde{a}_{ij}\cdot \tilde{a}_{ik}\right)\left(\tilde{a}_{ki}\cdot\tilde{a}_{ji}\right)+\left(\tilde{a}_{ij}\cdot\tilde{a}_{ki}\right)\left(\tilde{a}_{ik}\cdot\tilde{a}_{ji}\right)+\left(\tilde{a}_{(\leftrightarrow)}\right)\right]\notag\\
    & ~
    -2\sum_{\substack{i,j,k=1\\\{i,j,k\}\text{ Diff.}}}^N\Delta_{ij}^{-4}\Delta_{ik}^{-2}\left[\left(\Delta_{ij}\cdot\tilde{a}_{jk}\right)\left(\Delta_{ik}\cdot\tilde{a}_{kj}\right)+\left(\tilde{a}_{(\leftrightarrow)}\right)\right]\left(\tilde{a}_{ij}\cdot\tilde{a}_{ji}\right)\notag\\
    & ~
    -\sum_{\substack{i,j,k=1\\\{i,j,k\}\text{ Diff.}}}^N\Delta_{ij}^{-2}\Delta_{ik}^{-2}\Delta_{jk}^{-2}\left[\left(\Delta_{ij}\cdot\tilde{a}_{jk}\right)\left(\Delta_{ik}\cdot\tilde{a}_{ji}\right)\left(\tilde{a}_{ij}\cdot\tilde{a}_{kj}\right)+\left(\tilde{a}_{(\leftrightarrow)}\right)\right]\notag\\
    & ~
    -\frac{1}{4}\sum_{\substack{i,j,k=1\\\{i,j,k\}\text{ Diff.}}}^N\Delta_{ij}^{-4}\Delta_{ik}^{-4}\left[\left(\Delta_{ij}\cdot\tilde{a}_{jk}\right)^2\left(\Delta_{ik}\cdot\tilde{a}_{kj}\right)^2+\left(\tilde{a}_{(\leftrightarrow)}\right)\right]\notag\\
    & ~
    -\frac{1}{2}\sum_{\substack{i,j,k=1\\\{i,j,k\}\text{ Diff.}}}^N\Delta_{ij}^{-4}\Delta_{ik}^{-2}\Delta_{jk}^{-2}\left[\left(\Delta_{ij}\cdot\tilde{a}_{jk}\right)\left(\Delta_{ij}\cdot\tilde{a}_{ki}\right)\left(\Delta_{ik}\cdot\tilde{a}_{kj}\right)\left(\Delta_{kj}\cdot\tilde{a}_{ik}\right)+\left(\tilde{a}_{(\leftrightarrow)}\right)\right].
\end{align}
\begin{align}
\label{eq:T3_3}
    \mathcal{T}_{3\times3} = & ~ 
    \frac{1}{8}\sum_{\substack{i,j,k=1\\\{i,j,k\}\text{ Diff.}}}^N  \Delta_{ij}^{-2}\Delta_{ik}^{-2}\left[\left(\Delta_{ij}\cdot\tilde{a}_{jk}\right)\left(\Delta_{ik}\cdot\tilde{a}_{kj}\right)+\left(\tilde{a}_{(\leftrightarrow)}\right)\right]\notag\\
    & ~ \times\sum_{\substack{\ell,m,n=1\\\{\ell,m,n\}\text{ Diff.}}}^N\Delta_{\ell m}^{-2}\Delta_{\ell n}^{-2}\left[\left(\Delta_{\ell m}\cdot\tilde{a}_{mn}\right)\left(\Delta_{\ell n}\cdot\tilde{a}_{nm}\right)+\left(\tilde{a}_{(\leftrightarrow)}\right)\right].
\end{align}
\begin{align}
\label{eq:T4}
    \mathcal{T}_4 = & ~
    -\frac{1}{3}\sum_{\substack{i,j,k,\ell=1\\\{i,j,k,\ell\}\text{ Diff.}}}^N\Delta_{ij}^{-2}\Delta_{ik}^{-2}\Delta_{i\ell}^{-2}\left[\left(\Delta_{ij}\cdot\tilde{a}_{jk}\right)\left(\Delta_{ik}\cdot\tilde{a}_{k\ell}\right)\left(\Delta_{i\ell}\cdot\tilde{a}_{\ell j}\right)+\left(\tilde{a}_{(\leftrightarrow)}\right)\right]\notag\\
    & ~ 
    -\sum_{\substack{i,j,k,\ell=1\\\{i,j,k,\ell\}\text{ Diff.}}}^N\Delta_{ij}^{-2}\Delta_{ik}^{-2}\Delta_{i\ell}^{-2}\left[\left(\Delta_{ij}\cdot\tilde{a}_{jk}\right)\left(\Delta_{ik}\cdot\tilde{a}_{k\ell}\right)\left(\tilde{a}_{\ell i}\cdot\tilde{a}_{ij}\right)+\left(\tilde{a}_{(\leftrightarrow)}\right)\right]\notag\\
    & ~
    -\frac{1}{2}\sum_{\substack{i,j,k,\ell=1\\\{i,j,k,\ell\}\text{ Diff.}}}^N\Delta_{ij}^{-4}\Delta_{ik}^{-2}\Delta_{i\ell}^{-2}\left[\left(\Delta_{ij}\cdot\tilde{a}_{jk}\right)\left(\Delta_{ij}\cdot\tilde{a}_{j\ell}\right)\left(\Delta_{ik}\cdot\tilde{a}_{kj}\right)\left(\Delta_{i\ell}\cdot\tilde{a}_{\ell j}\right)+\left(\tilde{a}_{(\leftrightarrow)}\right)\right]\notag\\
    & ~
    -\frac{1}{2}\sum_{\substack{i,j,k,\ell=1\\\{i,j,k,\ell\}\text{ Diff.}}}^N\Delta_{ij}^{-4}\Delta_{ik}^{-2}\Delta_{j\ell}^{-2}\left[\left(\Delta_{ij}\cdot \tilde{a}_{jk}\right)\left(\Delta_{ik}\cdot \tilde{a}_{kj}\right)\left(\Delta_{ij}\cdot \tilde{a}_{\ell i}\right)\left(\Delta_{\ell j}\cdot \tilde{a}_{i\ell}\right)+\left(\tilde{a}_{(\leftrightarrow)}\right)\right]\notag\\
    & ~
    -\frac{1}{4}\sum_{\substack{i,j,k,\ell=1\\\{i,j,k,\ell\}\text{ Diff.}}}^N\Delta_{ij}^{-2}\Delta_{ik}^{-2}\Delta_{k\ell}^{-2}\Delta_{j\ell}^{-2}\left[\left(\Delta_{ij}\cdot\tilde{a}_{jk}\right)\left(\Delta_{ik}\cdot\tilde{a}_{jk}\right)\left(\Delta_{\ell k}\cdot\tilde{a}_{kj}\right)\left(\Delta_{\ell j}\cdot\tilde{a}_{i\ell}\right)+\left(\tilde{a}_{(\leftrightarrow)}\right)\right].
\end{align}
\begin{align}
\label{eq:T5}
    \mathcal{T}_5=-\frac{1}{4}\sum_{\substack{i,j,k,\ell,m=1\\
        \{i,j,k,\ell,m\}\text{ Diff.}}}^N & ~ \Delta_{ij}^{-2}\Delta_{ik}^{-2}\Delta_{i\ell}^{-2}\Delta_{im}^{-2}\left[\left(\Delta_{ij}\cdot\tilde{a}_{jk}\right)\left(\Delta_{ik}\cdot\tilde{a}_{k\ell}\right)\left(\Delta_{i\ell}\cdot\tilde{a}_{\ell m}\right)\left(\Delta_{im}\cdot\tilde{a}_{mj}\right)\right.\notag\\
        & ~ \left.+\left(\tilde{a}_{(\leftrightarrow)}\right)\right].
\end{align}

\section{\texorpdfstring{$xy$}{xy} Integration}
\label{app:int-XY}
We evaluate the triple integral of Eq.~\eqref{eq:Z[p]-1}.
First, the angle integration can be carried out with the Kummer confluent hypergeometric function~\cite{NIST:DLMF,GradshteynRyzhik},
\begin{equation}
    \int_0^\pi d\theta \, \sin^{d-3}\theta\,e^{-A\,\sin^2\theta}=\sqrt{\pi}\frac{\Gamma\left(\frac{d-2}{2}\right)}{\Gamma\left(\frac{d-1}{2}\right)}\,{}_1F_1\left( \frac{d-2}{2};\frac{d-1}{2};-A \right).
\end{equation}
Plugging this result back into the first term and performing the $x,y$ integration via the standard Gaussian--Kummer moment identities~\cite{NIST:DLMF,GradshteynRyzhik} yields a Tricomi confluent hypergeometric (Kummer $U$) function,
\begin{equation}
\begin{split}
    &\int_0^\infty\int_0^\infty dx dy\,x^{d-2}y^{d-2}\,{}_1F_1\left( \frac{d-2}{2};\frac{d-1}{2};-\frac{4x^2y^2}{p^4}\right)\,e^{-(x^2+y^2)}\\
    =&2^{-d}\,\left(\Gamma\left(\frac{d-1}{2}\right)\right)^2\,p^{2(d-2)}\,U\left( \frac{d-2}{2},\frac{1}{2},\frac{p^4}{4}\right).
\end{split}
\end{equation}
For the second term, we similarly obtain
\begin{equation}
\begin{split}
    &\int_0^\infty\int_0^\infty dx dy\,x^{d}y^{d-2}\,{}_1F_1\left( \frac{d-2}{2};\frac{d-1}{2};-\frac{4x^2y^2}{p^4}\right)\,e^{-(x^2+y^2)}\\
    =&2^{-d}\,\Gamma\left(\frac{d-1}{2}\right)\Gamma\left(\frac{d+1}{2}\right)\,p^{2(d-2)}\,U\left( \frac{d-2}{2},-\frac{1}{2},\frac{p^4}{4}\right).
\end{split}
\end{equation}
For the last term, note that the $\theta$ integral carries a different power of $\sin\theta$,
\begin{equation}
    \int_0^\pi d\theta \, \sin^{d-1}\theta\,e^{-A\sin^2\theta}=\sqrt{\pi}\frac{\Gamma\left(\frac{d}{2}\right)}{\Gamma\left(\frac{d+1}{2}\right)}\,{}_1F_1\left( \frac{d}{2};\frac{d+1}{2};-A \right).
\end{equation}
Carrying out the $x,y$ integration, one finds
\begin{equation}
\begin{split}
    &\int_0^\infty\int_0^\infty dx dy\,x^{d}y^{d}\,{}_1F_1\left( \frac{d}{2};\frac{d+1}{2};-\frac{4x^2y^2}{p^4}\right)\,e^{-(x^2+y^2)}\\
   =& 2^{-d-2} \left(\Gamma\left( \frac{d+1}{2} \right)\right)^2\, p^{2d} \, U\left( \frac{d}{2},\frac{1}{2},\frac{p^4}{4}\right).
\end{split}
\end{equation}
Finally, combining the three terms and applying the standard contiguous relations of the Kummer $U$ function~\cite{NIST:DLMF} to express the result in terms of $U\!\left(\frac{d}{2},\frac{3}{2},\cdot\right)$ and $U\!\left(\frac{d}{2},\frac{1}{2},\cdot\right)$, we reproduce the closed form of Eq.~\eqref{eq:Z[p]}. The overall factor of $\pi$ arises within each term: the angular integral supplies one factor of $\sqrt{\pi}$, and the Legendre duplication formula, $\Gamma\!\left(\tfrac{d-2}{2}\right)\Gamma\!\left(\tfrac{d-1}{2}\right)=2^{3-d}\sqrt{\pi}\,\Gamma(d-2)$, supplies the second.

\section{Closed Forms and Asymptotics of \texorpdfstring{$I_\text{max}\,,I_\text{saddle}\,,I_\text{min}$}{Imax, Isaddle, Imin} at \texorpdfstring{$d=10$}{d=10}}
\label{app:I-integrals}

Here we present the closed forms of the three gauge-fixed copies $I_\text{max}$, $I_\text{saddle}$, $I_\text{min}$ defined in Sec.~\ref{sec:svd}, in the special case $d=10$. As explained there, the argument of each $I_k$ is the singular value identified with the diagonal separation, $a$ (resp.\ $b$, $c$) $=p$.

\begin{equation}
\begin{aligned}
I_\text{max}(a)
&= \frac{1}{a^2} \int_0^a db \int_0^b dc \;
b^7 c^7 (a^2 - b^2)(a^2 - c^2)(b^2 - c^2)\,
\exp\!\left[-\frac{a^2 b^2 + b^2 c^2 + c^2 a^2}{4}\right] \\
&= \frac{e^{-\frac{3}{4} a^4}}{a^4}
\Bigg[
384\, e^{\frac{a^4}{2}}
+ 8 \left(2 a^{16} + 107 a^{12} + 1524 a^8 + 5628 a^4 + 2304\right) e^{\frac{3 a^4}{4}} \\
&\quad
- 8 \left(a^{16} + 55 a^{12} + 834 a^8 + 3600 a^4 + 2352\right) \\
&\quad
+ 4 \sqrt{\pi}\, a^2 \left(2 a^{16} + 111 a^{12} + 1722 a^8 + 7980 a^4 + 7560\right)
e^{a^4}
\left[
\operatorname{erf}\!\left(\frac{a^2}{2}\right)
- \operatorname{erf}\!\left(a^2\right)
\right]
\Bigg].
\end{aligned}
\end{equation}

\begin{equation}
\begin{aligned}
I_\text{saddle}(b)
&= \frac{1}{b^2} \int_b^\infty da \int_0^b dc \;
a^7 c^7 (a^2 - b^2)(a^2 - c^2)(b^2 - c^2)\,
\exp\!\left[-\frac{a^2 b^2 + b^2 c^2 + c^2 a^2}{4}\right] \\
&= \frac{384 e^{-\frac{b^4}{4}}}{b^4}.
\end{aligned}
\end{equation}

\begin{equation}
\begin{aligned}
I_\text{min}(c)
&= \frac{1}{c^2} \int_c^\infty da \int_c^a db \;
a^7 b^7 (a^2 - b^2)(a^2 - c^2)(b^2 - c^2)\,
\exp\!\left[-\frac{a^2 b^2 + b^2 c^2 + c^2 a^2}{4}\right] \\
&= \frac{e^{-\frac{3}{4} c^4}}{c^4}
\Bigg[
8 \left(c^{16} + 55 c^{12} + 834 c^8 + 3600 c^4 + 2352\right) \\
&\quad
- 4 \sqrt{\pi}\, c^2
\left(2 c^{16} + 111 c^{12} + 1722 c^8 + 7980 c^4 + 7560\right)
e^{c^4} \operatorname{erfc}\!\left(c^2\right)
\Bigg].
\end{aligned}
\end{equation}
Substituting these closed forms, one verifies the decomposition identity Eq.~\eqref{eq:I123-equals-Z} directly against the closed form Eq.~\eqref{eq:Z[p]} at $d=10$, with the signs exactly those assigned in Sec.~\ref{sec:three-extrema}.

\subsection{Asymptotics}

Different gauge choices lead to different UV and IR asymptotics.

\begin{align}
I_\text{max}(p\to\infty)\;&\simeq\; \frac{2580480}{p^{16}}-\frac{278691840}{p^{20}} + \mathcal{O}\!\left(p^{-24}\right) + \frac{384 e^{-\frac{p^4}{4}}}{p^4}+e^{-\frac{3p^4}{4}}\!\left( -6-\frac{150}{p^4}+\mathcal{O}\!\left(p^{-8}\right) \right)\,, \\
I_\text{max}(p\to0)\;&\simeq\; \frac{p^{20}}{15840}-\frac{17 p^{24}}{640640} + \mathcal{O}\!\left(p^{28}\right)\,, \\
I_\text{saddle}(p\to\infty)\;&=\; \frac{384 e^{-\frac{p^4}{4}}}{p^4}\,, \\
I_\text{saddle}(p\to0)\;&\simeq\; \frac{384}{p^4}-96+\mathcal{O}\!\left(p^4\right)\,, \\
I_\text{min}(p\to\infty)\;&\simeq\; e^{-\frac{3p^4}{4}}\!\left(6+\frac{150}{p^4}+\mathcal{O}\!\left(p^{-8}\right)\right)\,, \\
I_\text{min}(p\to0)\;&\simeq\; \frac{18816}{p^4}-\frac{30240 \sqrt{\pi }}{p^2}+\mathcal{O}\!\left(p^0\right)\,.
\end{align}

These asymptotics follow directly from the $d=10$ closed forms; we sketch the mechanism. The saddle copy $I_\text{saddle}(p)=384\,e^{-p^4/4}/p^4$ is purely exponential times a power, so both limits are read off immediately: as $p\to\infty$ it is the exact $384\,e^{-p^4/4}/p^4$ tail, and as $p\to0$ the exponential expands to $1-\tfrac14 p^4+\dots$, giving $384/p^4-96+\mathcal{O}(p^4)$. For $I_\text{max}$ and $I_\text{min}$ the only transcendental pieces are the error functions; everything else is a finite polynomial times $e^{-\frac34 p^4}$, $e^{-\frac14 p^4}$, or a bare polynomial. The two regimes are then controlled by the complementary asymptotics
\begin{equation}\label{eq:erfc-erf-asy}
\operatorname{erfc}(p^2)\;\underset{p\to\infty}{\sim}\;\frac{e^{-p^4}}{\sqrt{\pi}\,p^2}\!\left(1-\frac{1}{2p^4}+\dots\right),
\qquad
\operatorname{erf}(p^2)\;\underset{p\to0}{\sim}\;\frac{2}{\sqrt{\pi}}\!\left(p^2-\frac{p^6}{3}+\dots\right).
\end{equation}

\emph{IR ($p\to\infty$).} In $I_\text{min}$ the $\operatorname{erfc}(p^2)$ piece carries an explicit prefactor $e^{c^4}=e^{p^4}$; using \eqref{eq:erfc-erf-asy} the product $e^{p^4}\operatorname{erfc}(p^2)\sim \tfrac{1}{\sqrt{\pi}\,p^2}(1-\tfrac{1}{2p^4}+\dots)$ is only an inverse power, so the $\sqrt{\pi}$ cancels and $I_\text{min}$ collapses to $e^{-\frac34 p^4}$ times a polynomial: the leading term is $e^{-\frac34 p^4}(6+150/p^4+\dots)$. The same $e^{p^4}[\operatorname{erf}(p^2/2)-\operatorname{erf}(p^2)]$ combination sits in $I_\text{max}$: writing $\operatorname{erf}(p^2/2)-\operatorname{erf}(p^2)=\operatorname{erfc}(p^2)-\operatorname{erfc}(p^2/2)$ and using~\eqref{eq:erfc-erf-asy}, the $\operatorname{erfc}(p^2)$ piece reproduces the $I_\text{min}$-type $e^{-\frac34 p^4}$ tail, while $e^{p^4}\operatorname{erfc}(p^2/2)\sim e^{\frac34 p^4}\times(\text{inverse powers})$ against the $e^{-\frac34 p^4}/p^4$ prefactor leaves the genuine algebraic fall-off. Collecting the surviving polynomial pieces gives the one-loop power law $2580480/p^{16}-278691840/p^{20}+\dots$, together with the subleading $384\,e^{-p^4/4}/p^4$ (the $I_\text{saddle}$-type tail) and the $e^{-\frac34 p^4}$ (the $I_\text{min}$-type tail). Thus the power-law $\mathcal{O}(p^{-2d+4})=\mathcal{O}(p^{-16})$ survives \emph{only} in the maximal copy $I_\text{max}$; the saddle and minimal copies are exponentially suppressed by $e^{-\frac14 p^4}$ and $e^{-\frac34 p^4}$ respectively.

\emph{UV ($p\to0$).} Here $\operatorname{erf}(p^2)\to0$ and $e^{\pm c p^4}\to1$, so $I_\text{min}$ is dominated by its bare polynomial bracket divided by $c^4=p^4$: the constant $8\cdot2352=18816$ gives $18816/p^4$, and the leading $\operatorname{erfc}$-correction $-4\sqrt{\pi}\,p^2\cdot 7560\cdot \operatorname{erfc}(p^2)/p^4 \to -4\cdot 7560\,\sqrt{\pi}/p^2 = -30240\sqrt{\pi}/p^2$ (using $\operatorname{erfc}(p^2)\to1$ as $p\to0$) supplies the next term, matching the expansions above. The $p^{-4}$ pole is therefore shared by $I_\text{saddle}$ and $I_\text{min}$. In $I_\text{max}$, by contrast, the analogous $1/p^4$ contributions cancel among the four brackets order by order (this is the content of the rescaled form Eq.~\eqref{eq:I1-rescaled}, $I_\text{max}\sim p^{2d}$ from $b=ax$, $c=ay$), leaving the finite, positive-power result $p^{20}/15840+\dots=\mathcal{O}(p^{2d})$.

These expansions are stated for $d=10$, where the closed forms are available. The qualitative structure---an algebraic IR fall-off $\mathcal{O}(p^{-2d+4})$ surviving only in the maximal (largest-singular-value) copy, while the saddle and minimal copies decay as $e^{-\kappa p^4}$, and a UV pole carried by the saddle/minimal copies versus a finite $\mathcal{O}(p^{2d})$ in the maximal copy---is dimension-independent: it reflects which integration region keeps the boundary at the diagonal scale (giving powers) versus which pushes it to infinity (giving Gaussian tails), and the rescaling argument $I_\text{max}\sim p^{2d}$ holds for all $d$.


\bibliographystyle{unsrt}
\bibliography{ref}

\end{document}